\documentclass[prd,showpacs,nofootinbib,preprint, 10 pt]{revtex4}
\usepackage{amsmath,graphicx,xcolor,epsfig,slashed}
\usepackage{appendix}
\usepackage{tabularx}
\usepackage{float}
\usepackage{pdfpages}
\usepackage{epstopdf}
\usepackage{subfigure}
\usepackage{caption}
\usepackage{multirow}
\usepackage{hyperref}

\begin{document}
\hfill CERN-TH-2024-148 \\[5mm]
\begin{center}
{\large Polarized and un-polarized $\mathcal{R}_{K^*}$ in and beyond the SM
}\\[5mm]

\setlength {\baselineskip}{0.2in}
{Ishtiaq Ahmed$^{1,2}$, Saba Shafaq$^3$, M. Jamil Aslam$^4$, Saadi Ishaq$^5$}\\[5mm]
$^1$~{\it National Center for Physics, Islamabad 44000, Pakistan.}\\
$^2$~{\it Theoretical Physics Department, CERN, CH-1211 Genève 23, Switzerland.}\\
$^3$~{\it Department of Physics, International Islamic University, Islamabad 44000, Pakistan.}\\
$^4$~{\it Department of Physics, Quaid-i-Azam University, Islamabad 45320, Pakistan.}\\
$^5$~{\it School of Natural Sciences, Department of Physics, National University of Sciences and Technology, Sector H-12, Islamabad, Pakistan.}
\\[5mm]
\end{center}
{\bf Abstract}\\[5mm]
The Standard Model (SM) is lepton flavor universal, and the recent measurements of lepton flavor universality in $B \to (K,K^*)\ell^{+}\ell^{-}$, for $\ell = \mu, \; e$, decays now lie close to the SM predictions. However, this is not the case for the $\tau$ to $\mu$ ratios in these decays, where there is still some window open for the new physics (NP), and to accommodate them various extensions to the SM are proposed. 
It will be interesting to identify some observables which are not only sensitive on the parametric space of such NP models but also have some discriminatory power. We find that the polarization of the $K^{*}$ may play an important role, therefore, we have computed the unpolarized and polarized lepton flavor universality ratios of $\tau$ to $\mu$ in $B\to K^{*}\ell^{+}\ell^{-}$, $\ell= \mu, \tau$ decays. The calculation shows that in most of the cases, the values of the various proposed observables fall within the current experimental sensitivity, and their study at some on going and future experiments will serve as a tool to segregate the variants of the NP models.

\maketitle

\section{Introduction}\label{intro}
Although the predictions of the Standard Model (SM) are consistent with most of the particle-physics data, it is still not enough to explain some important puzzles, \textit{e.g.}, the contents of the dark matter, dark energy, matter-antimatter asymmetry in the universe, hierarchy problem, neutrino oscillations, \textit{etc}. The search of the physics beyond the SM, known as the new physics (NP), is a great challenge for different experimental programs of the high energy physics, where the main challenge is to detect the particle contents of the various NP models. The flavor physics is an ideal platform to explore the NP in an indirect way, where the flavor changing neutral current (FCNC) transitions have their special place. These transitions are forbidden at tree level in the SM, and are loop suppressed due to the Glashow-Iliopoulos-Maiani (GIM) mechanism. Therefore, they are quite sensitive to the particles running in the loop which make them the ideal probe for the NP searches.

A pertinent feature of the SM is the lepton flavor universality (LFU), which states that different generations of leptons interact identically with the SM Higgs boson, 
\textit{i.e.,} they have universal Higgs couplings, and differ only by their masses. However, for the last few years it has come under scrutiny, \textit{e.g.}, the study of the deviation of the LFU ratio $R^{\mu e}_{H}\equiv \mathcal{B}(B\to H\mu^{+}\mu^{-})/\mathcal{B}(B\to He^{+}e^{-})$,  where $H=K,K^{*}, X_{s},\dots$, from SM result, \textit{i.e.,} $R_{H}\approx 1$, triggered a lot of interest in physics beyond the SM \cite{Hiller:2014ula}. Though these decays involve the hadronic contributions arising from the form factors, the dominant theoretical uncertainties from QCD cancel out in the ratio, and the QED uncertainties are controlled to contribute only $1\%$ in $R_{K^\ast}$ predictions. Also, the source of LFUV in SM is through the Higgs to lepton couplings, but these are too small to make any difference to $R_{H}$ \cite{Bordone:2016gaq}.

The observation of $R^{\mu e}_{K^\ast}$ at the LHCb \cite{LHCb:2020lmf, LHCb:2017avl} and Belle \cite{Belle:2019oag} have $2.1 - 2.4\sigma$ deviations from the corresponding SM prediction. A $3.7\sigma$ discrepancy from the SM prediction is observed at LHCb for $R^{\mu e}_{K^*}\equiv \mathcal{B}(B\to K^*\mu^{+}\mu^{-})/\mathcal{B}(B\to K^*e^{+}e^{-})$ in $1.1\leq s \leq 6$ GeV$^2$ bin \cite{LHCb:2021trn}. The experimental measurements of $R^{\mu e}_{K^*}$ in low and central $s$ bins are \cite{LHCb:2022qnv, LHCb:2022vje}:
\begin{eqnarray}
R_{K^*}^{\text{LHCb}} & = & 0.927^{+0.090}_{-0.082}\; (\text{stat}) ^{+0.029}_{-0.027}\;(\text{syst}),\quad\quad 0.045 \leq s \leq 1.1\; \text{GeV}^2\;,\notag\\
R_{K^*}^{\text{LHCb}} & = & 1.027^{+0.072}_{-0.068}\;(\text{stat})^{+0.027}_{-0.026}(\text{syst}),\quad\quad\quad 1.1 \leq s \leq 6\; \text{GeV}^2. \label{LHCb-RKs}
\end{eqnarray}
Though, the current experimental data of $R_{K,\; K^*}^{^{\mu e}}$ is consistent with the corresponding SM predictions \cite{ Bordone:2016gaq, Isidori:2022bzw, Nabeebaccus:2022pje,Isidori:2020acz}, the global analysis of $b\to s \ell^{+}\ell^{-}$ for $\left(\ell = e,\mu\right)$ data  does not rule out the possibility of having the adequate values of LFUV components of the NP couplings in a number of NP scenarios, see \textit{e.g.,} \cite{SinghChundawat:2022ldm, Alguero:2023jeh}. One can expect the similar role by studying the ratios involving $\tau - \mu$, namely, $R_{K,K^{*}}^{\tau\mu}$, where any deviations, surpassing the one arising due to the mass difference of the ratio of $\tau/\mu$, from the SM predictions will hint towards the LFUV involving second and third generation of charged leptons. Ref. \cite{Alok:2023yzg} demonstrate that these ratios can deviate from their SM predictions even when the new physics couplings are universal linking to mass-related effects associated with the involvement of $\tau$ and $\mu$ leptons. Extending the SM weak effective Hamiltonian with new vector and axial-vector couplings, the analysis of full angular distribution of $B\to (K,K^*)\ell^{+}\ell^{-}$, where $\ell =\; \mu,\; \tau$ has been done in Ref. \cite{Alok:2024cyq}, to find the most optimized LFUV observables.


In the case of LFUV ratio involving $\mu - e$, the deviations from the SM were analyzed in the model independent effective field theory (EFT) framework by encoding the short distance physics, both arising due to SM and NP, in the Wilson coefficients of higher dimensional operators \cite{Descotes-Genon:2015uva, Capdevila:2017bsm, Alguero:2018nvb, Alguero:2019ptt, Geng:2021nhg, Altmannshofer:2021qrr,  Hurth:2020ehu, Alok:2019ufo, Ciuchini:2020gvn, Datta:2019zca, Hurth:2020rzx}. This model independent EFT approach provides a useful guide for the construction of NP models that are viable to explain these anomalies.  Ref. \cite{Alguero:2021anc} present an up-to-date complete model-independent global fit analysis by including the recent LHCb measurements
of $R_{K,K_s,K^\ast}$, $B_s \to \phi \mu^{+}\mu^{-}$ and $B_s \to \mu^{+}\mu^{-}$, which now includes 254 observables, superseding their previous analyses 
\cite{Capdevila:2017bsm, Alguero:2018nvb, Alguero:2019ptt}. Their are now two main scenarios:
\begin{itemize}
    \item The Wilson coefficients $C_{9^\prime \mu}$ and $C_{10^\prime \mu}$ which correspond to the right-handed couplings remain a suitable option.
    \item The LFUV left-handed coupling $C_{9 \mu}^{V}\;= - C_{10\; \mu }^V$ accommodates the data better, if the LFU new-physics is allowed in the WC $C_9^{U}$.
    \item It is found that the LFUV observable $Q_5$ help us to distinguish the both types of scenarios. 
\end{itemize}
 
 In $B\to K^{\ast}\ell^{+}\ell^{-}$, $\ell, \tau,\; \mu$ decay, the final state vector meson $K^{\ast}$ could have longitudinal and transverse polarizations, and in this study we explore the LFUV ratio in this decay with a particular polarization of $K^{\ast}$ meson both in the SM and by including the above mentioned NP scenarios. We will see that, together with $Q_5$, these physical observables have the potential to distinguish various scenarios. 

The paper is organized as follows: The weak effective Hamiltonian (WEH) responsible for the $B\to K^\ast\ell^{+}\ell^{-}$ decay is discussed in Section \ref{EH}, whereas Section \ref{PObs} presents the expressions of the polarized LFUV observables in terms of the SM and NP Wilson coefficients. In Section \ref{phAnaly}, we have plotted these observables with the above mentioned NP scenarios and we have also tabulated the numerical values of these observables in various momentum transfer $\left(q^2 \equiv (p_{\ell^{+}}+p_{\ell^-})^2\right)$ bins in the same section. Finally, we conclude the study in the same Section.

\section{Effective Hamiltonian in the Standard Model and Beyond}\label{EH}
The weak effective Hamiltonian for the rare $B$ meson decays can be obtained by integrating out the heavy degrees of freedom, such as the $W$-boson, $t$-quark and the Higgs boson \cite{Grinstein:1987vj, Buchalla:1995vs}. This approach is known as the operator product expansion (OPE), where the short distance effects render in the Wilson coefficients $\mathcal{C}_i$, leaving the operators $\mathcal{O}_i$ describing the physics at a long distance. By implementing this, the weak effective Hamiltonian can be written as :
\begin{eqnarray}
 H_{eff}=-\frac{4 G_{F}}{\sqrt{2}}\lambda_{t}\left[\sum_{i=1}^{6}C_{i}(\mu)O_{i}(\mu)+\sum_{i=7,9,10}C_{i}(\mu)O_{i}(\mu)
+C_{i}^{\prime}(\mu)O_{i}^{\prime}(\mu) \right].\label{H1}
\end{eqnarray}
In Eq. (\ref{H1}) $\lambda_{t}=V_{tb}V_{ts}^{\ast}$ are the CKM matrix elements, $G_{F}$ is the Fermi coupling constant, $C_{i}$ are 
the Wilson coefficients, and $O_{i}$ are the SM with operators with $V-A$ structure. For $B\to K^\ast \ell^{+}\ell^{-}$ decays in the SM, the operators $O_{7,\; 9,\; 10}$ and their corresponding WCs $C_{7,\;, 9\; 10}$ will contribute. These operators have the form 
\begin{eqnarray}
 O_{7} &=&\frac{e}{16\pi ^{2}}m_{b}\left( \bar{s}\sigma _{\mu \nu }P_{R}b\right) F^{\mu \nu }\,,  \notag \\
O_{9} &=&\frac{e^{2}}{16\pi ^{2}}(\bar{s}\gamma _{\mu }P_{L}b)(\bar{\ell}\gamma^{\mu }\ell)\,,  \label{op-form} \\
O_{10} &=&\frac{e^{2}}{16\pi ^{2}}(\bar{s}\gamma _{\mu }P_{L}b)(\bar{\ell} \gamma ^{\mu }\gamma _{5} \ell)\,.  \notag
\end{eqnarray}
Specifically, the operator $O_{7}$ describe the interaction of $b$ and $s$ quarks with the emission of a photon, whereas $O_{9,\; 10}$ correspond to the interaction of these quarks with charged leptons through (almost) same Yukawa couplings.
 
 In Eq. (\ref{H1}), the operators $O_{i}^{\prime}$ are the chirality flipped operators, \textit{i.e.}, with weak interaction structure $V+A$. In the SM, the WCs $C_{9,\; 10}^\prime$ are zero, where as $C_{7}^\prime$ is non-zero - but suppressed by a factor $m_{s}/m_{b}$. In contrast with $O_{9,\; 10}$, the NP operators $O_{9,\; 10}^\prime$ add different contributions to the transitions when final state leptons are the muons or the electrons.
 

The WCs given in Eq.(\ref{H1}) encode the short distance (high momentum) contributions and these are calculated using the perturbative approach. The contributions from current-current,  QCD penguins and chromomagnetic operators $O_{1-6,8}$, \textit{i.e.,} 
\begin{eqnarray}
O_{1}&=&\left(\bar{s}_ic_j\right)_{V-A}\left(\bar{c}_j b_i\right)_{V-A},\;\;\quad\quad\quad\quad\quad O_{2}=\left(\bar{s}_ic_i\right)_{V-A}\left(\bar{c}_j b_j\right)_{V-A},\notag\\
O_{3}&=&\left(\bar{s}_ib_i\right)_{V-A}\sum_q\left(\bar{q}_jq_j\right)_{V-A},\quad\quad\quad\quad O_{4}=\left(\bar{s}_ib_j\right)_{V-A}\sum_q\left(\bar{q}_jq_i\right)_{V-A},\notag\\
O_{5}&=&\left(\bar{s}_ib_i\right)_{V-A}\sum_q\left(\bar{q}_jq_j\right)_{V+A},\quad\quad\quad\quad O_{6}=\left(\bar{s}_ib_j\right)_{V-A}\sum_q\left(\bar{q}_jq_i\right)_{V+A},\notag\\
O_{8}&=&\frac{g_s m_b}{8\pi^2}\bar{s}_i\sigma^{\mu\nu}\left(1+\gamma_5\right)T^a_{ij}b_jG^a_{\mu\nu},\label{O1to8}
\end{eqnarray}
have been unified in the WCs $C_{9}^{\text{eff}}$ and $C_{7}^{\text{eff}}$, and their explicit expressions are given as follows \cite{Beneke:2001at, Greub:2008cy}:
\begin{eqnarray}
 C_{7}^{\text{eff}}(q^{2})=C_{7}-\frac{1}{3}\left(C_{3}+\frac{4}{3}C_{4}+20C_{5}+\frac{80}{3}C_{6}\right)-\frac{\alpha_{s}}{4\pi}\bigg[\left(C_{1}-6C_{2})F^{(7)}_{1,c}(q^{2})+C_{8}F^{7}_{8}(q^{2}\right)\bigg]\notag\\
 C_{9}^{\text{eff}}(q^{2})=C_{9}+\frac{4}{3}\left(C_{3}+\frac{16}{3}C_{5}+\frac{16}{9}C_{6}\right)-h(0,q^{2})\left(\frac{1}{2}C_{3}+\frac{2}{3}C_{4}+8C_{5}+\frac{32}{3}C_{6}\right)\notag\\
-\left(\frac{7}{2}C_{3}+\frac{2}{3}C_{4}+38C_{5}+\frac{32}{3}C_{6}\right)h(m_{b},q^{2})+\left(\frac{4}{3}C_{1}+C_{2}+6C_{3}+60C_{5})h(m_{c},q^{2}\right)\notag\\
-\frac{\alpha_{s}}{4\pi}\bigg[C_{1}F^{(9)}_{1,c}(q^{2})+C_{2}F^{(9)}_{2,c}(q^{2})+C_{8}F^{(9)}_{8}(q^{2})\bigg]\label{WC3}
\end{eqnarray}
The WC given in Eq. (\ref{WC3}) involves the functions $h(m_{q},s)$ with $q=c,b$, functions $F^{7,9}_{8}(q^{2})$, and $F^{(7,9}_{1,c}(q^{2})$ are
defined in \cite{Beneke:2001at, Greub:2008cy,Bharucha:2015bzk,Alok:2017sui}.

The numerical values of Wilson coefficients $C_{i}$ for $i=1,...,10$ at $\mu\sim m_{b}$ scale are presented in Table \ref{wc table}.
\begin{table*}[tbh]
\caption{The Wilson coefficients $C_{i}$ at the scale $\mu\sim m_{b}$ in the SM.}
\begin{tabular}{cccccccccc}
\hline\hline
$C_{1}$&$C_{2}$&$C_{3}$&$C_{4}$&$C_{5}$&$C_{6}$&$C_{7}$&$C_{9}$&$C_{10}$
\\ \hline
  \  -0.263 \  &  \  1.011  \ & \ 0.005 \ &   \ -0.0806   \ &   \ 0.0004  \ &   \ 0.0009   \ &   \ -0.2923  \ &   \ 4.0749  \ &  \ -4.3085 \ \\
\hline\hline
\end{tabular}
\label{wc table}
\end{table*}
The WCs $C^\ell_{(9,10)}$ and $C^\ell_{(9^{\prime},10^{\prime})}$ written in Eq.(\ref{H1}) correspond to the new vector-axial vector operators, which can be expressed as \cite{Bharucha:2015bzk} :
\begin{eqnarray}
C_{(9,10)}^\mu&=&C^{\text{U}}_{(9,10)}+C^{\text{V}}_{(9\mu,10\mu)},\qquad\quad C_{(9,10)}^\tau=C^{\text{U}}_{(9,10)}, \notag \\
C_{(9^{\prime},10^{\prime})}^\mu&=&C^{\text{U}}_{(9^{\prime},10^{\prime})}+C^{\text{V}}_{(9\mu^{\prime},10\mu^{\prime})},\qquad
C_{(9^{\prime},10^{\prime})}^\tau=C^{\text{U}}_{(9^{\prime},10^{\prime})}.\label{WC1}
\end{eqnarray}
The WCs $C^{\text{U}}_{(9,10)}$ and $C^{\text{U}}_{(9^{\prime},10^{\prime})}$ are associated with $b\to s\ell^{+}\ell^{-},(\ell=e,\tau)$ transitions, whereas the WC $C^{\text{V}}_{(9,10)}$ and $C^{\text{V}}_{(9^{\prime},10^{\prime})}$ contributes only to $b\to s\mu^{+}\mu^{-}$ transition.
In ref. \cite{Alguero:2021anc}, a particular NP scenario affecting the muon is considered, and the NP WCs are constrained by fitting the full data set available for all 254 observables, and also by restricting to 24 LFUV observables. In the present study, we consider the values obtained from the full data set and the three prominent 1D NP scenarios and eight $D > 1$ scenarios are summarized in Table \ref{wc table1} and \ref{wc table2}.

\begin{table}[hbt!]
\caption{Allowed NP solutions assuming NP couplings to be universal. Here $\Delta \chi^2=\chi^2_{SM}-\chi^2_{bf}$ where $\chi^2_{bf}$ is the $\chi^2$ at the best fit point and $\chi^2_{SM}$ corresponds to SM which is $\chi^2_{SM}\sim 184$ \cite{Alguero:2021anc}.}
\begin{center}
\begin{tabular}{ |c|c|c|c| } 
\hline\hline
Scenarios & WCs & 1$\sigma$ range & $\Delta\chi^2$ \\
\hline
S-I & $C_9^U$ & -1.165 $\pm$ 0.165 & 33.64\\ 
\hline
S-II & $C_9^U=-C_{10}^U$ & -0.50 $\pm$ 0.12 & 18.85\\ 
\hline
S-III & $C_9^U=-C^{ U}_{9\prime}$ & -0.88 $\pm$ 0.16& 26.92\\ 
\hline\hline
\end{tabular}
\label{wc table1}
\end{center}
\end{table}

\begin{table}[hbt!]
\caption{Allowed LFUV NP solutions \cite{Alguero:2023jeh} where V-VIII are introduced in ref. \cite{Alguero:2018nvb}, X, XI, and XIII are appeared in some Z-prime models ref. \cite{Alguero:2019ptt} while IX is motivated by two higgs doublet models ref. \cite{Crivellin:2019}.  
}
\begin{center}
\begin{tabular}{ |c|c|c|c||c|c|c|c| } 
\hline\hline
Scenarios & WCs & 1$\sigma$ range & pull & Scenarios & WCs & 1$\sigma$ range & pull \\
\hline
\multirow{3}{*}{S-V} & $C_{9\mu}^V$ & [-1.43, -0.61] &  & \multirow{3}{*}{S-IX} & \multirow{2}{*}{$C_{9\mu}^V = -C_{10\mu}^V$} & \multirow{2}{*}{[-0.29, -0.13]} &\\ 
& $C_{10\mu}^V$ & [-0.75, 0.00] & 5.8& &\multirow{2}{*}{$C_{10}^U$}  & \multirow{2}{*}{[-0.23, 0.11]} & 2.7\\
& $C_9^U=C_{10}^U$ & [-0.16, 0.58]& & & & &\\
\hline \multirow{2}{*}{S-VI} & $C_{9\mu}^V = -C_{10\mu}^V$ & [-0.34, -0.20] & \multirow{2}{*}{4.1} & \multirow{2}{*}{S-X} & $C_{9\mu}^V$ & [-0.81, -0.50] & \multirow{2}{*}{4.1}\\ 
& $C_9^U=C_{10}^U$ & [-0.53, -0.29]&  & & $C_{10}^U$ & [0.08, 0.18]& \\
\hline
\multirow{2}{*}{S-VII} & $C_{9\mu}^V$ & [-0.39, -0.02]& \multirow{2}{*}{4.0} & \multirow{2}{*}{S-XI} & $C_{9\mu}^V$ & [-0.84, -0.52]& \multirow{2}{*}{4.1}\\ 
& $C_{9}^U$ & [-1.21, -0.72] & & & $C_{10\prime}^{ U}$ & [-0.15, -0.09]&\\
\hline
\multirow{4}{*}{S-VIII} & \multirow{2}{*}{$C_{9\mu}^V = -C_{10\mu}^V$} & \multirow{2}{*}{[-0.14, -0.02]}&\multirow{4}{*}{5.6} & \multirow{4}{*}{S-XIII} & $C_{9\mu}^V$ & [-0.97, -0.60] & \multirow{4}{*}{3.8} \\ 
& \multirow{2}{*}{$C_{9}^U$}  & \multirow{2}{*}{[-1.27, -0.91]}& & & $C_{9\prime\mu}^{ V}$ & [0.10, 0.57] & \\
 & & & & & $C_{10}^U$ & [-0.04, 0.26] &\\
 & & & & &  $C_{10\prime}^{ U}$ & [-0.03, 0.30]& \\
\hline\hline
\end{tabular}
\label{wc table2}
\end{center}
\end{table}

\section{Physical Observables}\label{PObs}
In Section \ref{intro}, we emphasized that the LFU, defined by
\begin{equation}
R_{K^{\ast}}^{\mu e} = \frac{\mathcal{B}\left(B\to K^{\ast}\mu^{+}\mu^{-}\right)}{\mathcal{B}\left(B\to K^{\ast}e^{+}e^{-}\right)}, \label{LFU}
\end{equation}
is an important observable in establishing the NP present in these FCNC decays. The purpose here is to see if there exist some observables that could be used to segregate the effects of above mentioned NP scenarios. To do so, we considered the polarization of  the final state meson and defined the following ratios :
\begin{equation}
R^{\tau\ell}=\frac{\mathcal{B}(B\to M\tau^+\tau^-)}{\mathcal{B}(B\to M\ell^+\ell^-)}\qquad R^{\tau}=\frac{\mathcal{B}(B\to M\tau^+\tau^-)}{\mathcal{B}(B\to M^\prime\tau^+\tau^-)}\label{eqnRR},
\end{equation}
where $M=K^*,K^*_L,K^*_T$, with the subscript $L\left(T\right)$ designating the longitudinal(transverse) polarization of the final state vector meson. In $R^{\tau}$ the meson in the numerator and denominator have difference in their polarizations. It is important to emphasize that the ratios remain the same whether the lighter lepton is an electron or muon. However, we have present here the results by consider the light lepton to be the muon. The different combinations will give
\begin{eqnarray}
\mathcal{R}_{K^*}^{\tau\mu}&=&\frac{\mathcal{B}(B\to K^*\tau^+\tau^-)}{\mathcal{B}(B\to K^*\mu^+\mu^-)};\quad\mathcal{R}_{K^{*\tau}_L}^{\tau\mu}=\frac{\mathcal{B}(B\to K^*_L\tau^+\tau^-)}{\mathcal{B}(B\to K^*\mu^+\mu^-)}; \quad 
\mathcal{R}_{K^{*\tau}_T}^{\tau\mu}=\frac{\mathcal{B}(B\to K^*_T\tau^+\tau^-)}{\mathcal{B}(B\to K^*\mu^+\mu^-)};\notag \\
\mathcal{R}_{K^{*\mu}_L}^{\tau\mu}&=&\frac{\mathcal{B}(B\to K^*\tau^+\tau^-)}{\mathcal{B}(B\to K^*_L\mu^+\mu^-)};\quad \mathcal{R}_{K^{*\mu}_T}^{\tau\mu}=\frac{\mathcal{B}(B\to K^*\tau^+\tau^-)}{\mathcal{B}(B\to K^*_T\mu^+\mu^-)}; \quad 
  \mathcal{R}_{K^*_{LL}}^{\tau\mu}=\frac{\mathcal{B}(B\to K^*_L\tau^+\tau^-)}{\mathcal{B}(B\to K^*_L\mu^+\mu^-)};\notag \\
\mathcal{R}_{K^*_{TT}}^{\tau\mu}&=&\frac{\mathcal{B}(B\to K^*_T\tau^+\tau^-)}{\mathcal{B}(B\to K^*_T\mu^+\mu^-)};\quad \mathcal{R}_{K^*_{LT}}^{\tau\mu}=\frac{\mathcal{B}(B\to K^*_L\tau^+\tau^-)}{\mathcal{B}(B\to K^*_T\mu^+\mu^-)}; \quad \mathcal{R}_{K^*_{TL}}^{\tau\mu}=\frac{\mathcal{B}(B\to K^*_T\tau^+\tau^-)}{\mathcal{B}(B\to K^*_L\mu^+\mu^-)}; \notag\\
\mathcal{R}_{K^{*}_L}^{\tau}&=&\frac{\mathcal{B}(B\to K^*_L\tau^+\tau^-)}{\mathcal{B}(B\to K^*\tau^+\tau^-)};\quad \mathcal{R}_{K^{*}_{T}}^{\tau}=\frac{\mathcal{B}(B\to K^*_T\tau^+\tau^-)}{\mathcal{B}(B\to K^*\tau^+\tau^-)};\quad \mathcal{R}_{K^{*}_{LT}}^{\tau}=\frac{\mathcal{B}(B\to K^*_L\tau^+\tau^-)}{\mathcal{B}(B\to K^*_T\tau^+\tau^-)}.\label{ratios1}
\end{eqnarray}
Using the four-fold decay distribution in $B\to K^{*}\tau^{+}\tau^{-}$ the observables, $\mathcal{R}_{K^*}^{\tau\mu}$, $\mathcal{R}_{K^*_L}^\tau$ (longitudinal helicity fraction), and $\mathcal{R}_{K^*_T}^\tau$ (transverse helicity fraction) defined in Eq. (\ref{ratios1}) are studied phenomenologically in \cite{SinghChundawat:2022ldm, Alok:2023yzg, Alok:2024cyq}. However, the other observables associated with the polarized final state meson are new and interesting to explore theoretically and experimentally.

The SM values of the branching ratios, $\mathcal{B}(B\to (K^*,K^*_{L,T})\ell^+\ell^-)$, appear in the above ratios can be written as
\begin{eqnarray}
\mathcal{B}\left(B\to \left(K^*,\; K^{*}_{L,\; T}\right)\ell^{+}\ell^{-}\right)= \frac{G_F^2 \left|V_{tb}V^*_{ts}\right|^2}{2^{11} \pi^{5}m_B^3} \mathcal{D}_{j}^{\ell}\left(s\right), \label{eqn}
\end{eqnarray}
where $j$ represents $K^*,\; K^*_{L},\; K^*_{T}$ and $\ell = \mu,\tau$. The values of $\mathcal{D}_{j}^{\ell}\left(s\right)$, in the $14\leq s\leq s_{max} \text{GeV}^2$ bin, are appended in Table \ref{tableIV}. By using these values one can easily find the SM values of the ratios given in Eq. (\ref{ratios1}), and these are tabulated in Table \ref{SM ratios}. 

\begin{table*}[h]
\caption{The SM values of $\mathcal{D}_j^\ell\times10^{4}$ appear in Eq. (\ref{eqn}), in the $14\leq s\leq s_{max} \text{GeV}^2$ bin.}
\scalebox{1.2}{
\renewcommand{\arraystretch}{1.5}
\begin{tabular}{|c|c|c|c|c|c|}
\hline\hline
 $\mathcal{D}_{K^*}^\mu$&$\mathcal{D}_{K^*_L}^\mu$& $\mathcal{D}_{K^*_T}^\mu$&$\mathcal{D}_{K^*}^\tau$&$\mathcal{D}_{K^*_L}^\tau$&$\mathcal{D}_{K^*_T}^\tau$
\\ &&&&&\\[-1.5em] \hline
    $78.19^{+11.79}_{-10.81}$  &     $29.15^{+6.74}_{-6.04}$   &  $49.03^{+5.05}_{-4.77}$  &  $32.22^{+6.12}_{-5.38}$   &  $14.01^{+3.72}_{-4.37}$      & $18.21^{+1.75}_{-1.66}$  \\
\hline\hline
\end{tabular}}
\label{tableIV}
\end{table*}

 \begin{table*}[tbh]
\caption{The SM values of $\mathcal{R}_i^{\tau\mu(\tau)}$ in $14\leq s\leq s_{max} \text{GeV}^2$ bin,  where 
$\mathcal{R}_{K^{*}_L}^{\tau SM}+\mathcal{R}_{K^{*}_T}^{\tau SM}=1$.}
\scalebox{1.2}{
\renewcommand{\arraystretch}{1.5}
\begin{tabular}{|c|c|c|c|c|c|}
\hline\hline
$\mathcal{R}_{K^*}^{\tau\mu SM}$&$\mathcal{R}_{K_L^{*\tau}}^{\tau\mu SM}$& $\mathcal{R}_{K_T^{*\tau}}^{\tau\mu SM}$&$\mathcal{R}_{K_L^{*\mu}}^{\tau\mu SM}$&$\mathcal{R}_{K_T^{*\mu}}^{\tau\mu SM}$&$\mathcal{R}_{K_{LL}^*}^{\tau\mu SM}$
\\ &&&&&\\[-1.5em] \hline
    $0.412\pm0.014$&$0.179^{+0.025}_{-0.026}$&$0.233^{-0.011}_{+0.013}$&$1.105^{-0.037}_{+0.056}$&$0.657^{+0.052}_{-0.051}$&$0.480^{+0.031}_{-0.035}$ \\ \hline
   $\mathcal{R}_{K_{TT}^*}^{\tau\mu SM}$&$\mathcal{R}_{K_{LT}^*}^{\tau\mu SM}$&$\mathcal{R}_{K_{TL}^*}^{\tau\mu SM}$&$\mathcal{R}_{K^*_L}^{\tau SM}$&$\mathcal{R}_{K^*_T}^{\tau SM}$&$\mathcal{R}_{K^*_{TL}}^{\tau SM}$
\\ &&&&&\\[-1.5em] \hline
    $0.371^{-0.002}_{+0.003}$&$0.625^{-0.069}_{+0.091}$&$0.286^{+0.054}_{-0.053}$&$0.435^{+0.044}_{-0.051}$&$0.565^{-0.044}_{+0.051}$&$0.769^{0.151}_{0.147}$    \\
\hline\hline
\end{tabular}}
\label{SM ratios}
\end{table*}

 As we are dealing with the NP scenarios that affect the muon and tau modes, therefore, the unpolarized and polarized branching ratios, after addition of NP, in terms of the new WCs can be written as
\begin{eqnarray}
\mathcal{B}\left(B\to \left(K^*,\; K^{*}_{L,\; T}\right)\mu^{+}\mu^{-}(\tau^{+}\tau^{-})\right)= \frac{G_F^2 \left|V_{tb}V^*_{ts}\right|^2}{2^{11} \pi^{5}m_B^3} \mathcal{N}_{j}^{\mu(\tau)}\left(s\right). \label{eqnn}
\end{eqnarray}
The expressions of  $\mathcal{N}_{K^*}^{\mu(\tau)}$, $\mathcal{N}_{K^*_L}^{\mu(\tau)}$ and $\mathcal{N}_{K^*_T}^{\mu(\tau)}$, after integration over $s$ in the $14\leq s\leq s_{max} \text{GeV}^2$ bin, can be written in terms of new WCs which are given in Eq. (\ref{Nregion3}). Writing

\begin{eqnarray}\label{Nregion3}
\mathcal{N}_{K^*}^\mu&\simeq&\mathcal{D}_{K^*}^\mu+\mathcal{N}_{K^*}^{\prime\mu},\qquad\mathcal{N}_{K^*_L}^\mu\simeq\mathcal{D}_{K^*_L}^\mu+\mathcal{N}_{K^*_L}^{\prime\mu},\qquad\mathcal{N}_{K^*_T}^\mu\simeq\mathcal{D}_{K^*_T}^\mu+\mathcal{N}_{K^*_T}^{\prime\mu},\notag \\
\mathcal{N}_{K^*}^\tau&\simeq&\mathcal{D}_{K^*}^\tau+\mathcal{N}_{K^*}^{\prime\tau},\qquad\mathcal{N}_{K^*_L}^\tau\simeq\mathcal{D}_{K^*_L}^\tau+\mathcal{N}_{K^*_L}^{\prime\tau},\qquad\mathcal{N}_{K^*_T}^\tau\simeq\mathcal{D}_{K^*_T}^\tau+\mathcal{N}_{K^*_T}^{\prime\tau},\label{eq12}
\end{eqnarray}
with 
\begin{eqnarray}\label{Nregion3}
\mathcal{N}_{K^*}^{\prime\mu}&\simeq&\big[2.7 C_{10'}^{\mu2}-4. C_{10}^{\mu } C_{10'}^{\mu }+17. C_{10'}^{\mu }+2.7 C_{10}^{\mu2}+22. C_7^{\mu2}+2.7 C_{9'}^{\mu2}+2.7 C_9^{\mu2}-23. C_{10}^{\mu
   }-38. C_7^{\mu }\notag \\
   &&+15. C_7^{\mu } C_{9'}^{\mu }-14. C_{9'}^{\mu }-11. C_7^{\mu } C_9^{\mu }-4. C_{9'}^{\mu } C_9^{\mu }+19. C_9^{\mu }\big]\times10^{4},\notag \\
\mathcal{N}_{K^*_L}^{\prime\mu}&\simeq&\big[0.95 C_{10'}^{\mu2}-1.9 C_{10}^{\mu } C_{10'}^{\mu }+8.2 C_{10'}^{\mu }+0.95 C_{10}^{\mu2}+8.6 C_7^{\mu2}+0.95 C_{9'}^{\mu2}+0.95 C_9^{\mu2}-8.2
   C_{10}^{\mu }\notag \\
   &&-20. C_7^{\mu }+5.7 C_7^{\mu } C_{9'}^{\mu }-6.5 C_{9'}^{\mu }-5.7 C_7^{\mu } C_9^{\mu }-1.9 C_{9'}^{\mu } C_9^{\mu }+6.5 C_9^{\mu }\big]\times10^{4},\notag \\
\mathcal{N}_{K^*_T}^{\prime\mu}&\simeq&\big[1.4 C_{10'}^{\mu2}-2.8 C_{10}^{\mu } C_{10'}^{\mu }+12. C_{10'}^{\mu
   }+1.4 C_{10}^{\mu2}+13. C_7^{\mu2}+1.8 C_{9'}^{\mu2}+1.8
   C_9^{\mu2}-12. C_{10}^{\mu }\notag \\
   &&-18. C_7^{\mu }+9.6 C_7^{\mu }
   C_{9'}^{\mu }-7.4 C_{9'}^{\mu }-5.1 C_7^{\mu } C_9^{\mu }-2.1
   C_{9'}^{\mu } C_9^{\mu }+12. C_9^{\mu }\big]\times10^{4}\notag,\\
\mathcal{N}_{K^*}^{\prime\tau}&\simeq&\big[0.52 C_{10'}^{\tau2}-1.05 C_{10}^{\tau } C_{10'}^{\tau }+4.48 C_{10'}^{\tau }+0.52 C_{10}^{\tau2}+14.40 C_7^{\tau2}+1.80 C_{9'}^{\tau2}+1.80 C_9^{\tau2}-45.31 C_{10}^{\tau
   }\notag \\
   &&-25.53 C_7^{\tau }+10.14 C_7^{\tau } C_{9'}^{\tau }-9.44 C_{9'}^{\tau }-7.31 C_7^{\tau } C_9^{\tau }-2.70 C_{9'}^{\tau } C_9^{\tau }+12.51 C_9^{\tau }\big]\times10^{4},\notag \\
\mathcal{N}_{K^*_L}^{\prime\tau}&\simeq&\big[0.35 C_{10'}^{\tau2}-0.69 C_{10}^{\tau } C_{10'}^{\tau }+2.98 C_{10'}^{\tau }+0.35 C_{10}^{\tau2}+5.71 C_7^{\tau2}-4.31 C_{9'}^{\tau2}+4.31 C_9^{\tau2}-2.98
   C_{10}^{\tau }\notag \\
   &&-12.96 C_7^{\tau }+3.80 C_7^{\tau } C_{9'}^{\tau }-4.31 C_{9'}^{\tau }-3.80 C_7^{\tau } C_9^{\tau }-1.26 C_{9'}^{\tau } C_9^{\tau }+4.31 C_9^{\tau }\big]\times10^{4},\notag \\
\mathcal{N}_{K^*_T}^{\prime\tau}&\simeq&\big[1.78 C_{10'}^{\tau2}-0.36 C_{10}^{\tau } C_{10'}^{\tau }+1.51 C_{10'}^{\tau
   }+1.78 C_{10}^{\tau2}+8.70 C_7^{\tau2}+1.17 C_{9'}^{\tau2}+1.17
   C_9^{\tau2}-15.54 C_{10}^{\tau }\notag \\
   &&-12.57 C_7^{\tau }+6.34 C_7^{\tau }
   C_{9'}^{\tau }-5.12 C_{9'}^{\tau }-3.51 C_7^{\tau } C_9^{\tau }-1.43
   C_{9'}^{\tau } C_9^{\tau }+8.20 C_9^{\tau }\big]\times10^{4}\label{eq13},
\end{eqnarray}
and $\mathcal{D}^{\ell}_j$ is given in Table \ref{tableIV}. 
Using Eqs. (\ref{eq12}) and (\ref{eq13}) and with some manipulation we are able to rewrite the LFVU ratios in terms of two  components given as follows:
\begin{eqnarray}
\mathcal{R}_{K^*}^{\tau\mu}&=&\big[\mathcal{R}_{K^*}^{\tau\mu SM}+\mathcal{R}_{K^*}^{\tau\mu \prime}\big];\quad\mathcal{R}_{K^{*\tau}_L}^{\tau\mu}=\big[\mathcal{R}_{K^{*\tau}_L}^{\tau\mu SM}+\mathcal{R}_{K^{*\tau}_L}^{\tau\mu \prime}\big];
\quad \mathcal{R}_{K^{*\tau}_T}^{\tau\mu}=\big[\mathcal{R}_{K^{*\tau}_T}^{\tau\mu SM}+\mathcal{R}_{K^{*\tau}_T}^{\tau\mu \prime}\big];\notag \\
\mathcal{R}^{\tau\mu}_{K^{*\mu}_L}&=&\big[\mathcal{R}_{K^{*\mu}_L}^{\tau\mu SM}+\mathcal{R}_{K^{*\mu}_L}^{\tau\mu \prime}\big];\quad \mathcal{R}^{\tau\mu}_{K^{*\mu}_T}=\big[\mathcal{R}_{K^{*\mu}_T}^{\tau\mu SM}+\mathcal{R}_{K^{*\mu}_T}^{\tau\mu \prime}\big];
\quad\mathcal{R}_{K^*_{LL}}^{\tau\mu}=\big[\mathcal{R}_{K^*_{LL}}^{\tau\mu SM}+\mathcal{R}_{K^*_{LL}}^{\tau\mu \prime}\big];
\notag \\
\mathcal{R}^{\tau\mu}_{K^*_{TT}}&=&\big[\mathcal{R}^{\tau\mu SM}_{K^*_{TT}}+\mathcal{R}^{\tau\mu \prime}_{K^*_{TT}}\big];\quad \mathcal{R}^{\tau\mu}_{K^*_{LT}}=\big[\mathcal{R}_{K^*_{LT}}^{\tau\mu SM}+\mathcal{R}_{K^*_{LT}}^{\tau\mu \prime}\big]; 
\quad
\mathcal{R}_{K^*_{TL}}^{\tau\mu}=\big[\mathcal{R}_{K^*_{TL}}^{\tau\mu SM}+\mathcal{R}_{K^*_{TL}}^{\tau\mu \prime}\big];\notag \\
\mathcal{R}_{K^*_{L}}^{\tau }&=&\big[\mathcal{R}_{K^*_{L}}^{\tau SM}+\mathcal{R}_{K^*_{L}}^{\tau \prime}\big];\quad\quad
\mathcal{R}_{K^*_{T}}^{\tau }=\big[\mathcal{R}_{K^*_{T}}^{\tau SM}+\mathcal{R}_{K^*_{T}}^{\tau \prime}\big];\quad\quad
\mathcal{R}_{K^*_{TL}}^{\tau }=\big[\mathcal{R}_{K^*_{TL}}^{\tau SM}+\mathcal{R}_{K^*_{TL}}^{\tau \prime}\big].\label{ratios}
\end{eqnarray}
These expressions are written in such a way that, the first term has purely the SM contributions, whereas the second term encapsulate the NP as well as SM contribution given as $\mathcal{R}_i^{\tau\mu \prime}$ and $\mathcal{R}_i^{\tau \prime}$, where $i=K^*$, $K^*_L$, $K^*_T$, $K^*_{LT}$, $K^*_{TL}$, $K^*_{LL}$, $K^*_{TT}$. The later contributions to the LFUV ratios, appearing in Eq. (\ref{ratios}) becomes :
\begin{equation}
    \mathcal{R}_{K^*_{(\alpha)}}^{\tau\mu(\tau) NP}=\frac{\mathcal{D}_{{K^*_{(\alpha)}}}^\mu \mathcal{N}_{{K^*_{(\alpha)}}}^{\prime\tau}-\mathcal{D}_{{K^*_{(\alpha)}}}^\tau \mathcal{N}_{{K^*_{(\alpha)}}}^{\prime\mu}}{\mathcal{D}_{{K^*_{(\alpha)}}}^\mu(\mathcal{D}_{{K^*_{(\alpha)}}}^\mu+\mathcal{N}_{{K^*_{(\alpha)}}}^{\prime\mu})}; \qquad \mathcal{R}_{K^*_{(\alpha\beta)}}^{\tau\mu(\tau) NP}=\frac{\mathcal{D}_{{K^*_{(\beta)}}}^\mu \mathcal{N}_{{K^*_{(\alpha)}}}^{\prime\tau}-\mathcal{D}_{{K^*_{(\alpha)}}}^\tau \mathcal{N}_{{K^*_{(\alpha)}}}^{\prime\mu}}{\mathcal{D}_{{K^*_{(\beta)}}}^\mu(\mathcal{D}_{{K^*_{(\beta)}}}^\mu+\mathcal{N}_{{K^*_{(\beta)}}}^{\prime\mu})}\qquad\alpha,\beta=L, T
\end{equation}

There are three prominent one-dimensional scenarios: (i) S-I: $C_9^\mu$, (ii) S-II: $C_9^\mu$= $-C_{10}^\mu$ and (iii)  S-III: $C_9^\mu$= $-C_{9^\prime}^\mu$, where all the other WCs are set to be zero. The 1 $\sigma$ ranges of the new WCs in these scenarios are given in Table \ref{wc table}. For these scenarios, $C_{(9^{(\prime)},10^{(\prime)})}^V=0$, so from Eq. (\ref{WC1}), one can notice that $C_{(9^{(\prime)},10^{(\prime)})}^\mu=C_{9^{(\prime)},10^{(\prime)}}^\tau$. Therefore, the set of expression, $\mathcal{N}_{i}^{\mu(\tau)}$, given in Eq. (\ref{eq13}), for 1D scenarios can be written in the following general form by defining  $C_9^\mu\equiv C_9^U$
\begin{eqnarray}
\mathcal{N}_{j}^{\prime\mu(\tau)}&=&\mathcal{A}C_{9}^{U}+\mathcal{B}(C_{9}^{U})^2.\label{1DGEQ}
\end{eqnarray}

The coefficients $\mathcal{A}$ and $\mathcal{B}$ contain the contribution from SM WCs, and the form factors. Using the numerical values of the various inputs parameters, these are calculated in Table \ref{TableV}. Here, we have also included the uncertainties coming through the form factors and other input parameters. 
\begin{table*}[tbh]
\caption{\label{tab:table3}The numerical values of $\mathcal{A}$ and $\mathcal{B}$ for 1D NP scenarios appered in Eq. (\ref{1DGEQ}).}
\renewcommand{\arraystretch}{1}
\centering
\scalebox{1.1}{
\begin{tabular}{c|ccc|ccc}
\hline\hline &&&&&&\\[-0.9em]\\[-0.9em]
 \multirow{3}{*}{ $\mathcal{N}_j$} &\multicolumn{3}{c|}{ $\mathcal{A}\times10^{4}$} &  \multicolumn{3}{c}{ $\mathcal{B}\times10^{4}$}  \\
  &&&&&&\\[-0.9em]\cline{2-7}\\[-0.8em]
  &S-I    &S-II &S-III    &S-I    &S-II &S-III    \\ 
 &&&&&&\\[-0.9em]\cline{1-7}\\[-0.9em]
 $\mathcal{N}_{K^*}^\mu$  &\hspace{0.5cm} $18.77^{+2.72}_{-2.50}$\hspace{0.5cm}    &$39.23^{+5.99}_{-5.50}$\hspace{0.5cm}  & $32.70^{+5.26}_{-4.82}$\hspace{0.5cm} &\hspace{0.5cm}$2.71^{+0.41}_{-0.37}$\hspace{0.5cm} &    $5.06^{+0.78}_{-0.72}$\hspace{0.5cm}  &$9.41^{+1.52}_{-1.39}$    \\
&&&&&&\\[-0.9em]\cline{1-7}\\[-0.9em]
 $\mathcal{N}_{K_L^*}^\mu$ &\hspace{0.5cm} $6.51^{+1.51}_{-1.35}$\hspace{0.5cm} & $14.72^{+3.38}_{-3.03}$\hspace{0.5cm} &   $13.02^{+3.02}_{-2.70}$\hspace{0.5cm} &\hspace{0.5cm}$0.95^{+0.22}_{-0.19}$\hspace{0.5cm} &$1.91^{+0.43}_{-0.39}$\hspace{0.5cm}  & $3.82^{+0.87}_{-0.78}$     \\
&&&&&&\\[-0.9em]\cline{1-7}\\[-0.9em]
  $\mathcal{N}_{K^*_T}^\mu$ &\hspace{0.5cm} $12.26^{+1.21}_{-1.15}$\hspace{0.5cm}     &$24.51^{+2.61}_{-2.47}$\hspace{0.5cm}   &$19.68^{+2.24}_{-2.12}$\hspace{0.5cm}     &\hspace{0.5cm}$1.76^{+0.19}_{-0.18}$\hspace{0.5cm}     &$3.15^{+0.35}_{-0.33}$\hspace{0.5cm}   &$5.59^{+0.65}_{-0.61}$     \\
&&&&&&\\[-0.9em]\cline{1-7}\\[-0.9em]
 $\mathcal{N}_{K^*}^\tau$ &\hspace{0.5cm}$12.51^{+1.83}_{-1.68}$\hspace{0.5cm}    &$17.04^{+3.19}_{-2.82}$\hspace{0.5cm}   &$21.94^{+3.53}_{-3.24}$\hspace{0.5cm}   &\hspace{0.5cm}$1.80^{+0.27}_{-0.25}$\hspace{0.5cm}     &$2.33^{+0.43}_{-0.38}$\hspace{0.5cm}   &$6.31^{+1.02}_{-0.93}$    \\
&&&&&&\\[-0.9em]\cline{1-7}\\[-0.9em]
   $\mathcal{N}_{K_L^*}^\tau$ &\hspace{0.5cm}$4.31^{+1.01}_{-0.91}$\hspace{0.5cm} & $7.29^{+2.20}_{-1.88}$\hspace{0.5cm}  &  $8.62^{+2.03}_{-1.81}$\hspace{0.5cm}   &\hspace{0.5cm}$0.63^{+0.15}_{-0.13}$\hspace{0.5cm}    &$0.98^{+0.28}_{-0.24}$\hspace{0.5cm}   &$2.53^{+0.58}_{-0.52}$    \\
  &&&&&&\\[-0.9em]\cline{1-7}\\[-0.9em]
   $\mathcal{N}_{K^*_T}^\tau$ & \hspace{0.5cm} $8.20^{+0.82}_{-0.78}$\hspace{0.5cm}   & $9.75^{+0.99}_{-0.94}$\hspace{0.5cm}   &$13.32^{+1.51}_{-1.43}$\hspace{0.5cm}   &\hspace{0.5cm}$1.17^{+0.13}_{-0.12}$\hspace{0.5cm}    & $1.35^{+0.15}_{-0.14}$\hspace{0.5cm}   &$3.78^{+0.43}_{-0.41}$      \\ \hline \hline   
\end{tabular}}\label{TableV}
\end{table*}

Similarly, for scenarios $D>1$, one can express $\mathcal{N}_{j}^{\prime\tau}$ in terms of the NP WCs as
\begin{eqnarray}
\mathcal{N}_{j}^{\prime\tau}&=&\mathcal{A}^\tau C_{XX}^{\tau}+\mathcal{B}^\tau(C_{XX}^{\tau})^2,\label{2DGEQ}
\end{eqnarray}
where $C_{XX}^\tau=C_{9^\prime}^\tau\equiv C_9^U$ for scenarios V, VI, VII, and VIII, $C_{XX}^\tau=C_{10}^\tau\equiv C_{10}^U$ for S-IX, S-X, $C_{XX}^\tau=C_{10^\prime}^\tau\equiv C_{10^\prime}^{U}$ with $\mathcal{A}^\tau=-\mathcal{A}^\tau$  for S-XI, and $C_{XX}^\tau=C_{10^\prime}^{U}-C_{10}^{U}$ with $\mathcal{A}^\tau=-\mathcal{A}^\tau$  for S-XIII.  The numerical values of $\mathcal{A}^\tau$ and $\mathcal{B}^\tau$ are given in Table \ref{Table2DAB}. 
\begin{table*}[tbh]
\caption{\label{tab:table3}The numerical values of $\mathcal{A}^\tau$ and $\mathcal{B}^\tau$ for $D>1$ NP scenarios appered in Eq. (\ref{2DGEQ}).}
\renewcommand{\arraystretch}{1}
\centering
\scalebox{1.2}{
\begin{tabular}{c|ccc|ccc}
\hline\hline &&&&&&\\[-0.9em]\\[-0.9em]
 \multirow{3}{*}{ $\mathcal{N}_j$} &\multicolumn{3}{c|}{$\mathcal{A}^\tau\times10^{4}$}&\multicolumn{3}{c}{ $\mathcal{B}^\tau\times10^{4}$}  \\
  &&&&&&\\[-1.15em]\cline{2-7}\\[-0.9em]
  &V \& VI&VII \& VIII&IX  \& XIII \& X&V \& VI&VII \& VIII &IX \& X \& XIII   \\
&&&&&&\\[-1.2em]\cline{1-7}\\[-0.9em]
 $\mathcal{N}_{K^*}^\tau$ &\hspace{0.5cm}$7.98^{+0.48}_{-0.55}$\hspace{0.5cm}   &$12.51^{+1.83}_{-1.61}$\hspace{0.5cm} &$-4.53^{+1.14}_{-1.36}$\hspace{0.5cm} &\hspace{0.5cm}$2.33^{+0.43}_{-0.38}$\hspace{0.5cm}  &$1.80^{+0.27}_{-0.25}$\hspace{0.5cm} &$0.52^{+0.16}_{-0.13}$\hspace{0.5cm}   \\
&&&&&&\\[-0.9em]\cline{1-7}\\[-0.9em]
   $\mathcal{N}_{K_L^*}^\tau$ &\hspace{0.5cm}$1.33^{+0.17}_{-0.07}$\hspace{0.5cm} &  $4.31^{+1.01}_{-0.91}$\hspace{0.5cm} &$-2.98^{+0.97}_{-1.18}$\hspace{0.5cm} & \hspace{0.5cm}$0.98^{+0.28}_{-0.24}$\hspace{0.5cm}    &$0.63^{+0.15}_{-0.13}$\hspace{0.5cm}     &$0.35^{+0.14}_{-0.11}$\hspace{0.5cm} \\
  &&&&&&\\[-0.9em]\cline{1-7}\\[-0.9em]
   $\mathcal{N}_{K^*_T}^\tau$ & \hspace{0.5cm} $6.64^{+0.64}_{-0.61}$\hspace{0.5cm} &$8.20^{+0.82}_{-0.78}$\hspace{0.5cm}  & $-1.55^{+0.17}_{-0.18}$\hspace{0.5cm}     & \hspace{0.5cm}$1.35^{+0.15}_{-0.14}$\hspace{0.5cm}    &$1.17^{+0.13}_{-0.12}$\hspace{0.5cm}      &$0.18^{+0.02}_{-0.019}$\hspace{0.5cm}      \\ \hline \hline   
\end{tabular}}\label{Table2DAB}
\end{table*}
Similarly, for $\mathcal{N}^{\prime\mu}_i$ in $D>1$ NP scenarios, we have used the Eq. (\ref{eq13}) with the conditions given in Table \ref{wc table2}. 
\begin{table*}[tbh]
\caption{\label{tab:table3}The numerical values of expressions $\mathcal{R}^{\tau\mu(\tau)}_{i}$ in the $14\leq s\leq s_{max} \text{GeV}^2$ bin for 1D NP scenarios.}
\renewcommand{\arraystretch}{1}
\centering
\scalebox{1}{
\begin{tabular}{c|ccc|c|ccc}
\hline\hline &&&&&&&\\[-0.9em]
$\mathcal{R}^{\tau\mu(\tau)}_{i}$   &S-I    &S-II &S-III&$\mathcal{R}^{\tau\mu(\tau)}_{i}$    &S-I    &S-II &S-III    \\ 
 &&&&&&&\\[-0.9em]\cline{1-8}\\[-0.9em]
 $\mathcal{R}_{K^*}^{\tau\mu}$  &\hspace{0.5cm} $(0.324-0.346)$\hspace{0.3cm}    &$(0.40-0.41)$\hspace{0.3cm}  & $(0.30-0.33)$\hspace{0.3cm} & $\mathcal{R}_{K_L^{*\tau}}^{\tau\mu }$&\hspace{0.3cm}$(0.162-0.166)$\hspace{0.3cm} &    $(0.177-0.178)$\hspace{0.3cm}  &$(0.143-0.153)$    \\
&&&&&&&\\[-0.9em]\cline{1-8}\\[-0.9em]
 $\mathcal{R}_{K_T^{*\tau}}^{\tau\mu }$ &\hspace{0.3cm} $(0.162-0.180)$\hspace{0.3cm} & $(0.227-0.230)$\hspace{0.3cm} &   $(0.155-0.177)$\hspace{0.3cm} &  $\mathcal{R}_{K_L^{*\mu}}^{\tau\mu }$&\hspace{0.3cm}$(0.850-0.915)$\hspace{0.3cm} &$(1.086-1.094)$\hspace{0.3cm}  & $(0.821-0.904)$     \\
&&&&&&&\\[-0.9em]\cline{1-8}\\[-0.9em]
 $\mathcal{R}_{K_T^{*\mu}}^{\tau\mu }$ &\hspace{0.3cm} $(0.526-0.560)$\hspace{0.3cm}     &$(0.643-0.649)$\hspace{0.3cm}   &$(0.469-0.521)$\hspace{0.3cm} & $\mathcal{R}_{K_{LL}^*}^{\tau\mu }$    &\hspace{0.3cm}$(0.423-0.437)$\hspace{0.3cm}     &$(0.475-0.477)$\hspace{0.3cm}   &$(0.394-0.418)$     \\
&&&&&&&\\[-0.9em]\cline{1-8}\\[-0.9em]
 $\mathcal{R}_{K_{TT}^*}^{\tau\mu }$ &\hspace{0.3cm}$(0.262-0.290)$\hspace{0.3cm}    &$(0.362-0.366)$\hspace{0.3cm}   &$(0.244-0.280)$\hspace{0.3cm} & $\mathcal{R}_{K_{LT}^*}^{\tau\mu }$  &\hspace{0.3cm}$(0.262-0.268)$\hspace{0.3cm}     &$(0.281-0.283)$\hspace{0.3cm}   &$(0.225-0.241)$    \\
&&&&&&&\\[-0.9em]\cline{1-8}\\[-0.9em]
    $\mathcal{R}_{K_{TL}^*}^{\tau\mu }$ &\hspace{0.3cm}$(0.423-0.474)$\hspace{0.3cm} & $(0.611-0.617)$\hspace{0.3cm}  &  $(0.428-0.486)$\hspace{0.3cm}& $\mathcal{R}_{K^*_L}^{\tau }$   &\hspace{0.3cm}$(0.480-0.500)$\hspace{0.3cm}    &$(0.436-0.437)$\hspace{0.3cm}   &$(0.463-0.479)$    \\
  &&&&&&&\\[-0.9em]\cline{1-8}\\[-0.9em]
  $\mathcal{R}_{K^*_T}^{\tau }$ & \hspace{0.3cm} $(0.500-0.520)$\hspace{0.3cm}   & $(0.563-0.564)$\hspace{0.3cm}   &$(0.521-0.537)$\hspace{0.3cm}   & $\mathcal{R}_{K^*_{TL}}^{\tau }$&\hspace{0.3cm}$(0.920-1.001)$\hspace{0.3cm}    & $(0.774-0.778)$\hspace{0.3cm}   &$(0.862-0.921)$. \\     \hline \hline   \end{tabular}}
  \label{TableRK}
\end{table*}
Finally, by using the constraints of NP WCs, summarized in Table \ref{wc table}, we have calculated the numerical values of $\mathcal{R}^{\tau\mu(\tau)}_{i}$, in the $14\leq s\leq s_{max} \text{GeV}^2$ bin in different NP scenarios, and registered them in  Tables \ref{TableRK} and \ref{TableVa}.

\section{Phenomenological Analysis}\label{phAnaly}
In this section, we present the phenomenological analysis of the twelve potential physical observables that are given in Eq. (\ref{ratios1}). For $R^{\tau \mu}_{K^{\ast}}$, the  SM prediction in $s\in [14,\;19]\; \text{GeV}^2$ bin is $0.41\pm 0.01$. Unlike the well measured $R^{\mu \,e}_{K^{\ast}}$, the experimental measurements of  $R^{\tau \mu}_{K^{\ast}}$ are currently missing due intricacy in reconstruction of tauons in final state. However, a similar study in the flavor changing charged current process governed by $b\to c \tau \nu_{\tau}$ using proton-proton $(pp)$ collision data corresponding to an integrated luminosity of $2fb^{-1}$ collected by the LHCb experiment during the periods 2015-16, the LFU ratios $R_{D(D^*)}$ were measured \cite{Chen:2024hln}. Therefore, quite optimistically in future we will be able to measure these asymmetries involving the $\tau$ leptons in the final state, and could scrutinize the SM further in the $\tau-\mu$ sector. 

\begin{figure}[H]
\centering
\includegraphics[width=3.5in,height=2.3in]{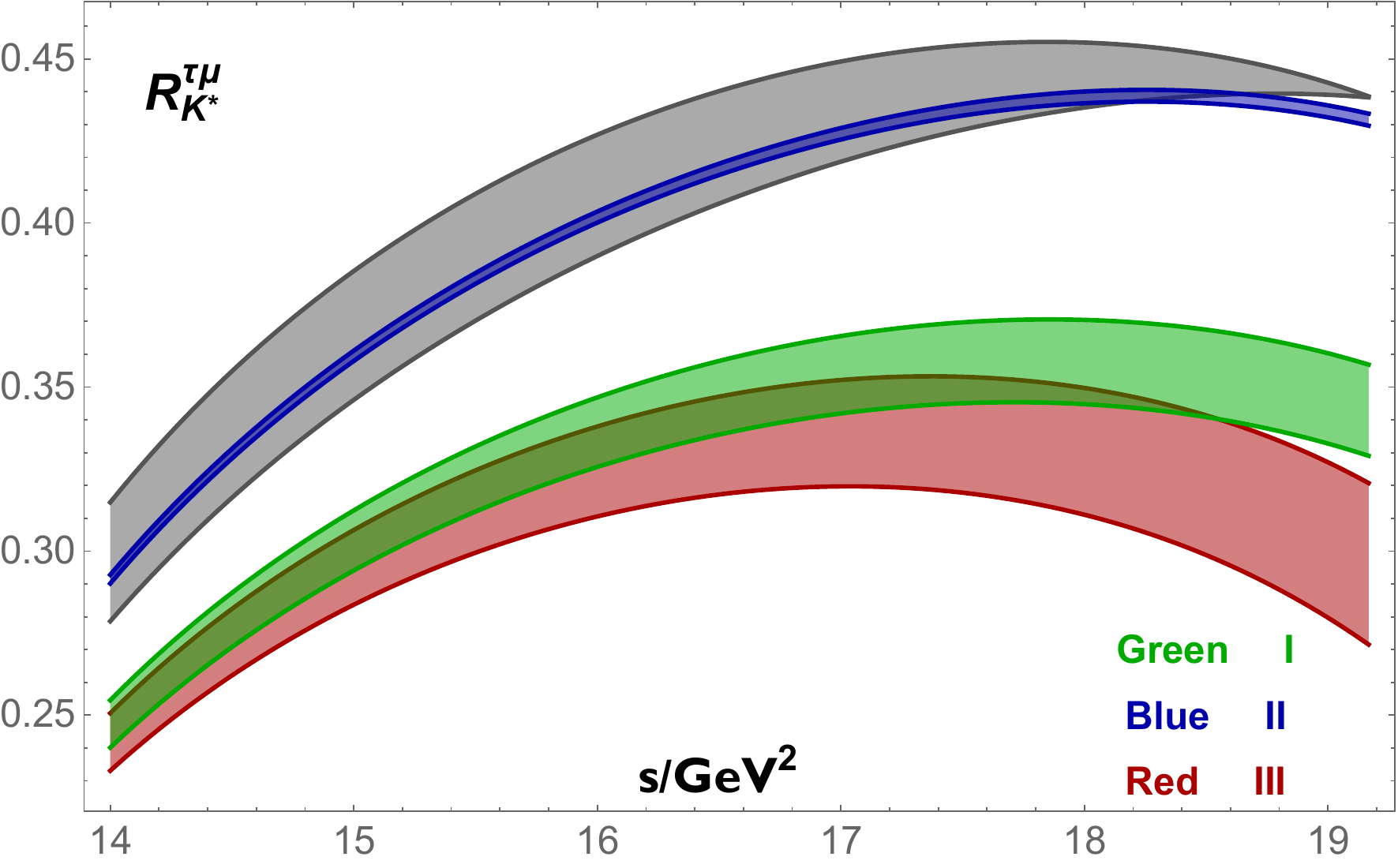}
\includegraphics[width=3.5in,height=2.3in]{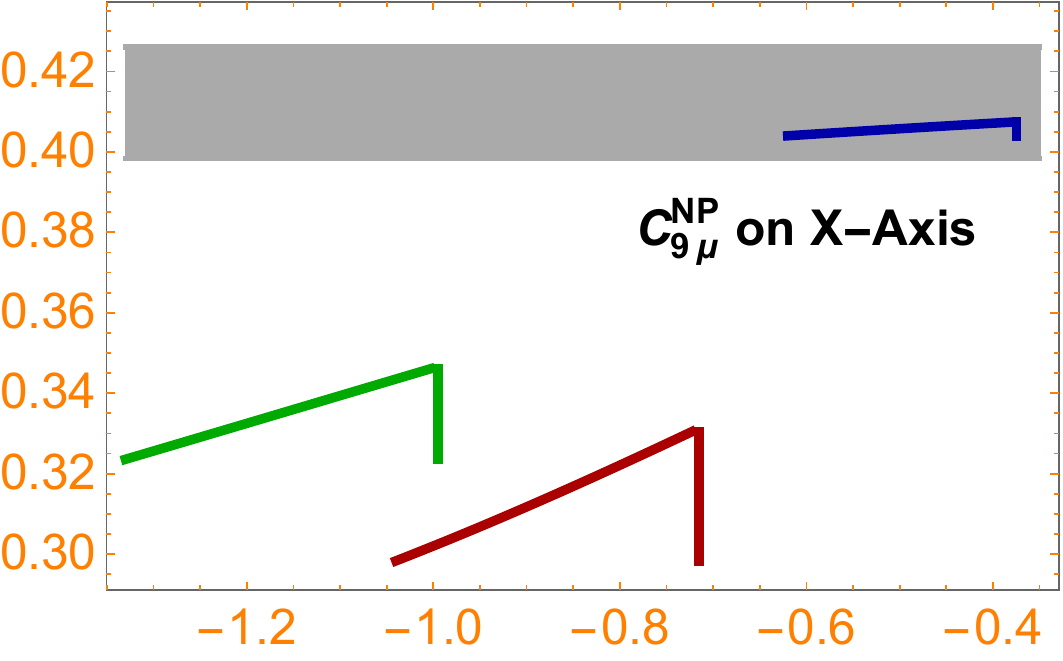}
\\
\caption{The 1 $\sigma$ plots for the $q^2$ distribution of $R_{K^{\ast}}$ in different 1D scenarios. The SM predictions shown in gray color with the width of the band represent the uncertainties due to form factors. The bars show the magnitude of the $R_{K^{\ast}}$ in each scenario.}
\label{1Dob1}
\end{figure}

\begin{figure}[H]
\centering
\includegraphics[width=3.4in,height=2.2in]{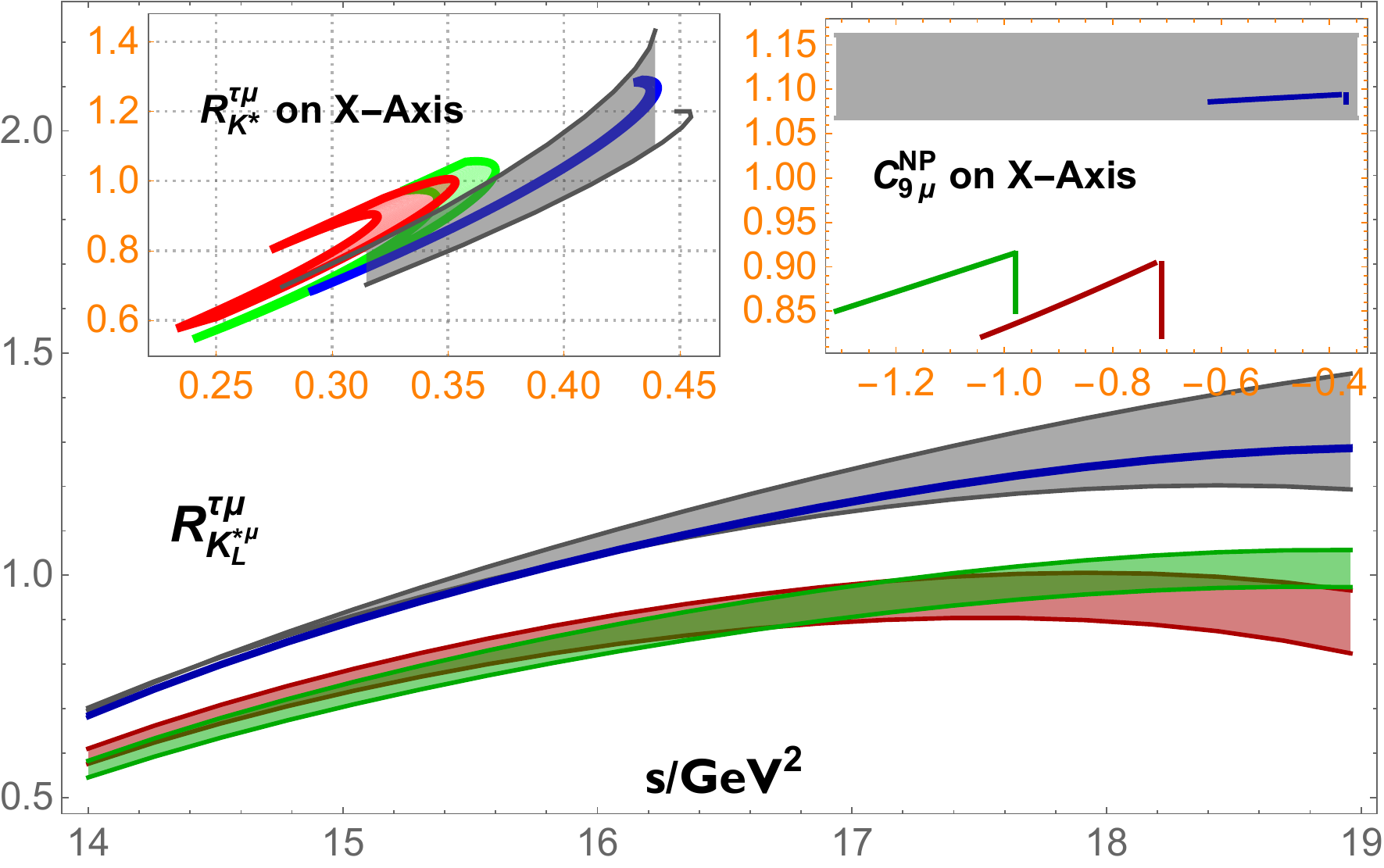}
\includegraphics[width=3.4in,height=2.2in]{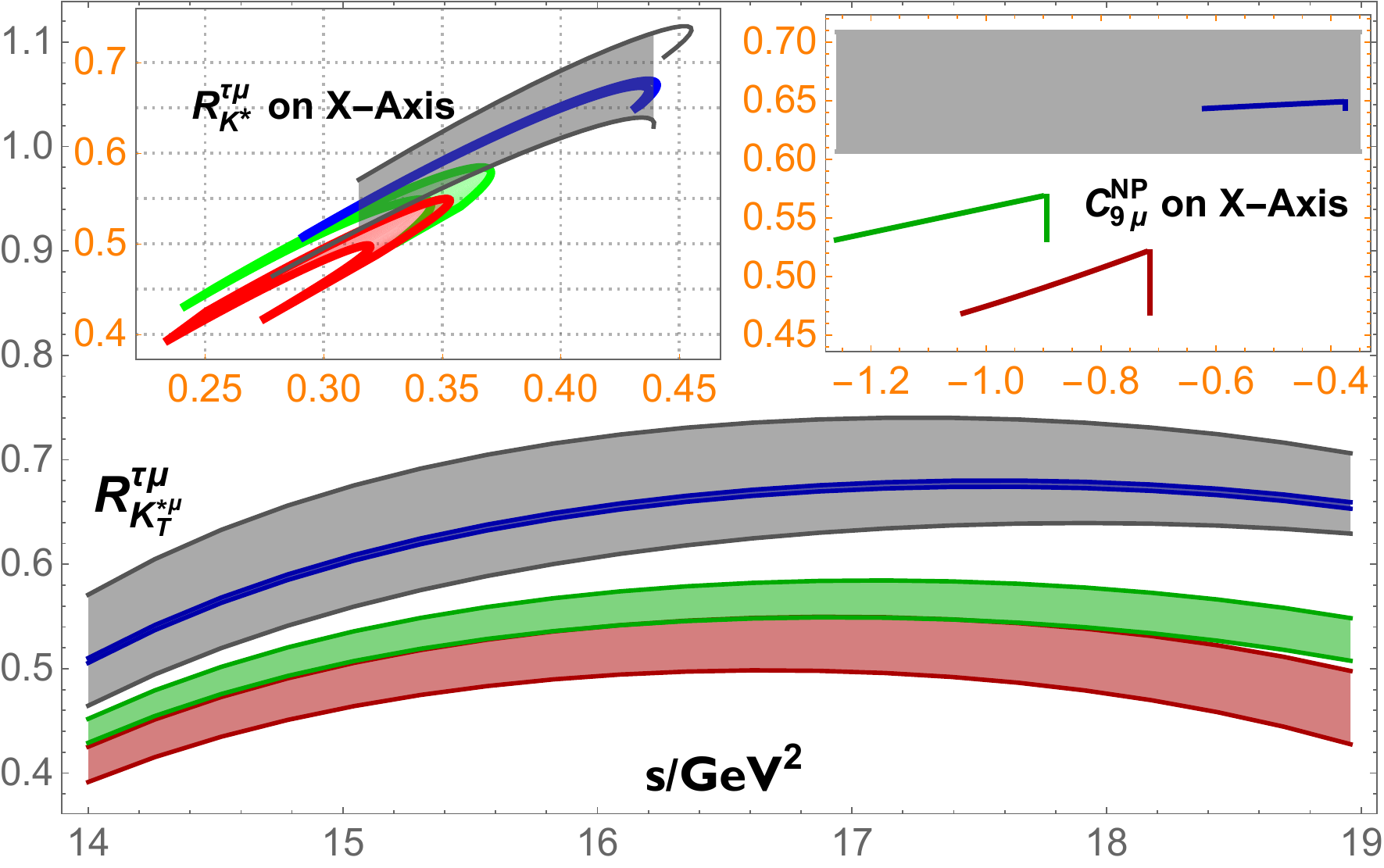}
\\
\includegraphics[width=3.4in,height=2.2in]{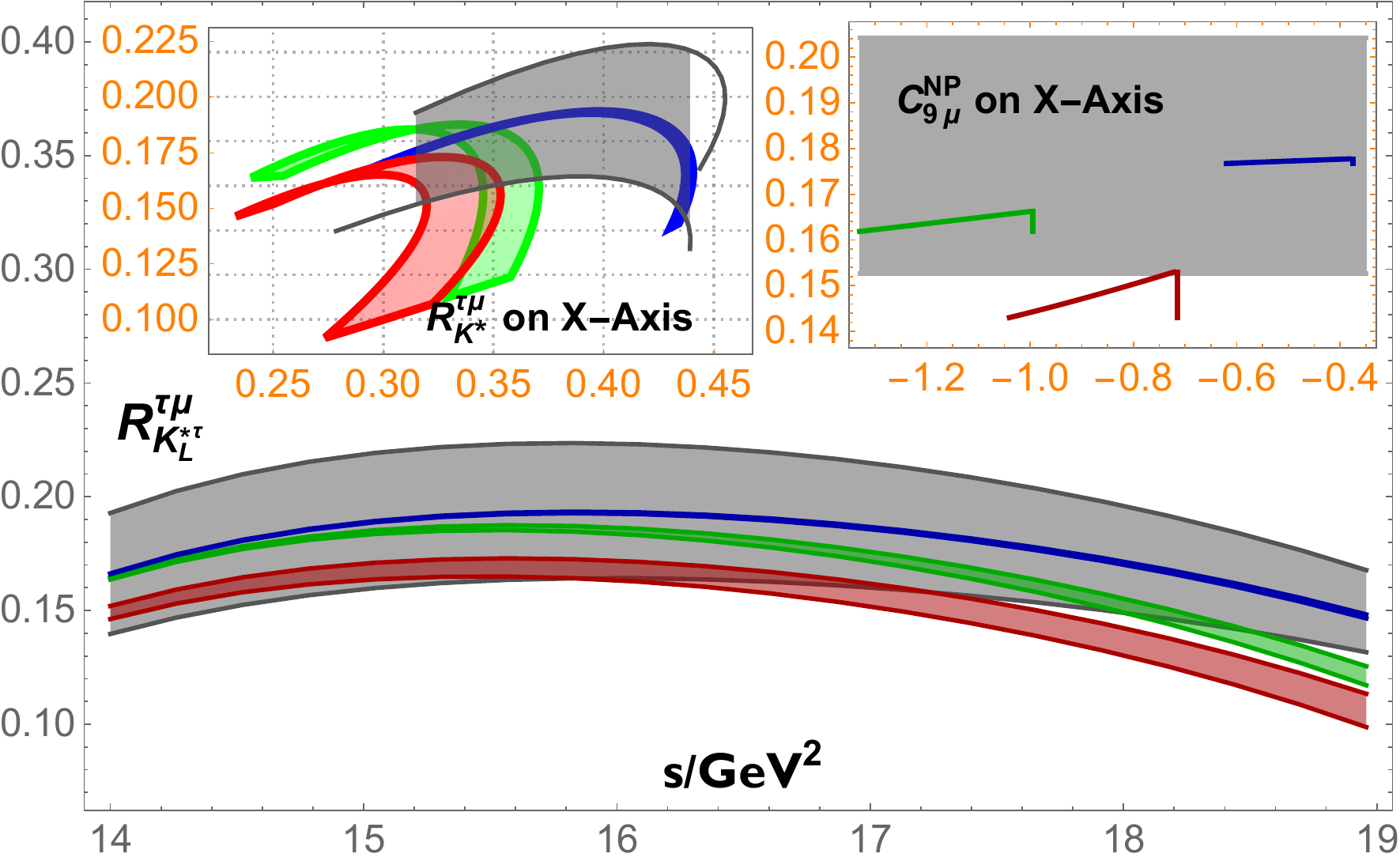}
\includegraphics[width=3.4in,height=2.2in]{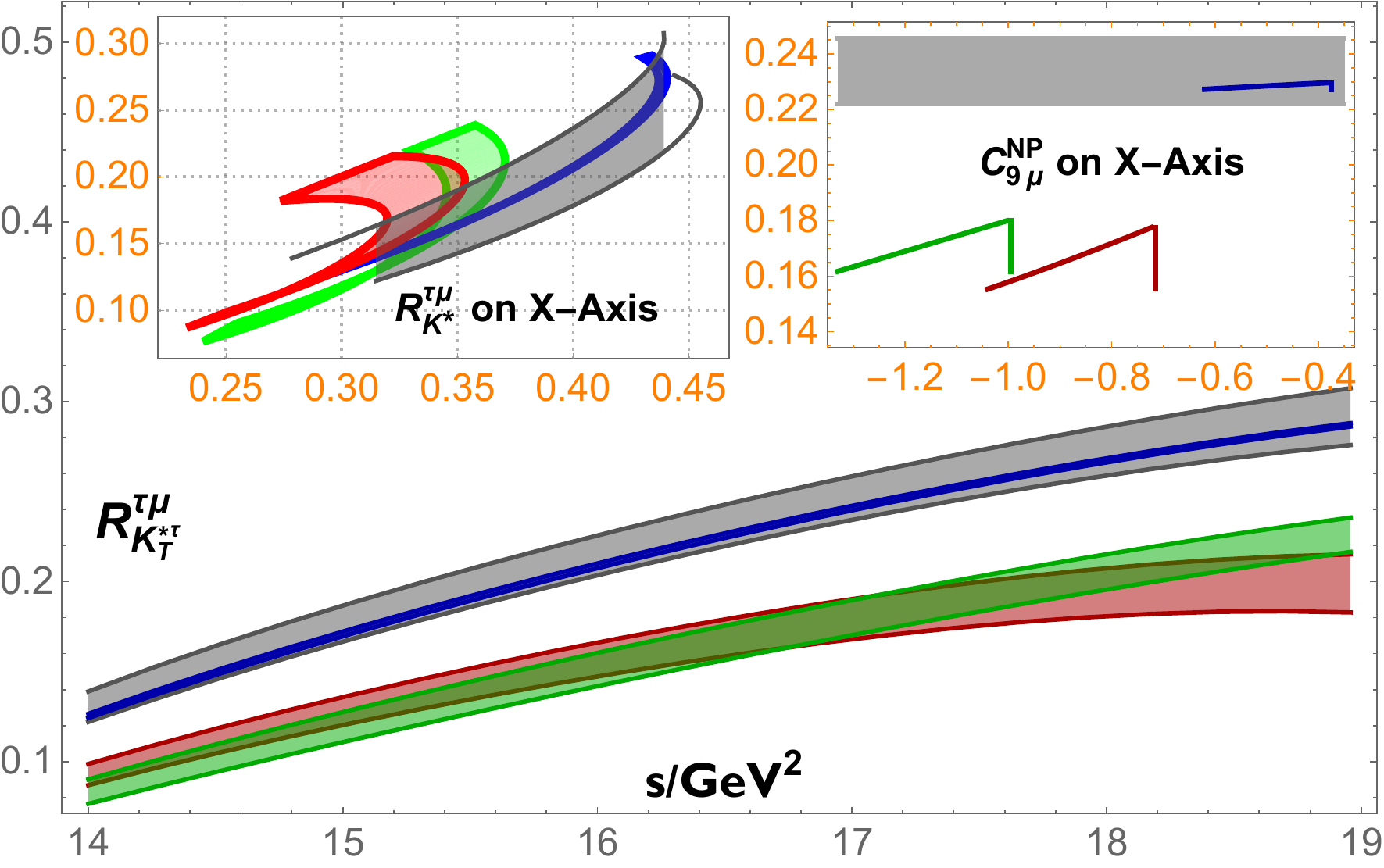}
\\
\includegraphics[width=3.4in,height=2.2in]{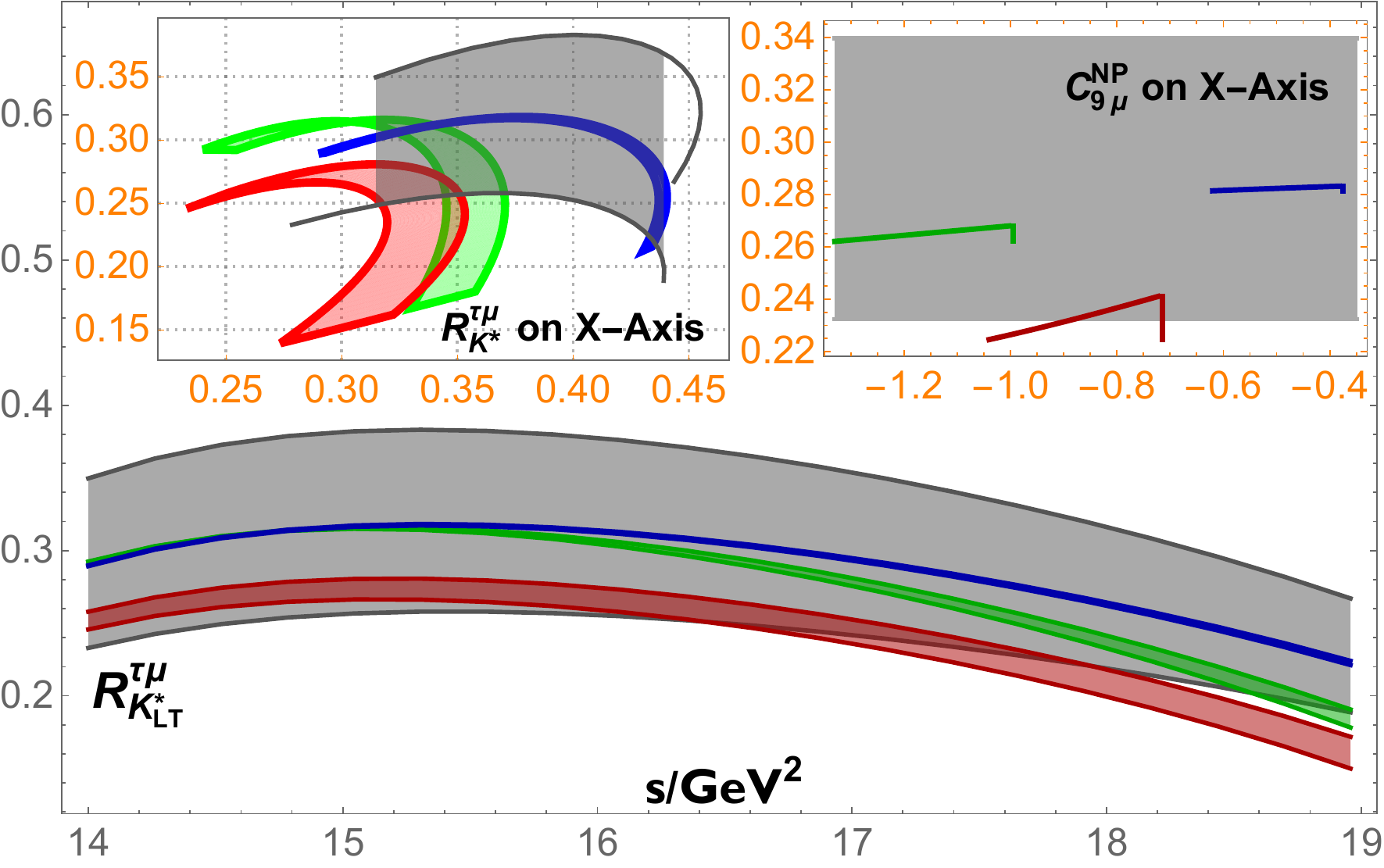}
\includegraphics[width=3.4in,height=2.2in]{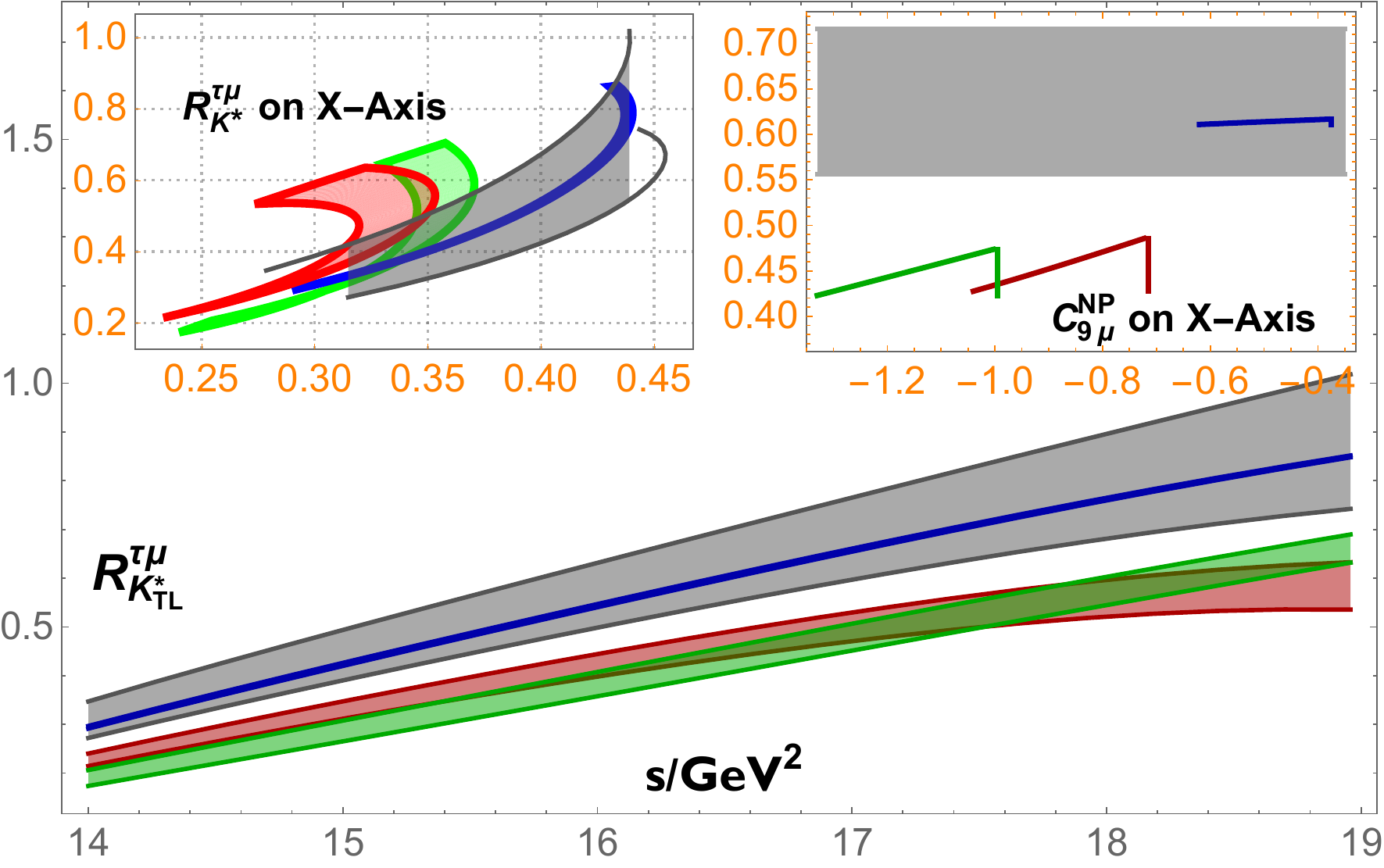}
\\
\caption{The plots of the observables ($\mathcal{O}_{i}$) with $s$ in the SM as well as different 1D scenarios where the width of the gray band represents the uncertainties due to form factors. The plot on the top left corner in the inset shows the correlation among the  $\mathcal{O}_{i}$ and $R^{\tau \mu}_{K^{\ast}}$ with $R^{\tau \mu}_{K^{\ast}}$ taken along the x-axis. The top right corner shows the magnitude of the $\mathcal{O}_{i}$ in each scenario in the form of bar plots.}
\label{1Dob1a}
\end{figure}

\begin{figure}[H]
\centering
\includegraphics[width=3.5in,height=2.3in]{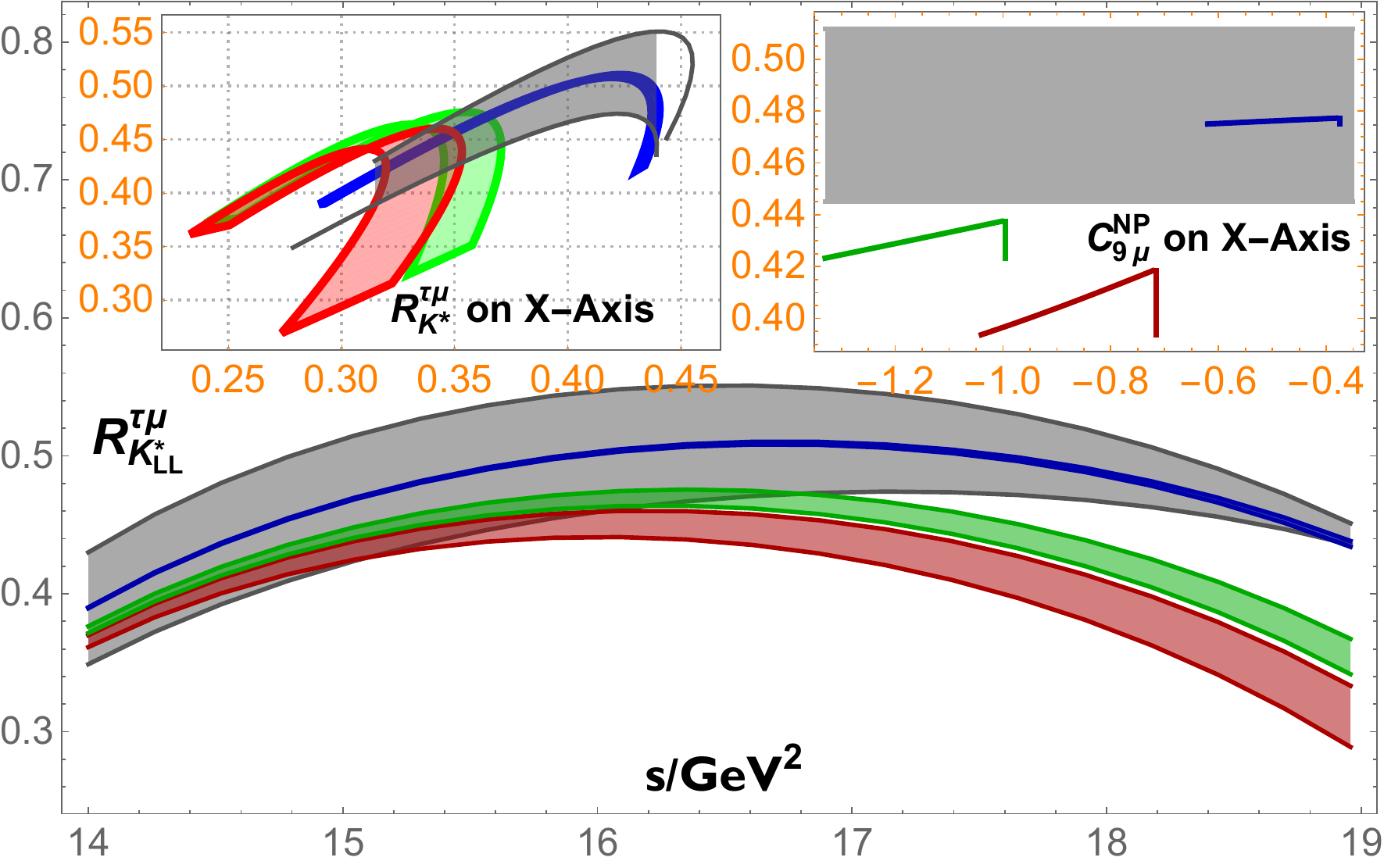}
\includegraphics[width=3.5in,height=2.3in]{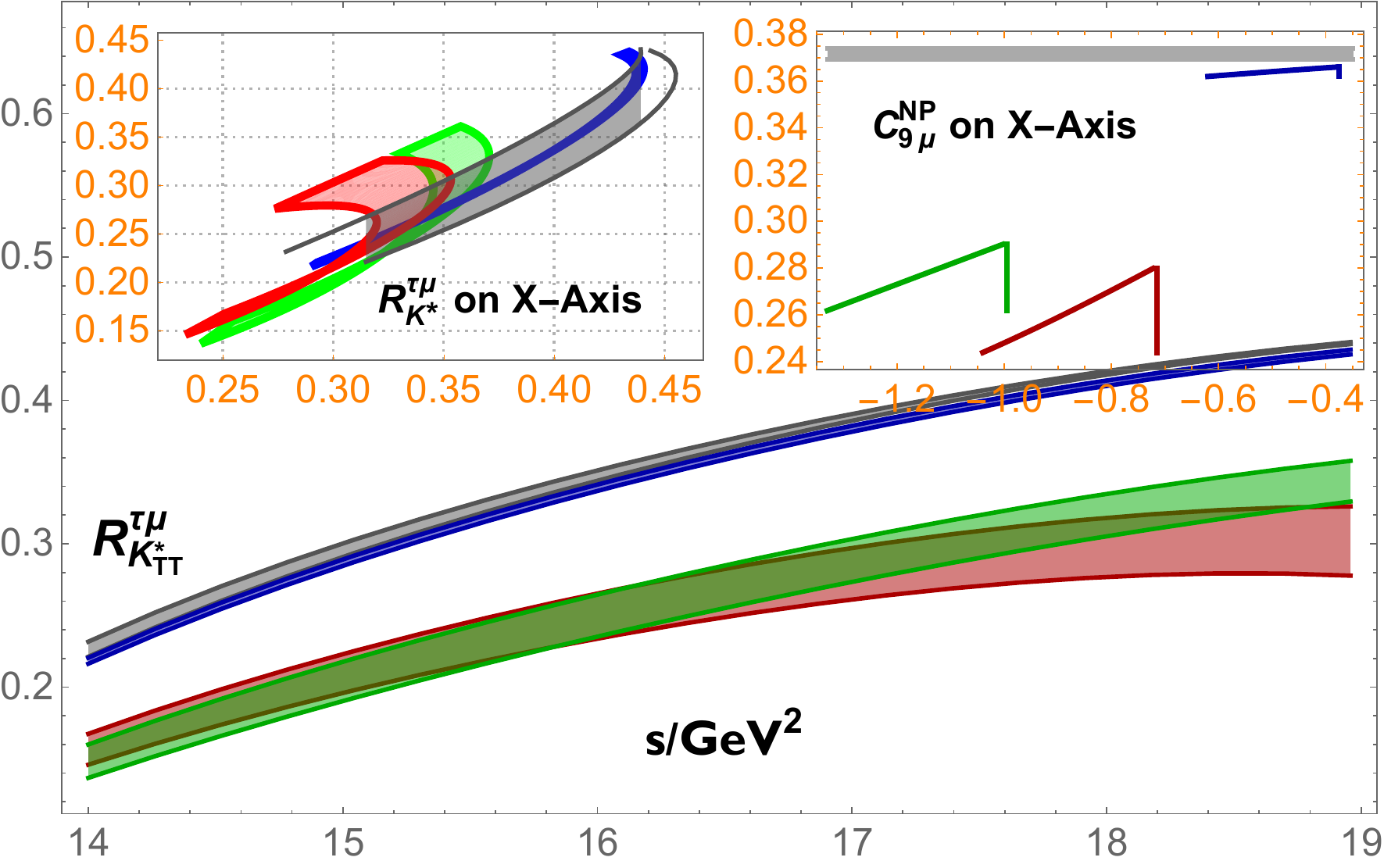}
\\
\includegraphics[width=3.5in,height=2.3in]{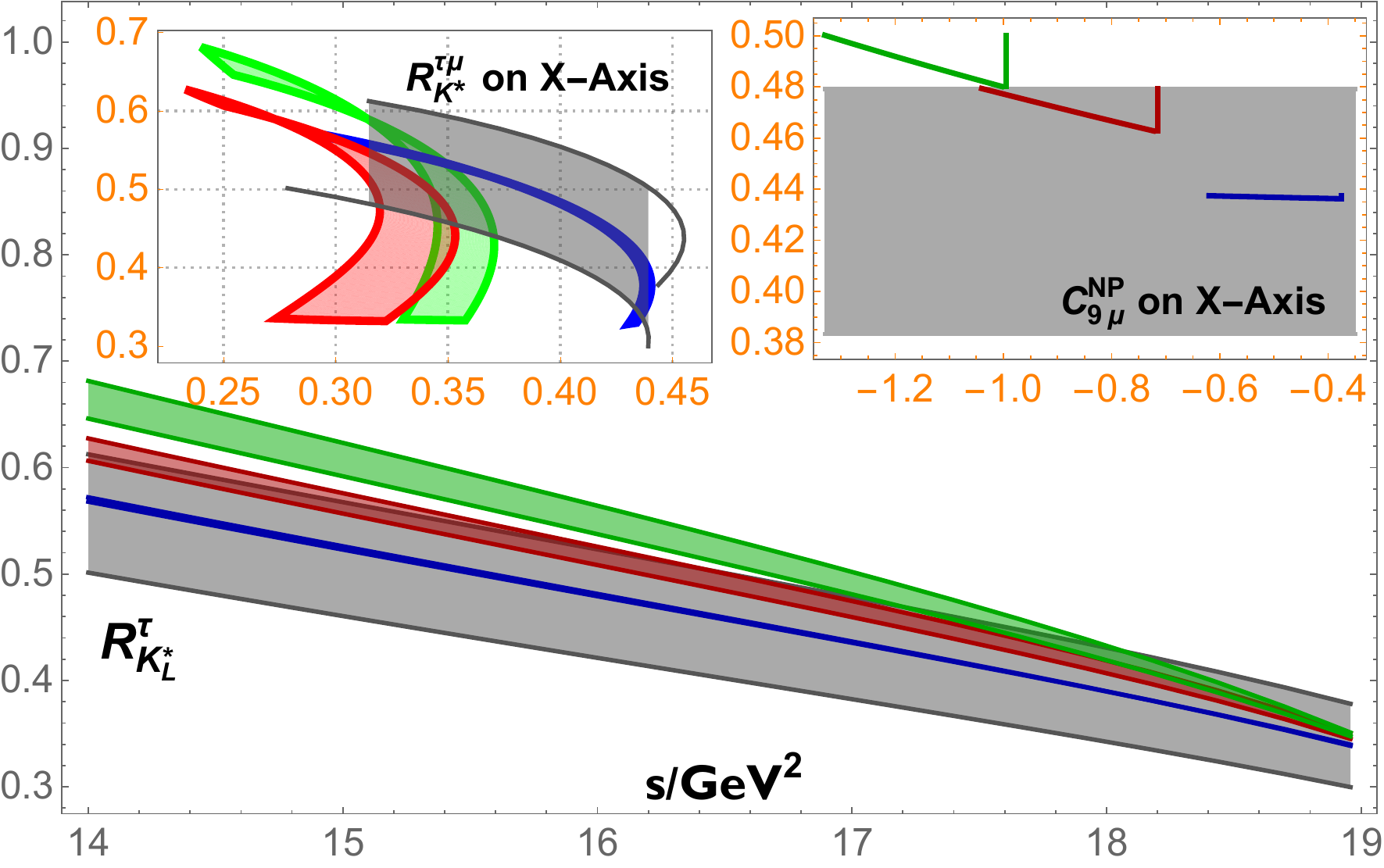}
\includegraphics[width=3.5in,height=2.3in]{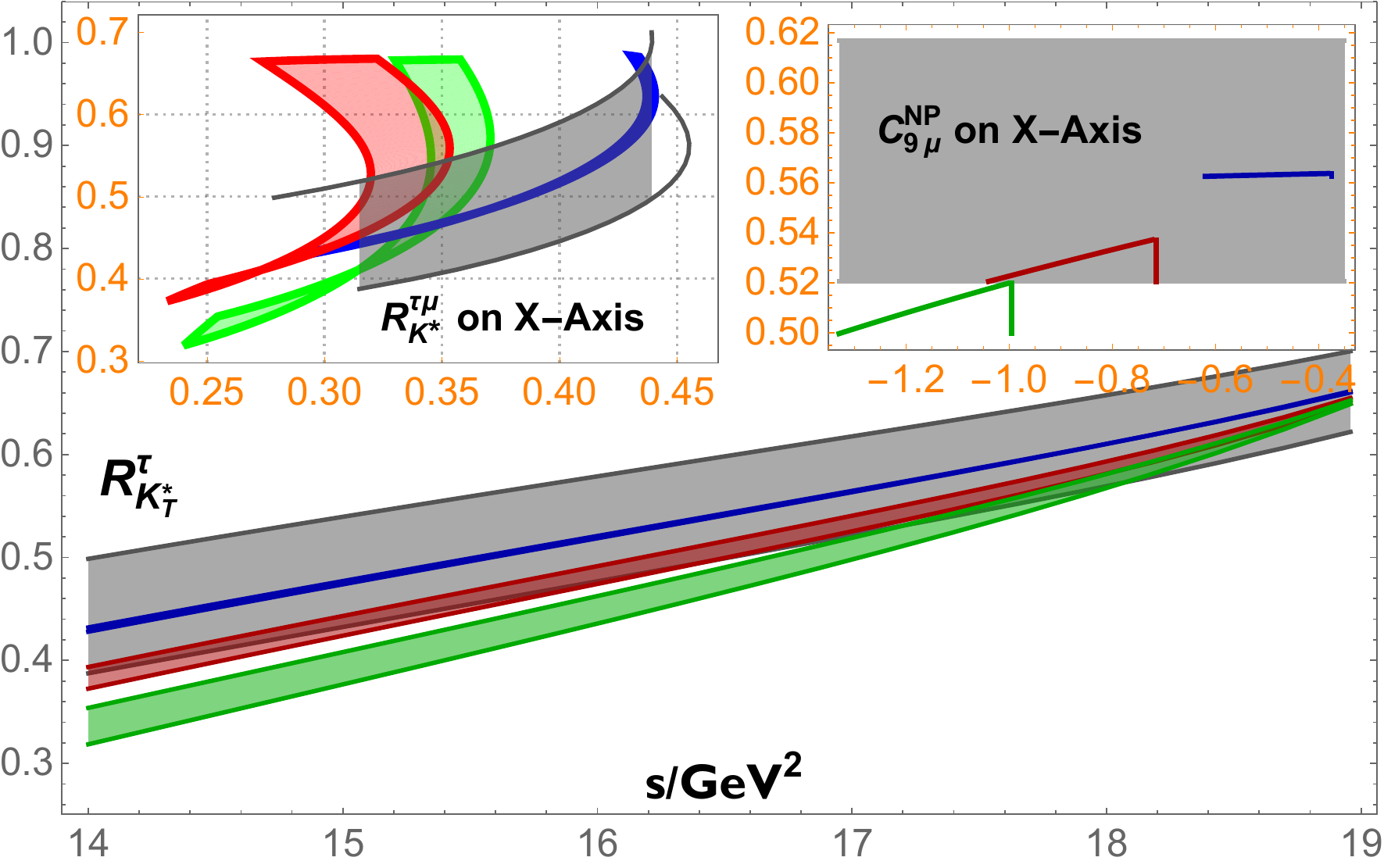}
\\
\includegraphics[width=3.5in,height=2.3in]{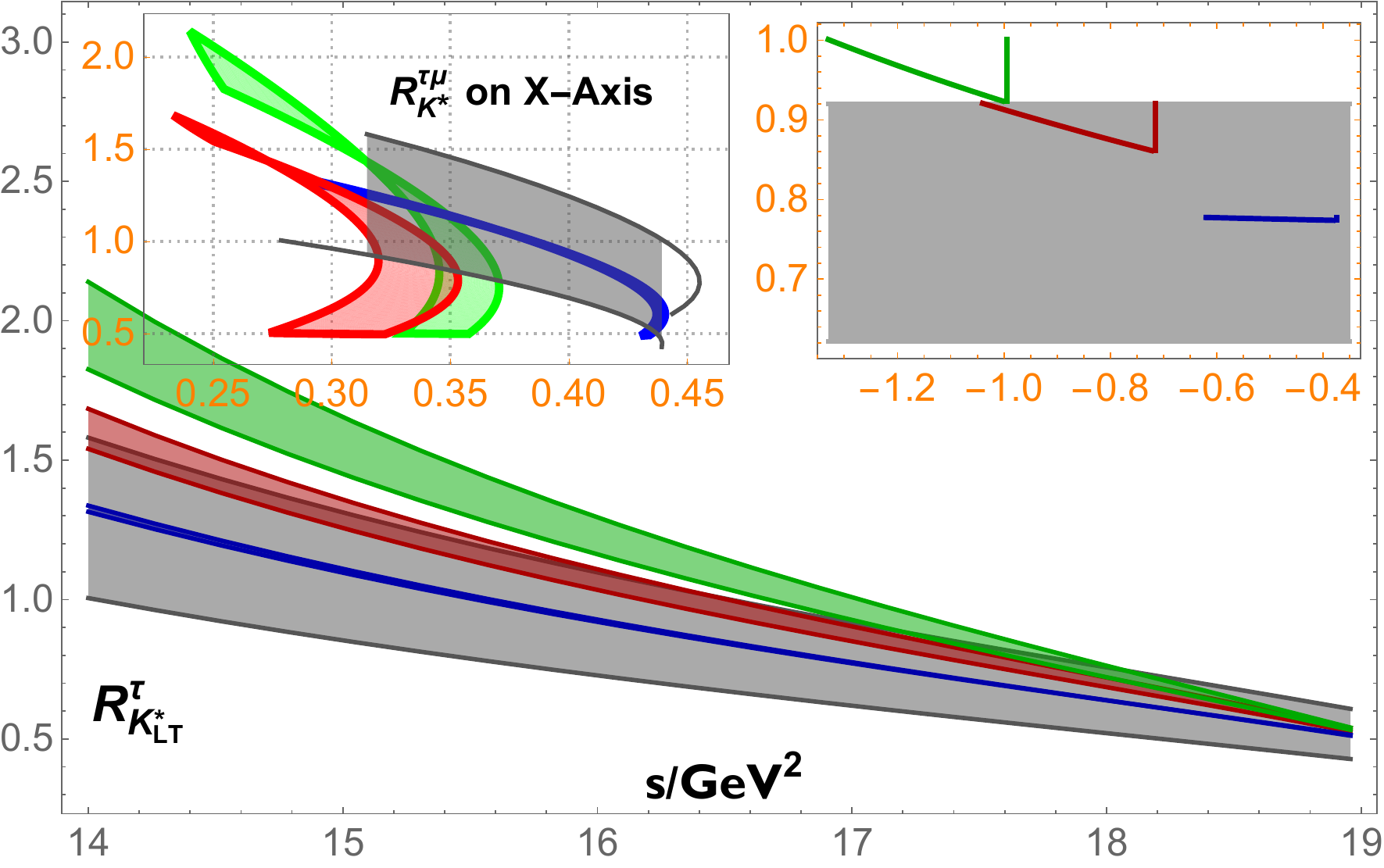}
\includegraphics[width=3.5in,height=2.3in]{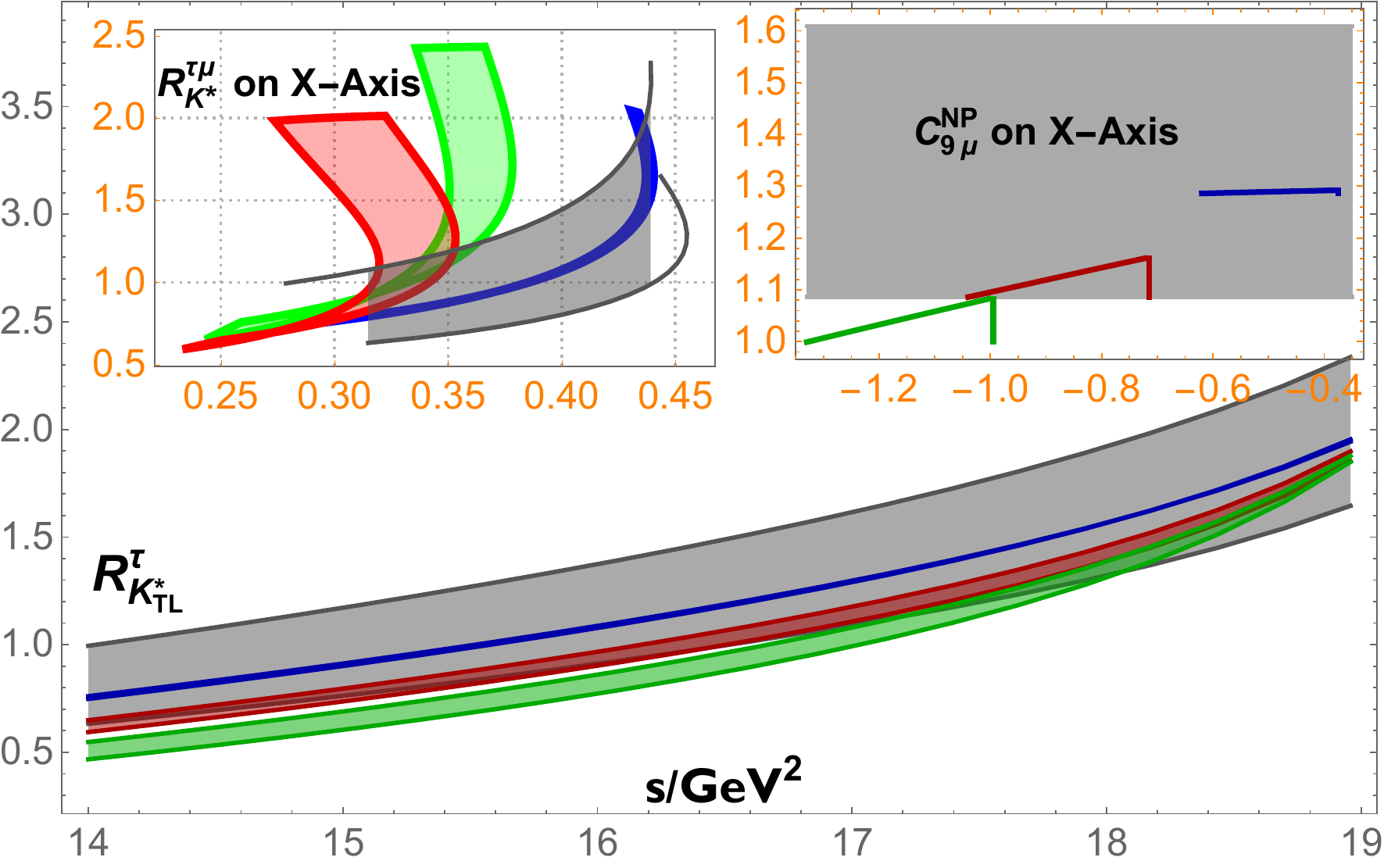}
\\
\caption{The plots of the observables ($\mathcal{O}_{i}$) as a function of $s$ in the SM as well as different 1D scenarios where the width of the gray band represents the uncertainties due to form factors. The plot on the top left corner in the inset shows the correlation among the  $\mathcal{O}_{i}$ and $R^{\tau \mu}_{K^{\ast}}$ with $R^{\tau \mu}_{K^{\ast}}$ taken along the x-axis. The top right corner shows the magnitude of the $\mathcal{O}_{i}$ in each scenario in the form of bar plots.}
\label{1Dob1b}
\end{figure}

Figs. \ref{1Dob1}, \ref{1Dob1a} and \ref{1Dob1b} show the $s\equiv q^2$ profile of $R^{\tau \mu (\tau)}_{i}$, where $i=K^*$, $K^*_L$, $K^*_T$, $K^*_{LT}$, $K^*_{TL}$, $K^*_{LL}$ and $K^*_{TT}$, for $D=1$.  Before starting the analysis, it is useful to mention here the SM results are plotted in gray color, where the band correspond to the uncertainties arising from the different input parameters; particularly the form factors. The other color bands show the variation in their values for the new WCs in $1\sigma$ range. In this case of NP, we have plotted the results for the central values of the form factors. 


\begin{itemize}
     \item  Fig. \ref{1Dob1} (left) illustrates the dependence of ratio $R^{\tau \mu}_{K^{\ast}}$ on new WCs as function of $s$. It can be noted that the second scenario (S-II) $(C^U_9=-C^U_{10})$, drawn in blue color, is precluded by uncertainties coming from SM; whereas, S-I and S-III predict values lower than that of SM. The results of these scenarios are well distinguished in almost all the $s$ range, particularly, in $s\in [18,19] \text{GeV}^2$ bin, where the results of S-III are significantly lower than S-I even. This is in line with the results given in Table \ref{TableRK}, where the values of $R^{\tau \mu}_{K^{\ast}}$ for S-I and S-III $\sim (0.33 - 0.35)$ is well segregated from the S-II result, i.e., $(0.40-0.41)$. To make it more visible, after integrating over $s$, these variations of $R^{\tau \mu}_{K^{\ast}}$ with $C^{\text{NP}}_{9\mu}$ are drawn in Fig. \ref{1Dob1} (right), where we can easily see that the results in three scenarios are well separated from each other.

    \item The polarized LFUV ratios $\mathcal{R}_{K^{*\mu}_{L,T}}^{\tau\mu}$, where ${K^{*\mu}_{L,T}}$ respresent the polarized vector meson when we have muons in the final state, are presented as first line of Fig. \ref{1Dob1a}. The corresponding SM results of these ratios are given in Table \ref{SM ratios}, where the maximum value for it is $1.105^{+0.056}_{-0.037}$, and it is shown by the gray band in the Fig.\ref{1Dob1a}. We can see that the values of new WCs data arising from all three scenarios destructively interfere with the SM contributions, hence decreasing the corresponding observable values. This can also be noticed from Table \ref{TableRK} where the maximum suppression is for the S-III for $\mathcal{R}_{K^{*\mu}_T}^{\tau \mu}$ case. In the case of LFUV ratios with polarized final state meson, these are the ratios of full to longitudinal or transverse polarized ratios, therefore, we can expect $\mathcal{R}_{K^*_L}^{SM}+\mathcal{R}_{K^*_T}^{SM} > 1$, and this can bee seen in Fig. \ref{1Dob1a}.
    
    
    \item In the second row of Fig. \ref{1Dob1a} we have presented the ratios $\mathcal{R}_{K^{*\tau}_{L,T}}^{\tau\mu}=\mathcal{B}(B\to K^*_{L,T}\tau^+\tau^-)/\mathcal{B}(B\to K^*\mu^+\mu^-)$. With the particular polarization of the $K^*$ in the numerator and the different impact of NP to $\mu$ and $\tau$, the value of $\mathcal{R}_{K^{*\tau}_{L,T}}^{\tau\mu}$ is expected to be less than one in the whole $s$ region, and it is evident from the second row of Fig. \ref{1Dob1a}. Also, the trend of longitudinal and transverse polarized LFUV ratios are opposite to each other. Once again, the range of S-II is masked by the uncertainty in the values of the SM - but the results of S-I and S-III are distinguishable from the SM, especially for $\mathcal{R}_{K^{*\tau}_{T}}^{\tau\mu}$.
    \item In Eq. (\ref{ratios1}) the ratios, $R^{\tau \mu}_{K^*_{LT}}$, $R^{\tau \mu}_{K^*_{TL}}$, $R_{K^*_{LL}}$ and $R_{K^*_{TT}}$ can be defined as the polarized lepton flavor universality  violation (PLFV) ratios due to a particular polarization of the final state meson in the numerator and denominator. It can be seen form the last line of Fig. \ref{1Dob1a} that the observables $R^{\tau \mu}_{K^*_{LT}}$ and $R^{\tau \mu}_{K^*_{TL}}$ have the same trend as $\mathcal{R}_{K^{*\tau}_{L}}^{\tau\mu}$ and  $\mathcal{R}_{K^{*\tau}_{T}}^{\tau\mu}$, respectively. 
    
    In line with the LFUV ratio: $\mathcal{R}_{K^*}^{SM}=\mathcal{B}(B\to K^*\tau^+\tau^-)/\mathcal{B}(B\to K^*\mu^+\mu^-)$, we expect that in SM $R_{K^*_{LL}}+R_{K^*_{TT}}\approx 1$, because the similar polarizations in the numerator and denominator do not change the total probability, and this can be seen in the first row of Fig. \ref{1Dob1b} at any value of $s$. We can also observe from Table \ref{SM ratios}, that this has good discriminatory power, and the S-III can be distinguished from the SM and other scenarios for $R_{K^*_{LL}}$ in the high $s$ range, \textit{i.e.}, $s\in[17,\; 19]\; \text{GeV}^{2}$ bin.
    \item Contrary to the LFUV ratios, the $R_{K^*_{(L,T)}}^{\tau}$ defined in  Eq. (\ref{ratios1}) correspond to the case when for each polarization of $K^*$ meson, the final state leptons are tauons only; therefore, we can say that these are just the helicity fractions. These are plotted in Fig. \ref{1Dob1b} (last two rows). We can see that the NP contributes equally in the the numerator and denominator, that is why we have $\mathcal{R}_{K^*_{L}}^{\tau}+\mathcal{R}_{K^*_{T}}^{\tau}\simeq1$. This can be observed from Fig. \ref{1Dob1b} and from the values tabulated in Table \ref{SM ratios}.  From these plots, one can notice that the scenario S-I is well distinct from the scenarios S-II and S-III, and the variation in the values of these ratios due to the new physics can also be seen from the Table \ref{TableRK}.
\end{itemize}
It is to mention here that along with the variations of $\mathcal{R}_{K^{*(\tau\mu)}_i}^{\tau,\mu}$, where $i$ represent the particular polarization of the vector meson in the numerator and denominator, we have also analyzed the correlation between the $\mathcal{R}_{K^{*(\tau,\mu)}_i}^{\tau\mu}$ and $\mathcal{R}_{K^*}^{\tau\mu}$ and have shown the results in Figs. \ref{1Dob1a} and \ref{1Dob1b} (left inset). The behaviour of the observables by taking into account the new WCs and the total magnitude over the range of $1\sigma$ is also given in the inset (right) of the corresponding observables with SM uncertainties plotted as the gray band. From these plots, one can not only correlate these physical observables, but can also discriminate between the three NP scenarios considered here.

\begin{figure}[H]
\centering
 \includegraphics[width=2.95in,height=2.3in]{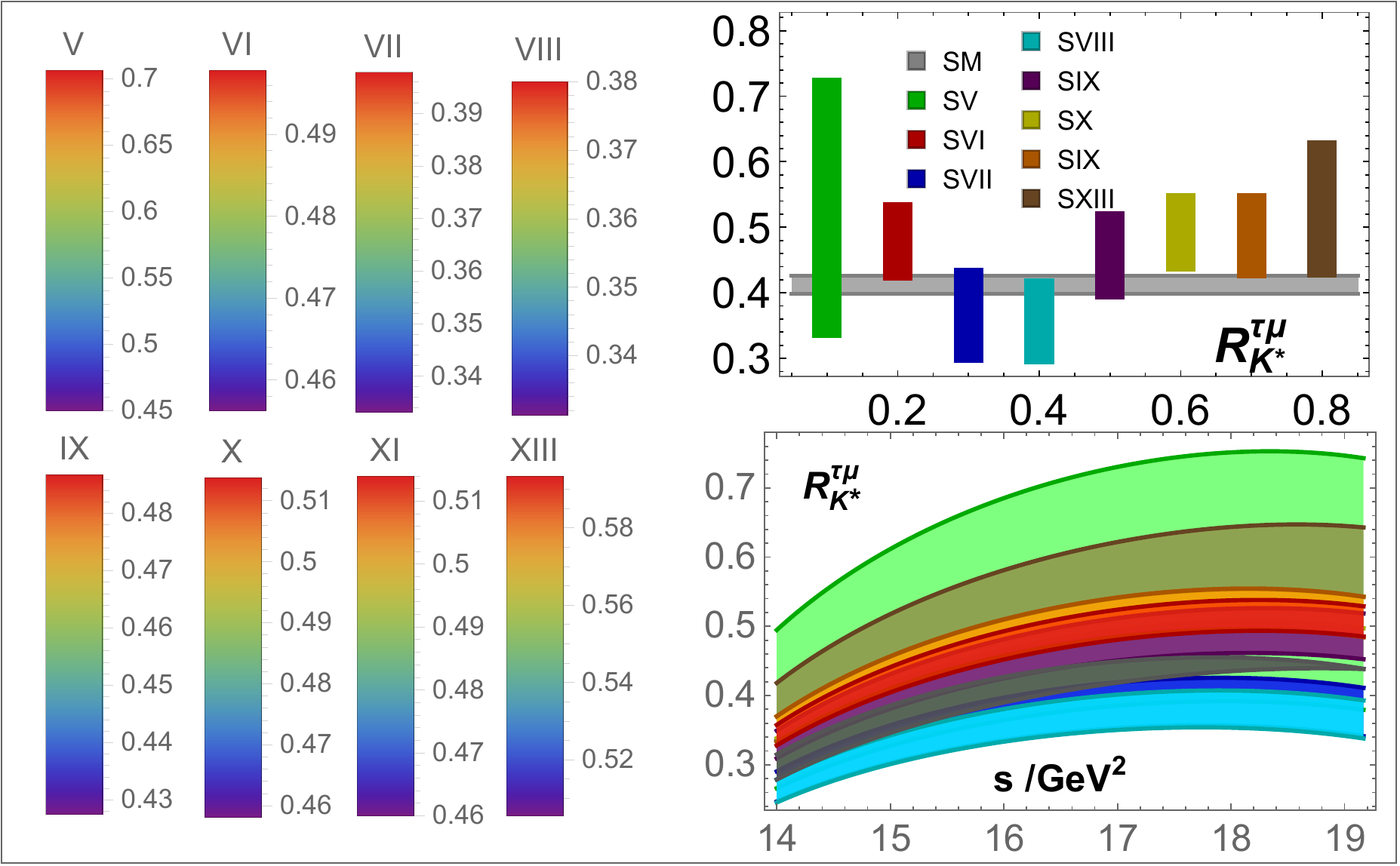}
\includegraphics[width=2.95in,height=2.3in]{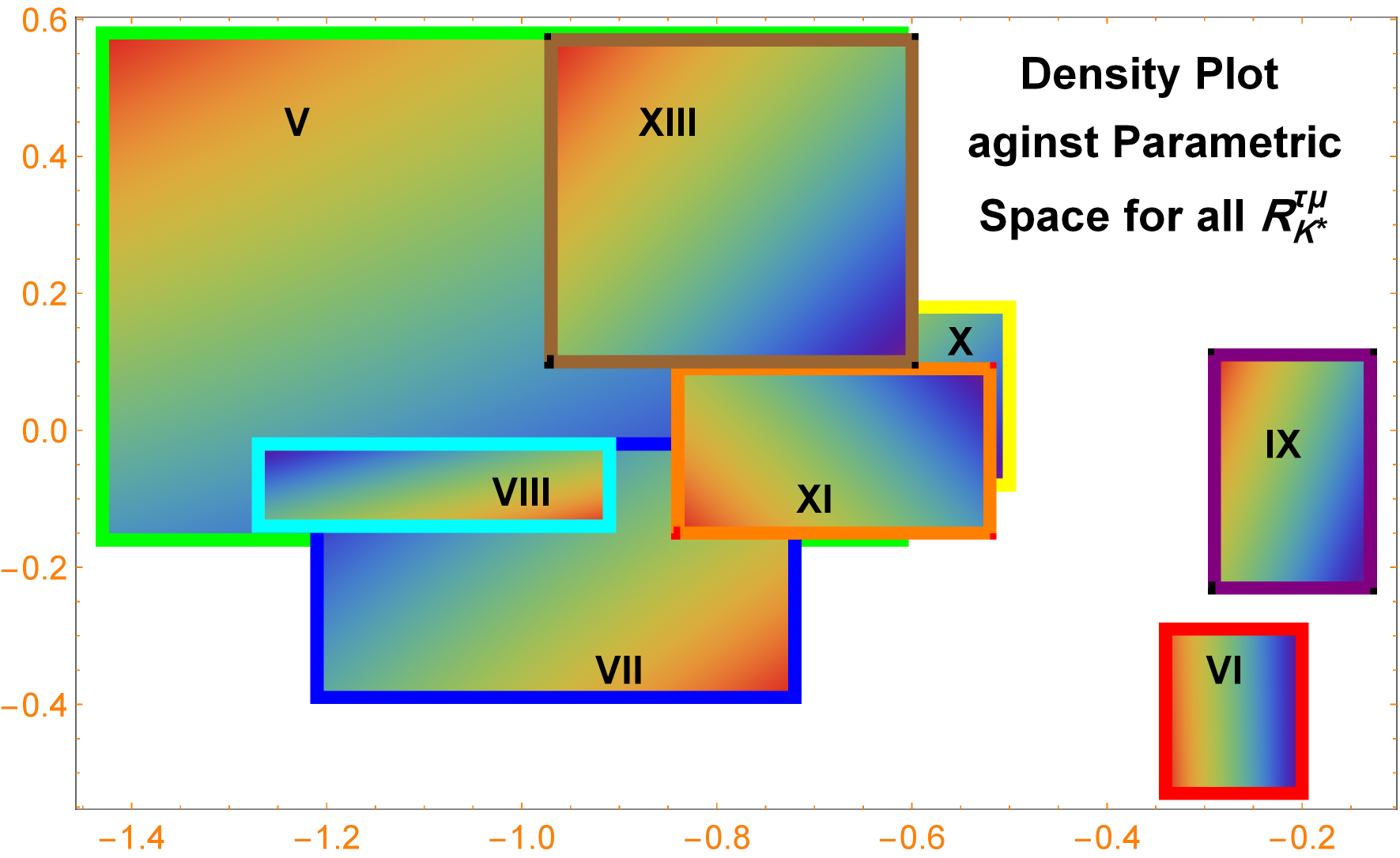}
\\
\includegraphics[width=2.95in,height=2.3in]{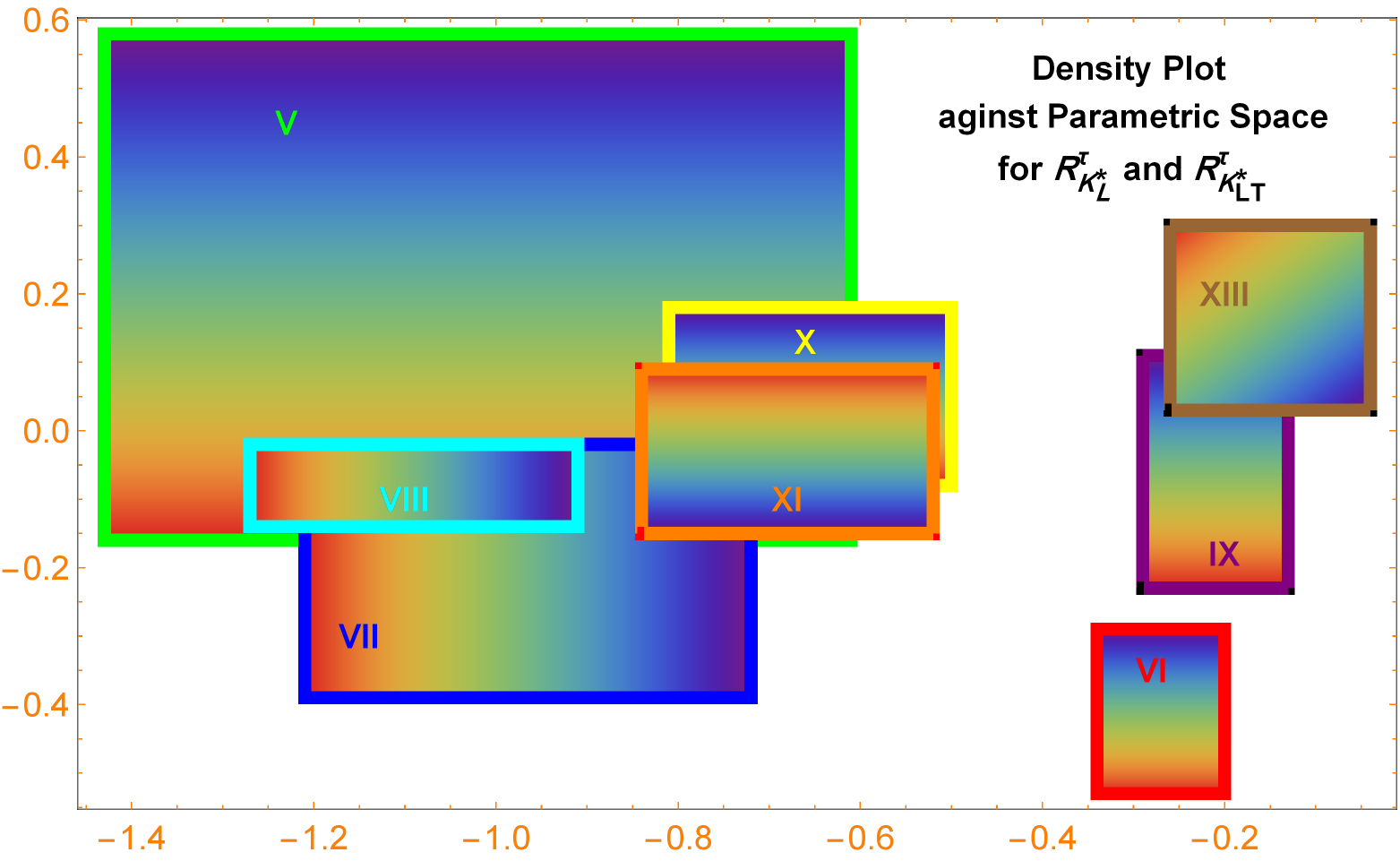}
\includegraphics[width=2.95in,height=2.3in]{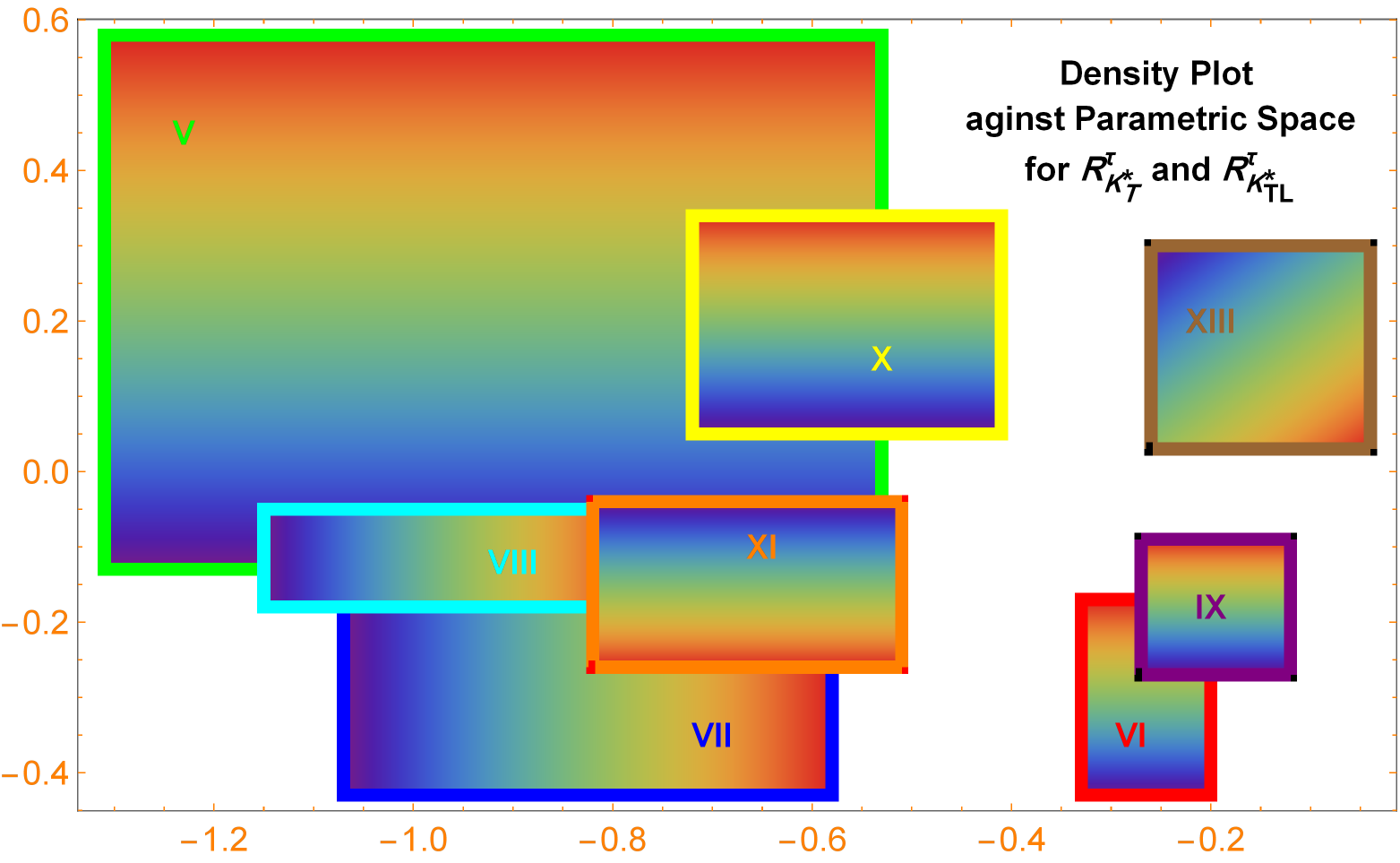}
\\
\caption{The first plot shows 1 $\sigma$ variations with $q^2$ of $\mathcal{R}_{K^*}$ in different $D>1$ NP scenarios. The SM predictions are given in gray color with the width of the band representing the uncertainties due to form factors while the bars show the magnitude of the  $\mathcal{R}_{K^*}$ in each scenario. The second, third, and fourth plots are the density plots against parametric space for all $
\mathcal{R}_{K^*}$, ($
\mathcal{R}_{K_L^*}^{\tau}$, $
\mathcal{R}_{K_{L,T}^*}^{\tau}$) and  ($
\mathcal{R}_{K_T^*}^{\tau}$, $
\mathcal{R}_{K_{T,L}^*}^{\tau}$) resp.}
\label{2Dob1}
\end{figure}

\begin{figure}[H]
\centering
\includegraphics[width=3.5in,height=2.3in]{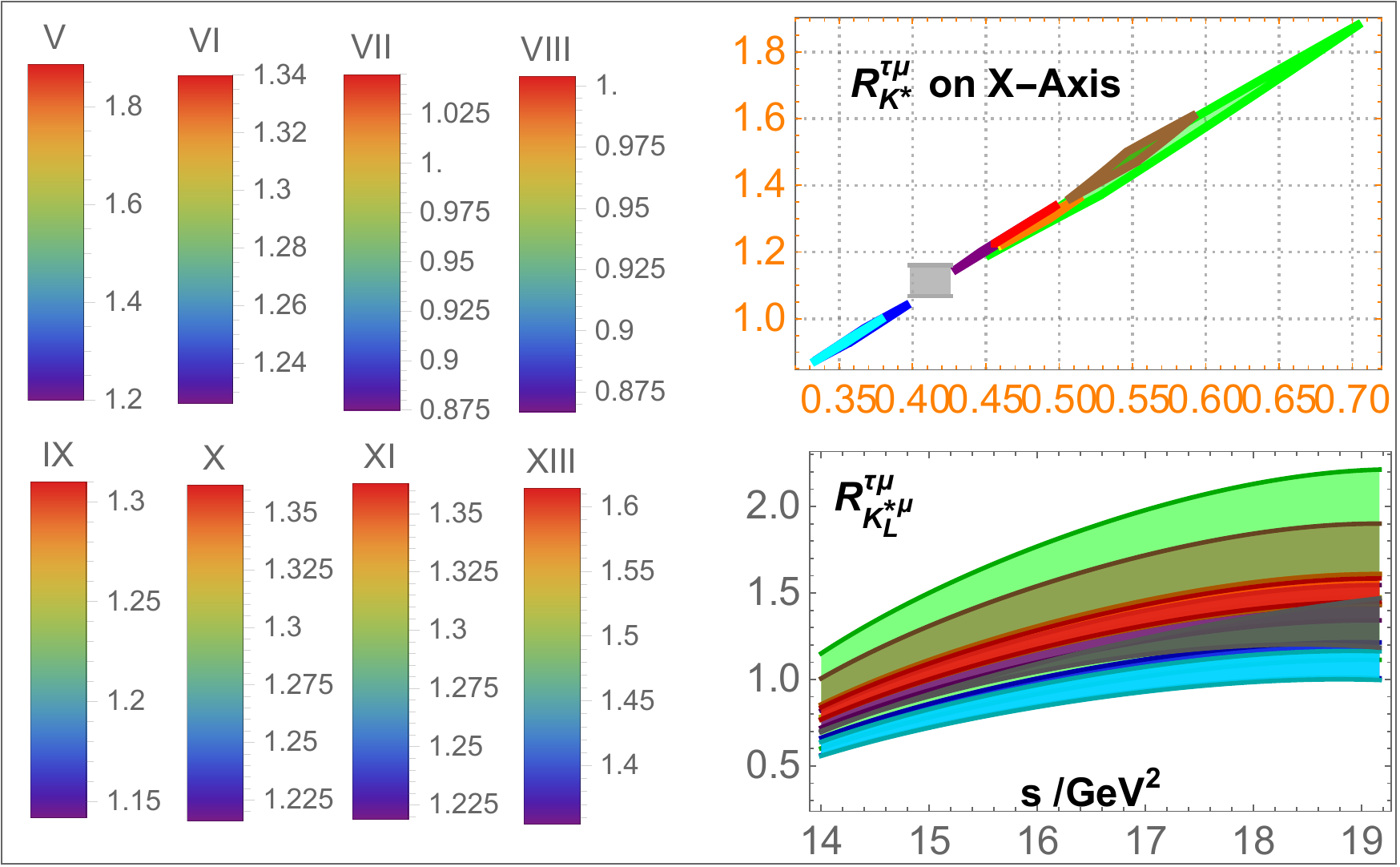}
\includegraphics[width=3.5in,height=2.3in]{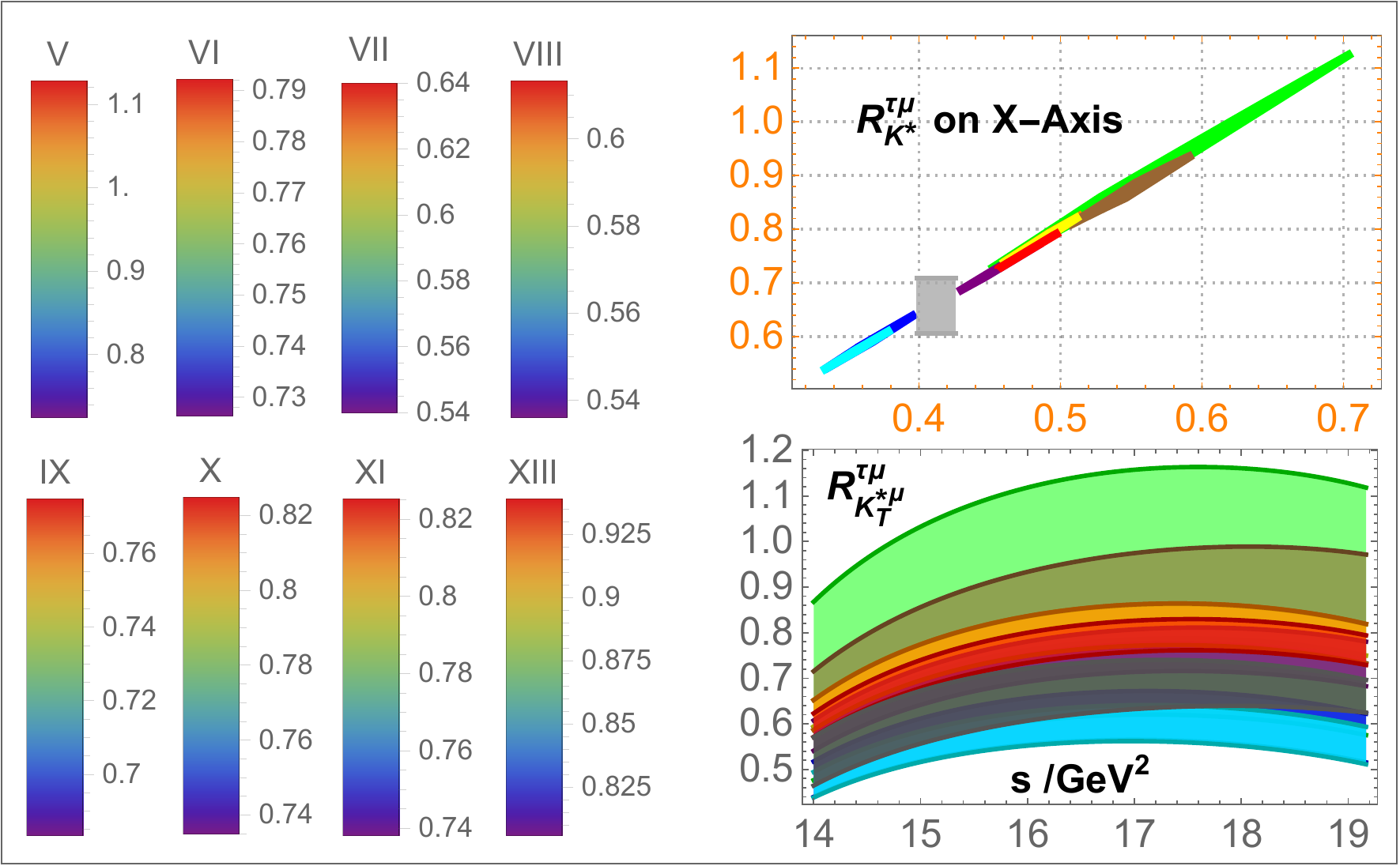}
\\
\includegraphics[width=3.5in,height=2.3in]{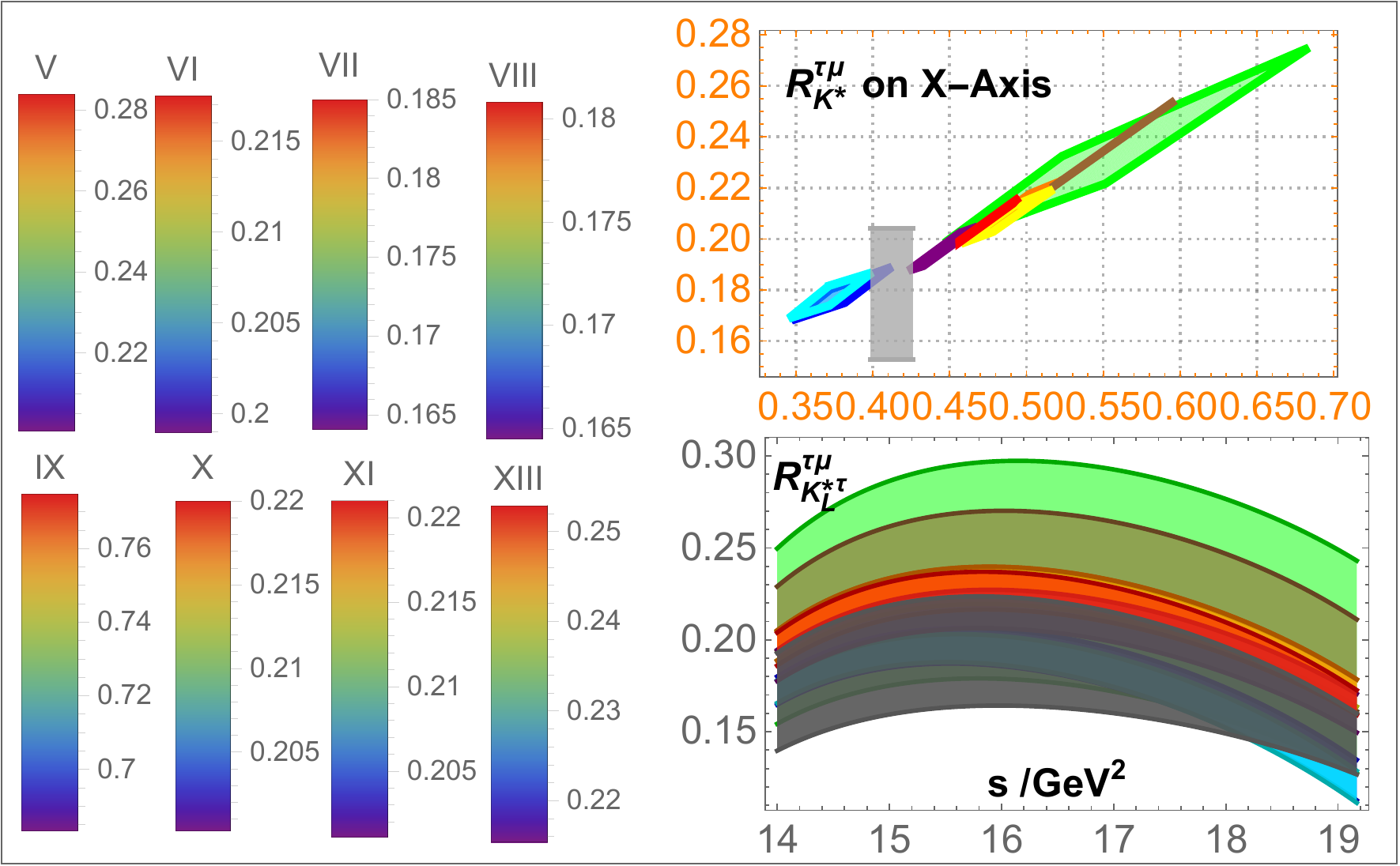}
\includegraphics[width=3.5in,height=2.3in]{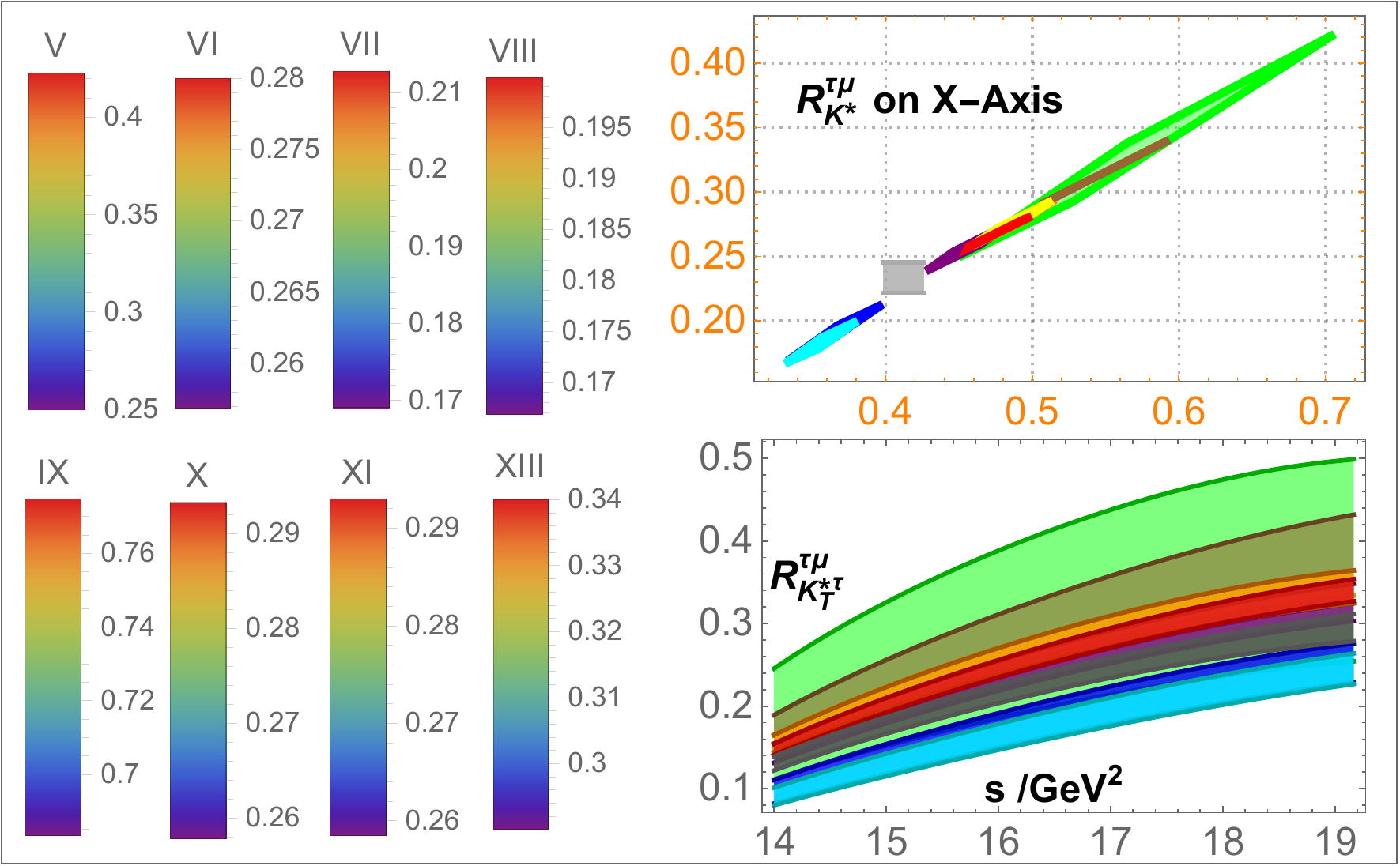}
\\
\includegraphics[width=3.5in,height=2.3in]{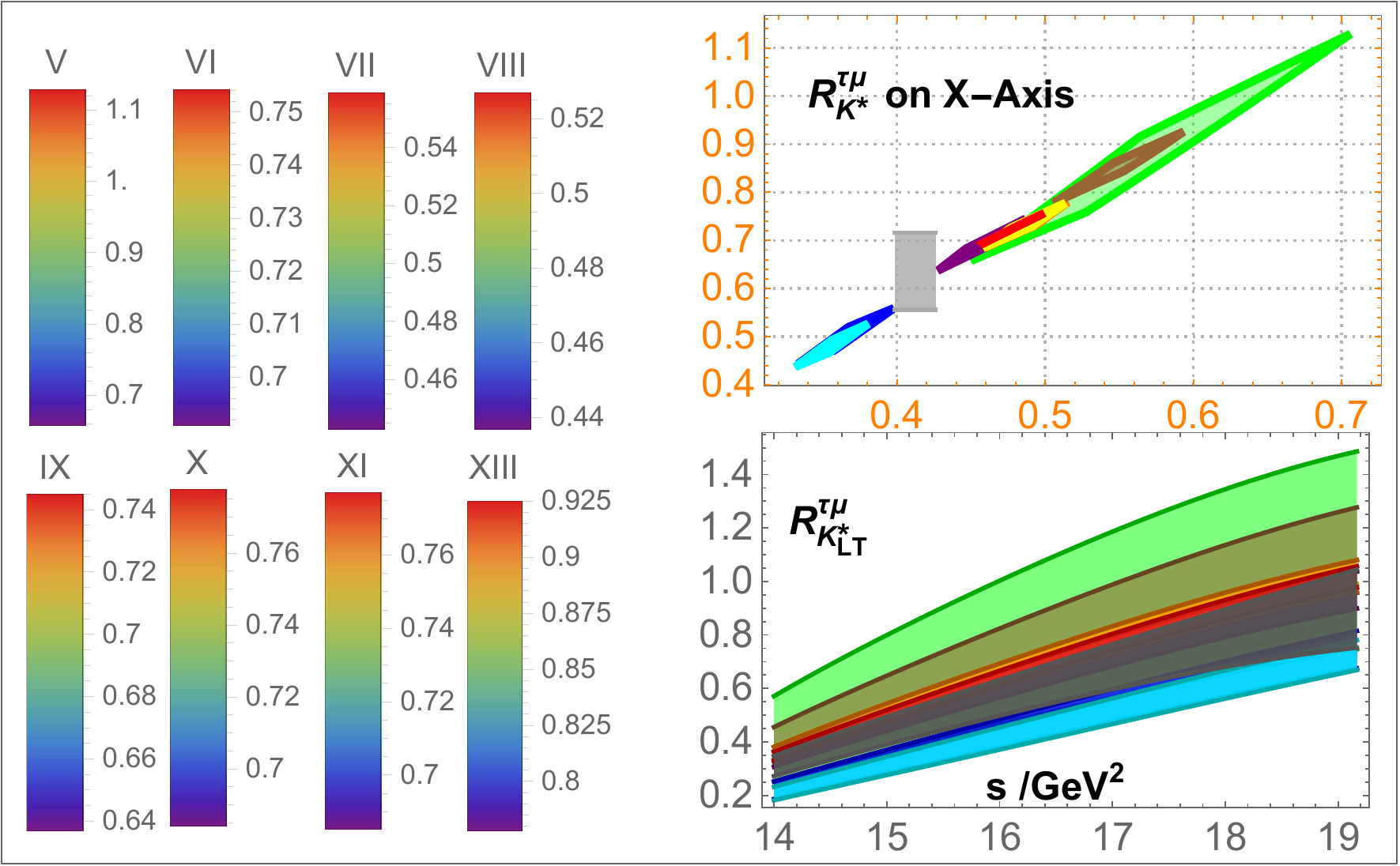}
\includegraphics[width=3.5in,height=2.3in]{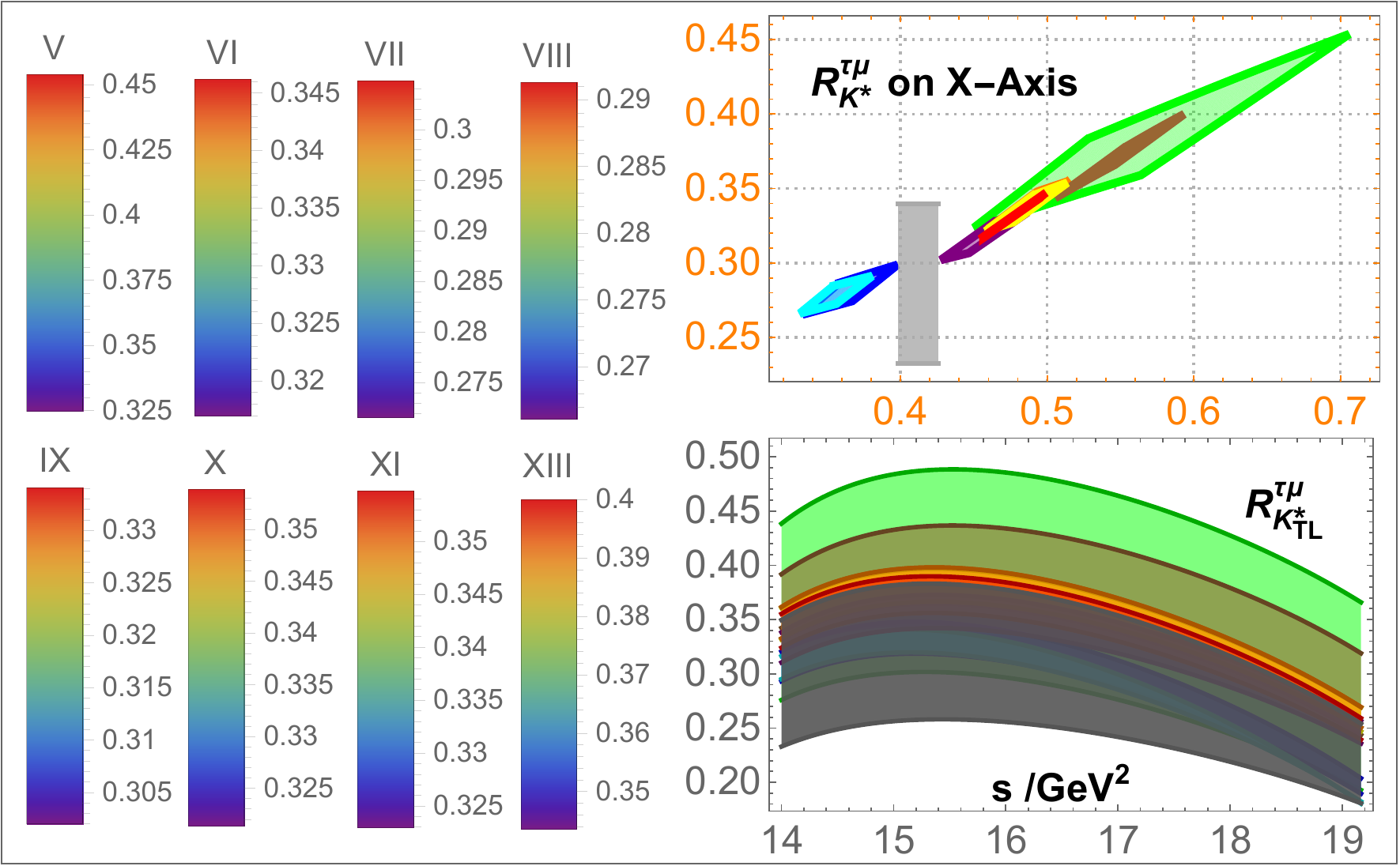}
\\
\caption{The  bar plots of the observables ($\mathcal{O}_{i}$) for different $D>1$ scenarios showing there magnitude. The plot on the top right corner in the inset shows the correlation among the  $\mathcal{O}_{i}$ and $R^{\tau \mu}_{K^{\ast}}$ taken along the x-axis. The bottom right corner shows the plots of the observables ($\mathcal{O}_{i}$) for the $q^2$ distribution in the SM as well as different NP scenarios where the width of the gray band represents the uncertainties due to form factors.}
\label{2Dob1a}
\end{figure}

\begin{figure}[H]
\centering
\includegraphics[width=3.5in,height=2.3in]{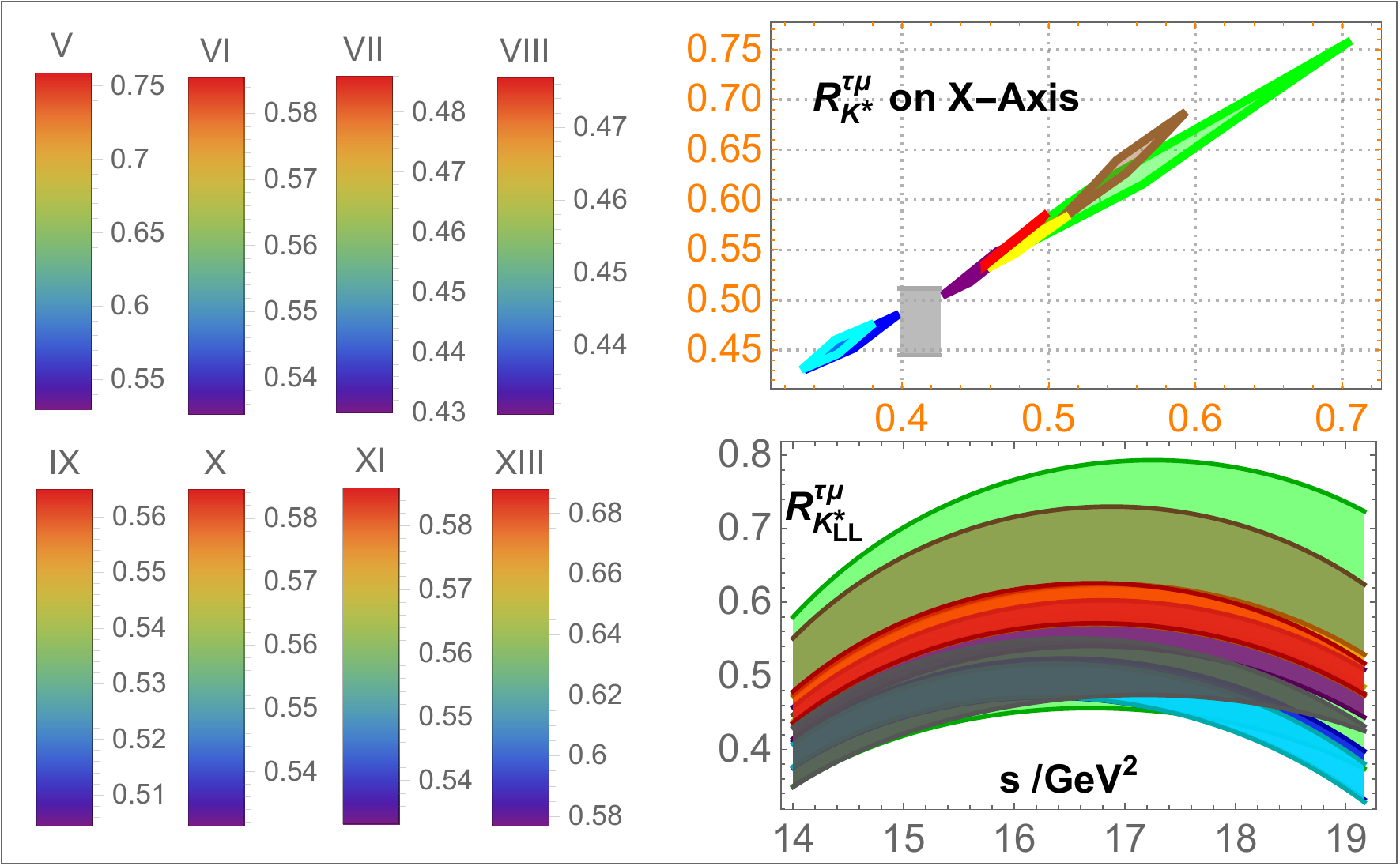}
\includegraphics[width=3.5in,height=2.3in]{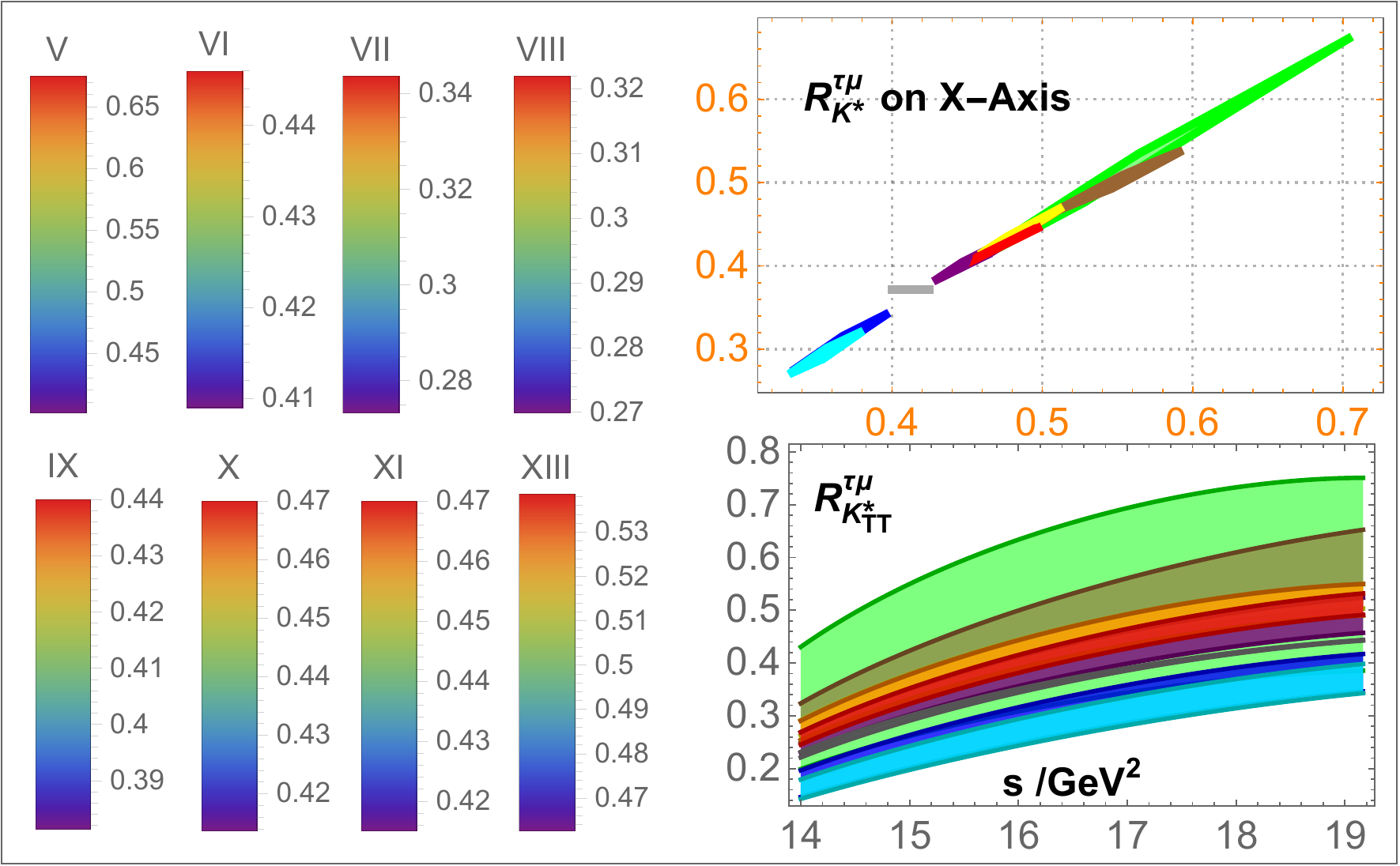}
\\
\includegraphics[width=3.5in,height=2.3in]{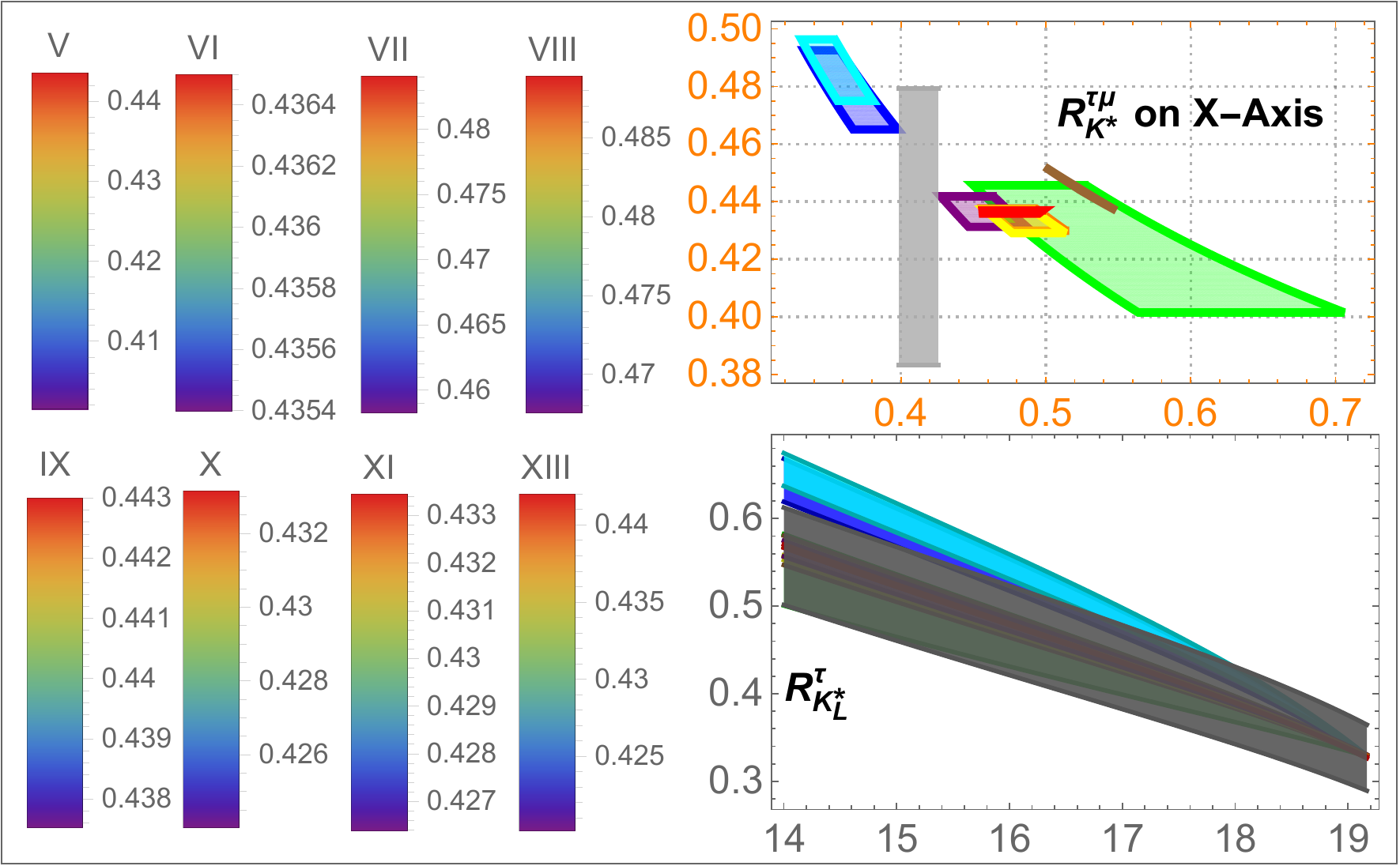}
\includegraphics[width=3.5in,height=2.3in]{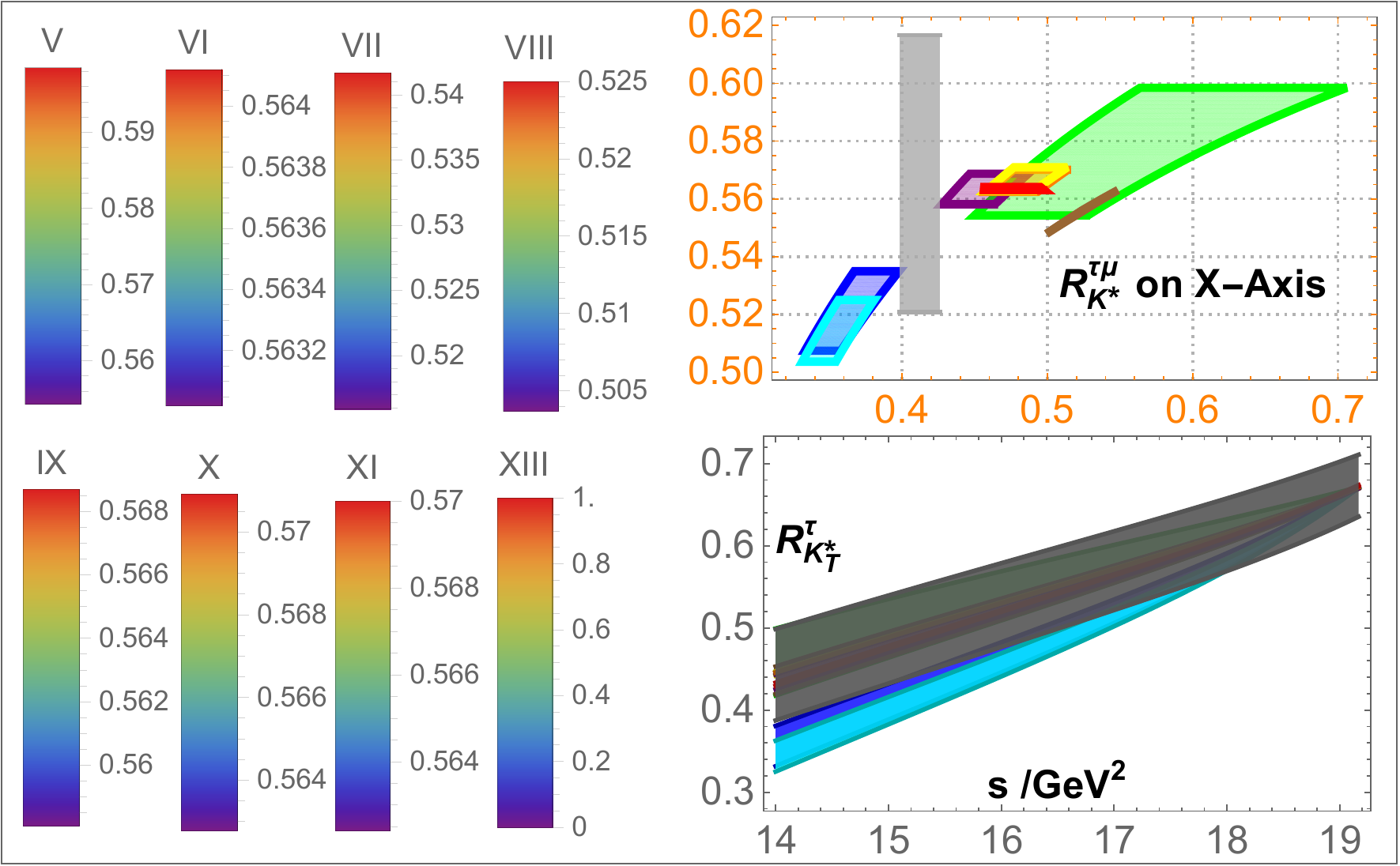}
\\
\includegraphics[width=3.5in,height=2.3in]{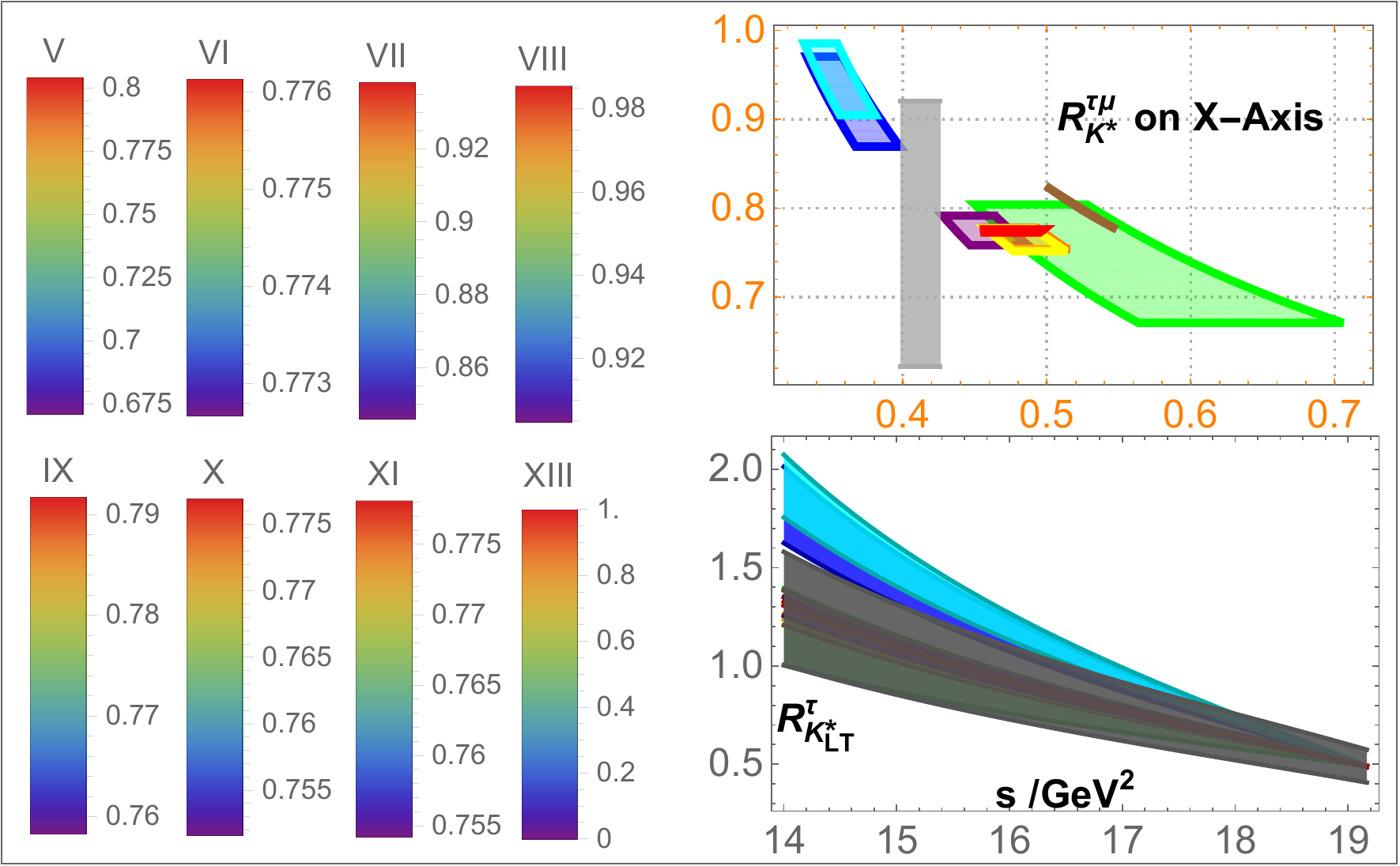}
\includegraphics[width=3.5in,height=2.3in]{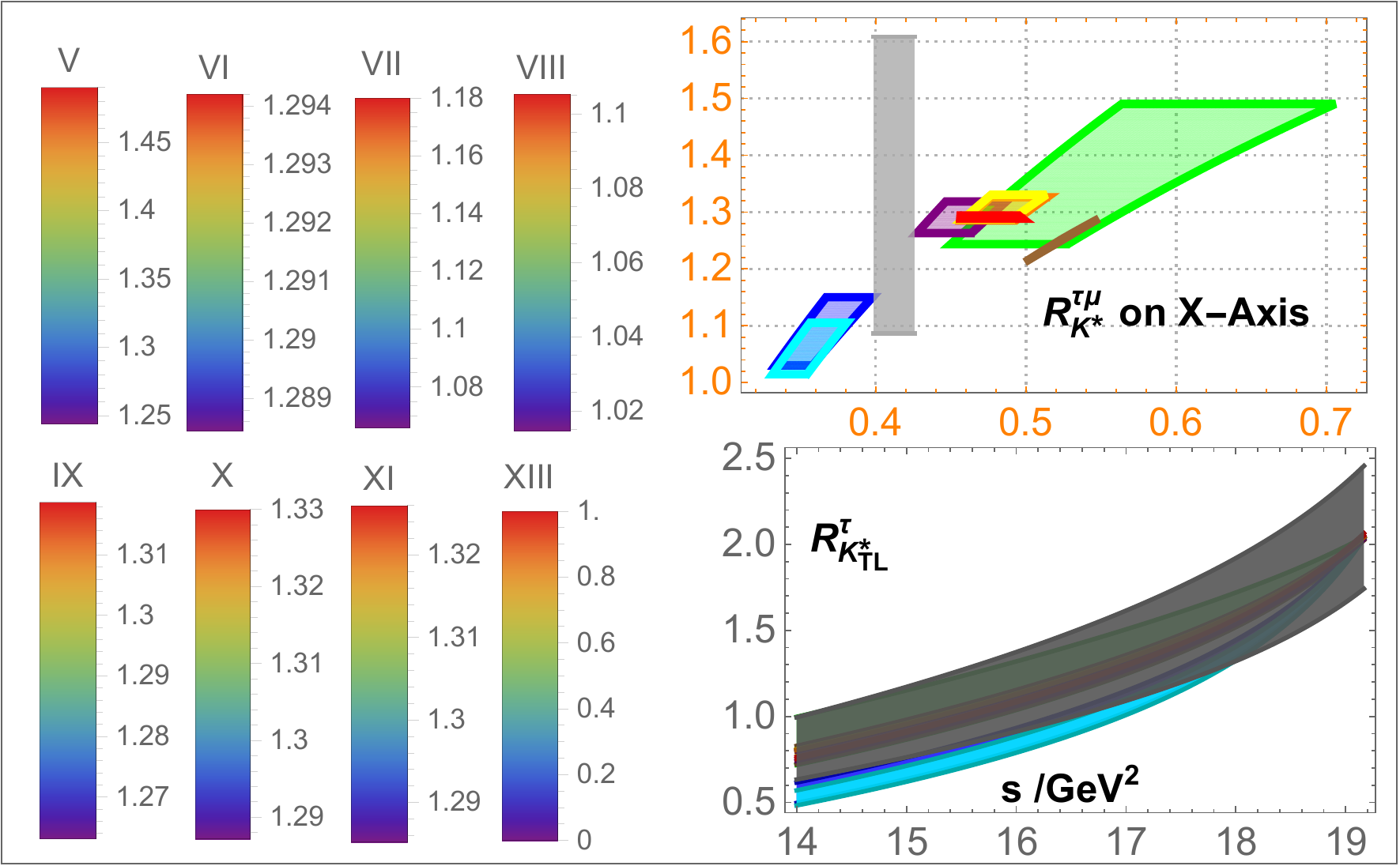}
\\
\caption{The  bar plots of the observables ($\mathcal{O}_{i}$) for different $D>1$ scenarios showing there magnitude. The plot on the top right corner in the inset shows the correlation among the  $\mathcal{O}_{i}$ and $R^{\tau \mu}_{K^{\ast}}$ taken along the x-axis. The bottom right corner shows the plots of the observables ($\mathcal{O}_{i}$) for the $q^2$ distribution in the SM as well as different NP scenarios where the width of the bands represents the uncertainties due to form factors.}
\label{2Dob1b}
\end{figure}

\begin{figure}[H]
\centering
\includegraphics[width=2.1in,height=1.5in]{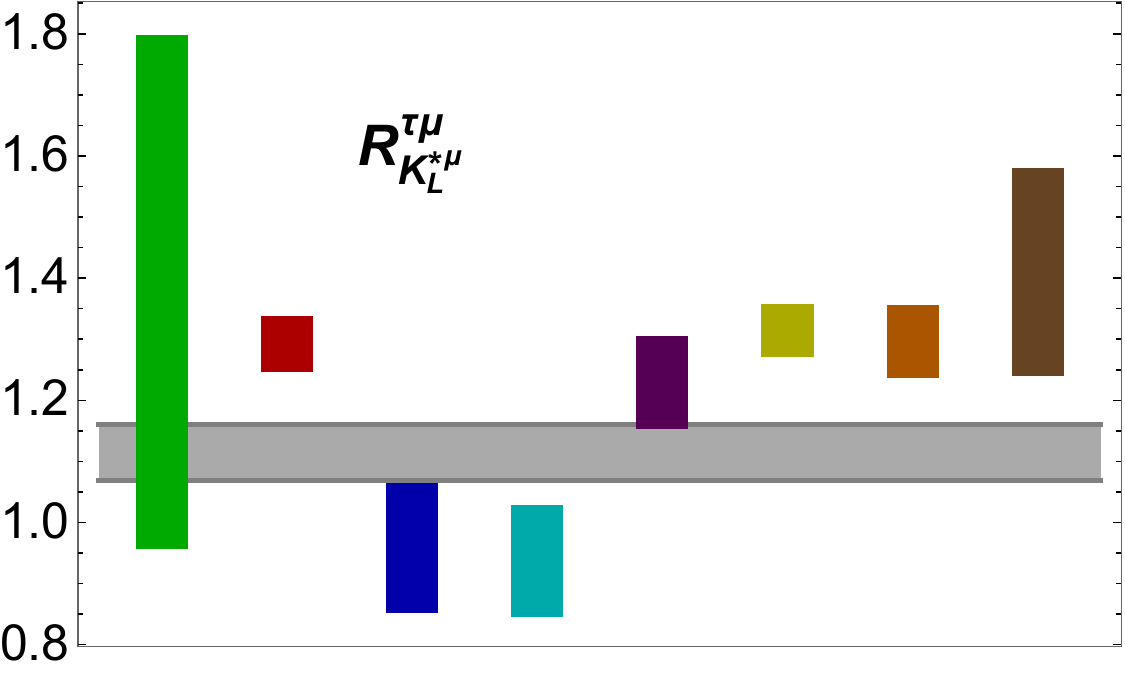}
\includegraphics[width=2.1in,height=1.5in]{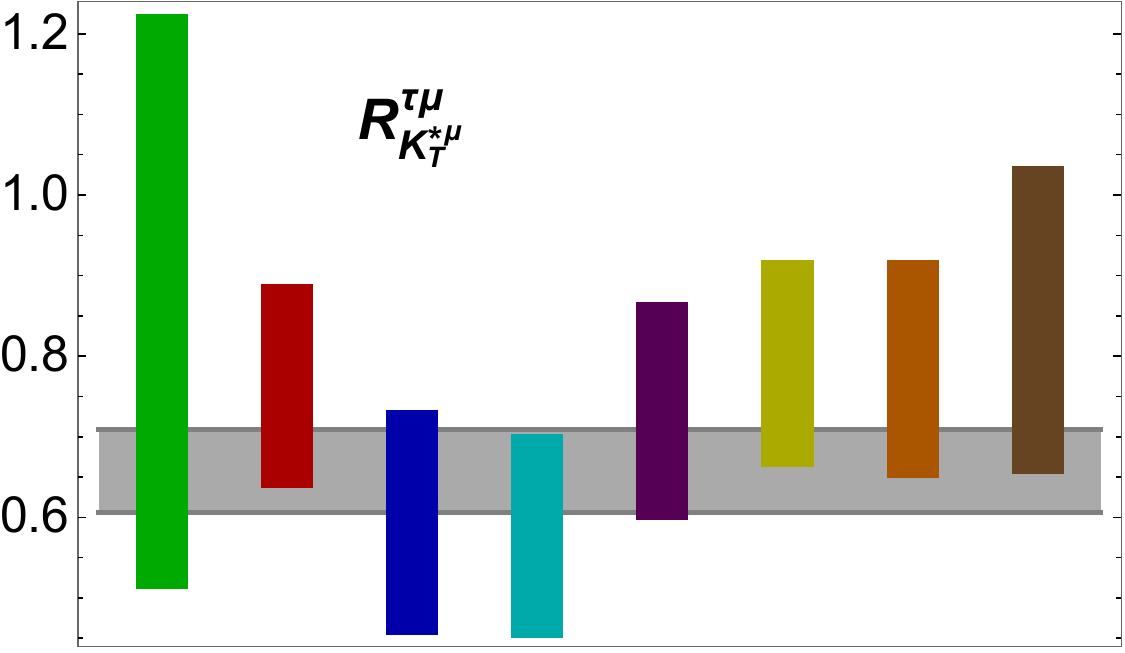}
\includegraphics[width=2.1in,height=1.5in]{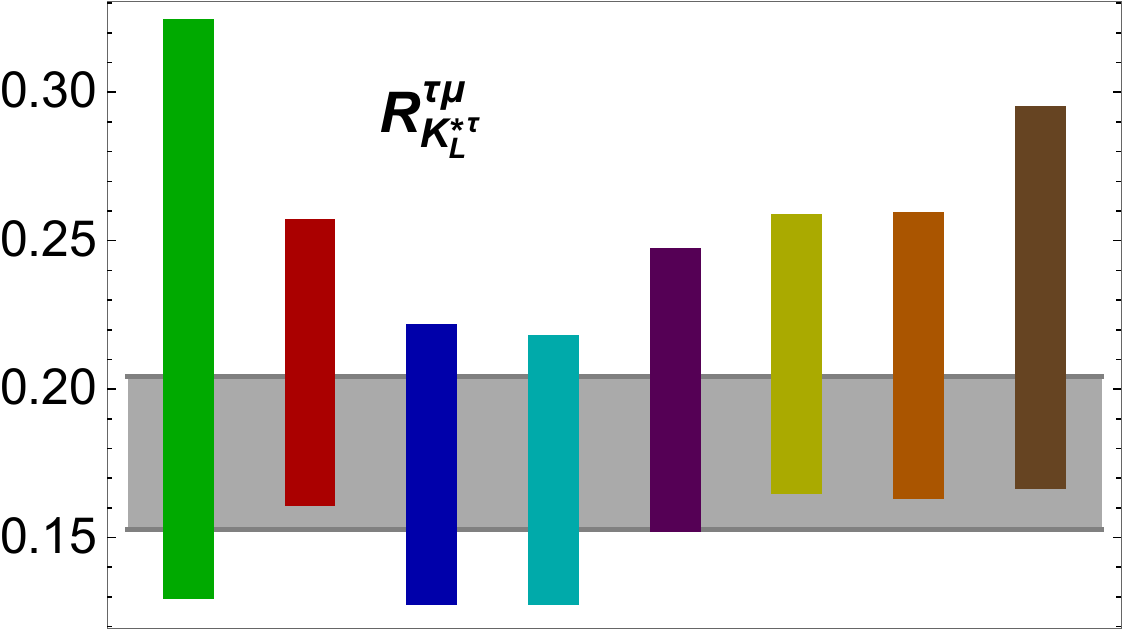}
\\
\includegraphics[width=2.1in,height=1.5in]{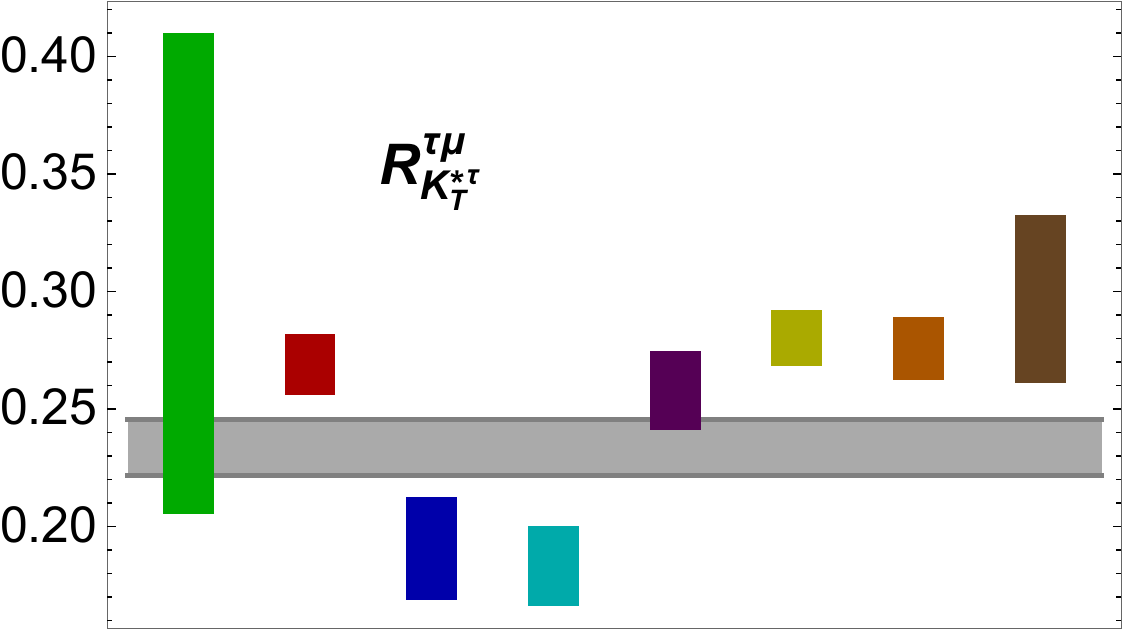}
\includegraphics[width=2.1in,height=1.5in]{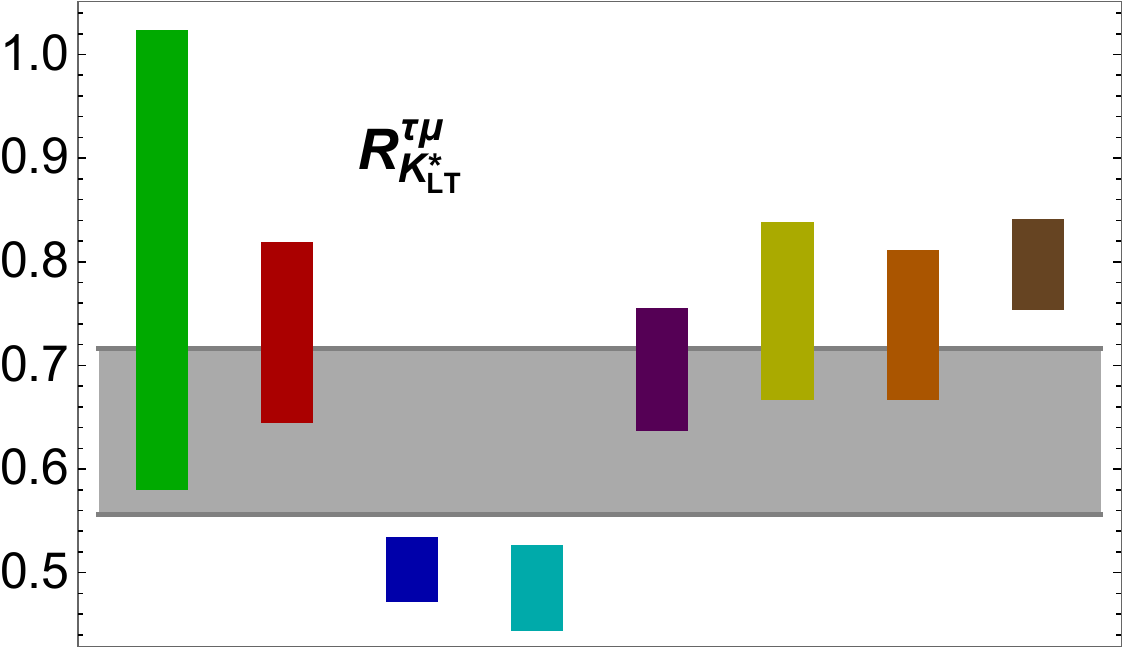}
\includegraphics[width=2.1in,height=1.5in]{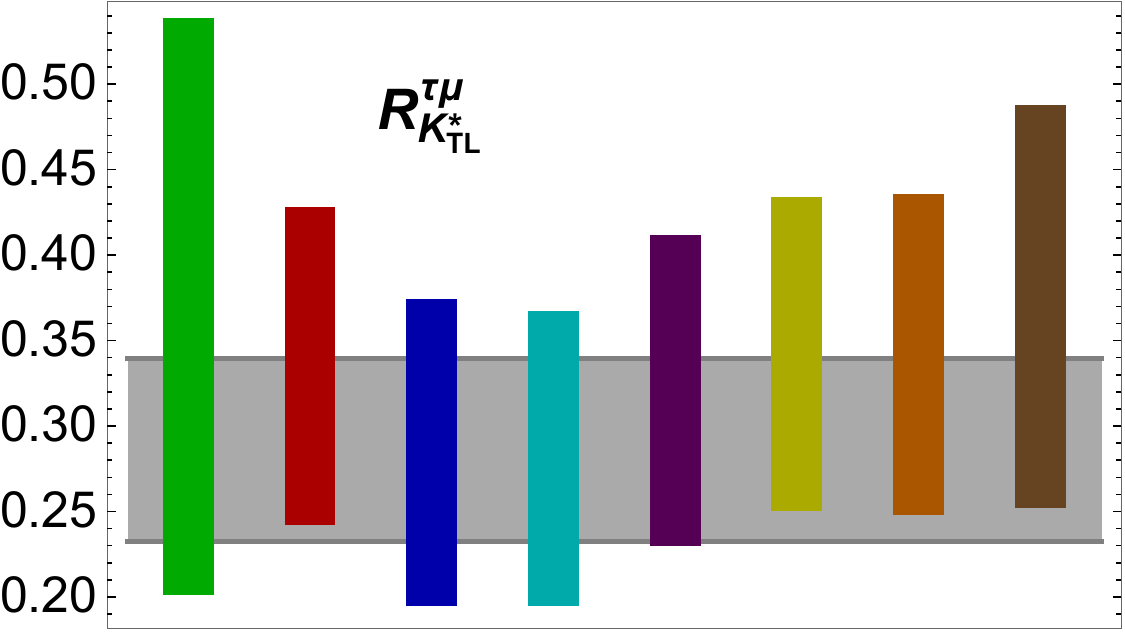}
\\
\includegraphics[width=2.1in,height=1.5in]{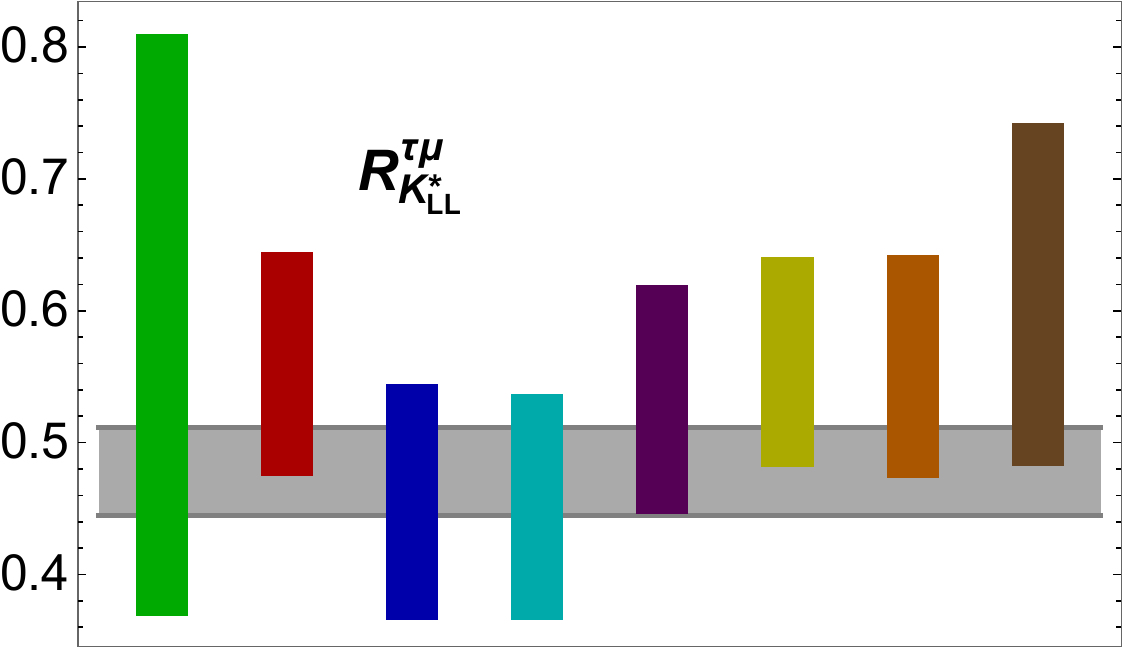}
\includegraphics[width=2.1in,height=1.5in]{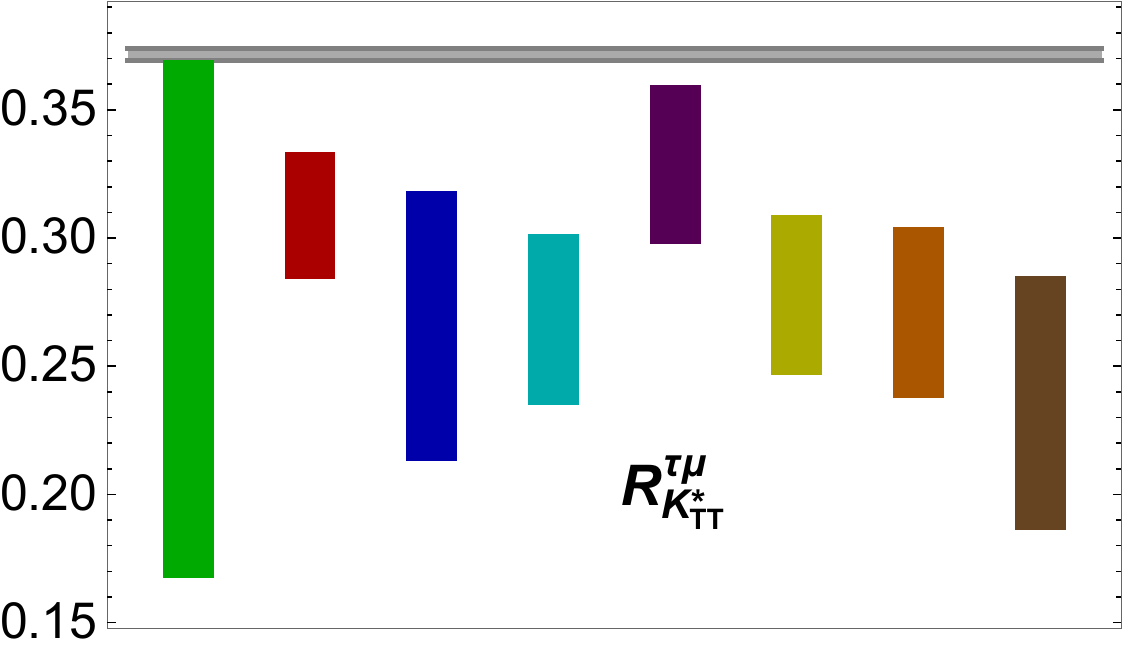}
\includegraphics[width=2.1in,height=1.5in]{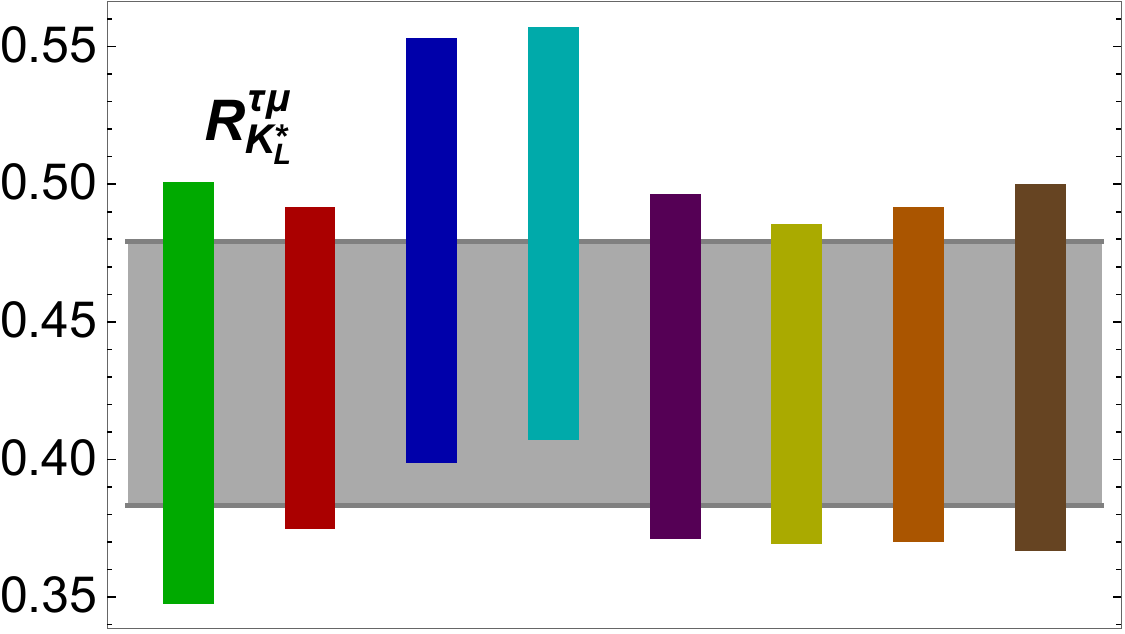}
\\
\includegraphics[width=2.1in,height=1.5in]{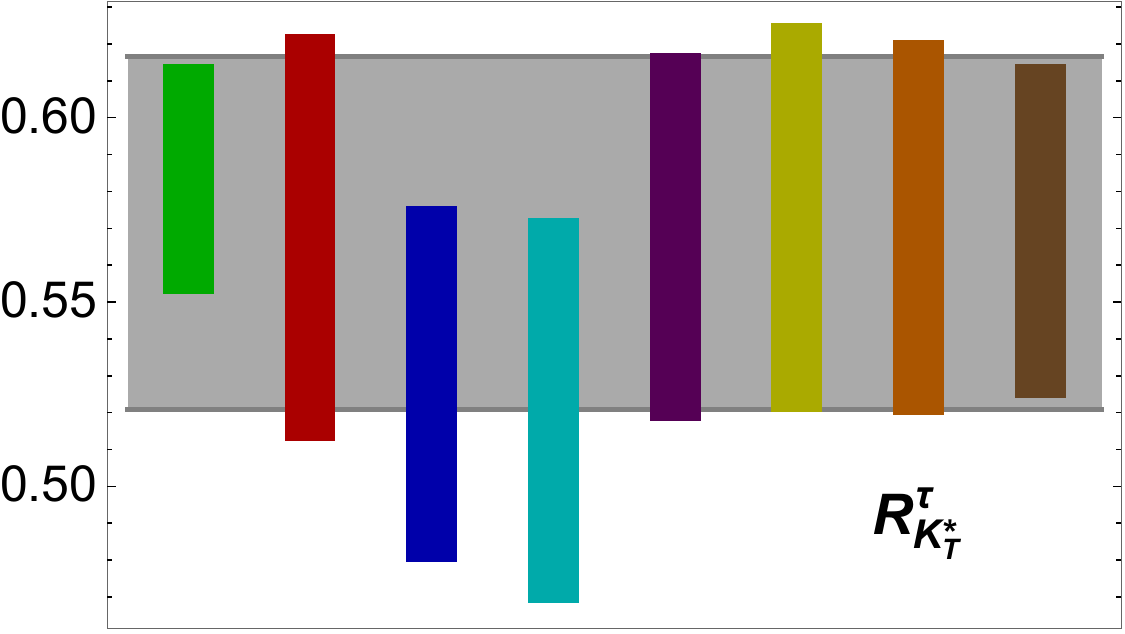}
\includegraphics[width=2.1in,height=1.5in]{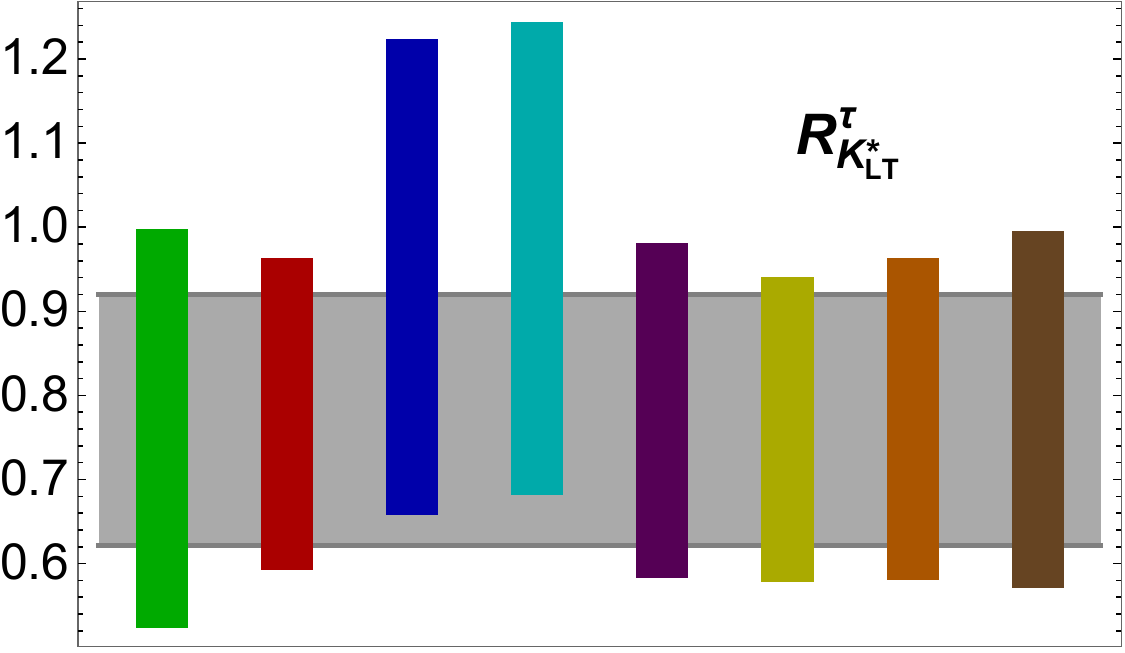}
\includegraphics[width=2.1in,height=1.5in]{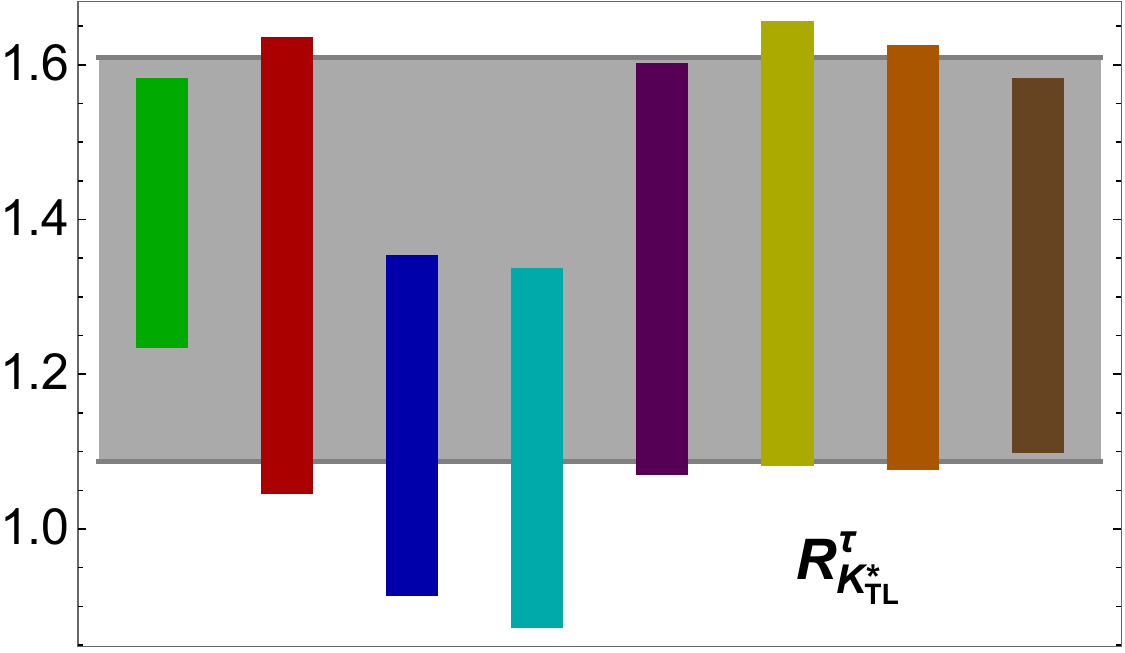}
\\
\caption{The bar plots of the $\mathcal{O}_{i}$ in different NP scenarios the gray band represent SM values with the uncertainties due to form factors while the bars show the magnitude of the  $\mathcal{O}_{i}$ in each scenario.}
\label{2Dob1c}
\end{figure}

The profiles of the above mentioned observables for the $D>1$ cases, \textit{i.e.}, for the scenarios mentioned in Table \ref{wc table2} are given in Figs. \ref{2Dob1}, \ref{2Dob1a}, \ref{2Dob1b} and 
Fig. \ref{2Dob1c}. Also, the the values after integrating over $s$ are given in Table \ref{TableVa}.

The first plot in Fig. \ref{2Dob1} represents the $1\sigma$ variations of $R_{K^*}^{\tau \mu}$ as a function of $s$ for different NP scenarios. The color coding for the various scenarios is shown in the top right inset showing by the single color bar plots for the same $1\sigma$ range. Here, we can see that all the NP scenarios are distinguishable from the SM and from each other, particularly for S-V and S-XIII, where the respective maximum values are $0.71$ and $0.63$ (c.f first row Table \ref{TableVa}).

The second plot of Fig. \ref{2Dob1} (top right panel) shows the density plot drawn for $R_{K^*}^{\tau \mu}$, whereas in the second row of Fig. \ref{2Dob1}, we have plotted the density profile for ($R_{K^*_{L}}^{\tau}$, $R_{K^*_{LT}}^{\tau}$) and ($R_{K^*_T}^{\tau }$,$R_{K^*_{TL}}^{\tau}$) against the available parametric space of different NP scenarios.  Here $x-$axis and $y-$axis show the available parametric space of WCs for all the considered NP scenarios. In these plots, the variation in the colors correspond to the variation in the magnitude of the these observables, and this is shown in the first plot of Fig. \ref{2Dob1} and also in  Figs. \ref{2Dob1a} and \ref{2Dob1b} by the multicolored bars. These density plots would be helpful to extract the precise parametric space of a particular NP scenario when the said observables will be measured precisely in future.

To make the analysis more clear, the variation of $R_{K^{*\mu}_{L}}^{\tau\mu}$ and $R_{K^{*\mu}_{T}}^{\tau\mu}$ is shown in first row of Fig. \ref{2Dob1a}. It can be seen that for S-V, S-VII, S-X, S-XI and S-XIII the predicted values are greater then SM predictions while for S-VI and S-VIII, the values are smaller. Once again, the scenario V is the one that has shown the deviations from the SM results.

 From the second row of Fig. \ref{2Dob1a} we can see that the  $R_{K^{*\tau}_{L}}^{\tau\mu}$ and $R_{K^{*\tau}_{T}}^{\tau\mu}$ has the same profile with $s$ as  $R_{K^{*}_{LL}}^{\tau\mu}$ and $R_{K^{*}_{TT}}^{\tau\mu}$ drawn in the first row of Fig. \ref{2Dob1b}. We can see that the values of $R_{K^{*\tau}_{L}}^{\tau\mu}$ and $R_{K^{*}_{LL}}^{\tau\mu}$ decreases with increasing $s$, where as $R_{K^{*\tau}_{T}}^{\tau\mu}$ and $R_{K^{*}_{TT}}^{\tau\mu}$ increses with $s$. This value is significantly large to be measured in some ongoing experiments like LHCb.

 In the last row of Fig. \ref{2Dob1a}, we have drawn the ratios $R_{K^{*}_{LT}}^{\tau\mu}$ and $R_{K^{*}_{TL}}^{\tau\mu}$ with $s$. We can see that the value of first increases with $s$, where as the second decreses by the same increment making their sum equal to $1$. The maximum value is for the scenarios V, whereas the minimum is in S-XIII.

 Fig. \ref{2Dob1b} shows the ratio $R^{\tau}_{K^*_{L,T}}$ and $R^{\tau}_{K^*_{(LT,TL)}}$ in last two rows. The observable $R^{\tau}_{K^*_{L,T}}$ and $R^{\tau}_{K^*_{T}}$ are just the longitudinal and transverse polarization fractions of the $K^*$ meson, discussed in the literature.  It is observed that except S-VI and S-VIII all other scenarios are precluded by SM uncertainties. These two, however, are also overlapping and can not be distinguish from one another. Now, in the  insets of Fig. \ref{2Dob1a} and Fig. \ref{2Dob1b} the correlation plots for different LFU violations rations are drawn. One can see that this enables us to see how different $R^{\tau\mu }_i$'s correlates with the $R^{\tau \mu}_{K^*}$ making them useful for the future experimental studies.

Finally, Fig. \ref{2Dob1c} shows the effect of the $1 \sigma$ values of new WCs along with SM predictions in bar chart form. It can bee seen from these plots that the LFUV ratios $R^{\tau \mu}_{K^{*\mu}_L}$, $R^{\tau \mu}_{K^{*\mu}_T}$, $R^{\tau \mu}_{K^{*\tau}_T}$, $R^{\tau \mu}_{K^*_{LL}}$, $R^{\tau \mu}_{K^*_{TT}}$ and $R^{\tau \mu}_{K^*_{TL}}$ show significant deviations from SM results, making them useful to hunt for the NP. The behaviour of $R^{\tau \mu}_{K^{*\tau}_{L}}$ and $R^{\tau \mu}_{K_{LT}*}$ is somewhat similar. For other observables like $R^{\tau}_{K^{*}_T}$ and $R^{\tau}_{K^{*}_{LT}}$ only S-VI and S-VIII lie outside the uncertainties band of the SM. While for $R^{\tau }_{K^{*}_{LT}}$ and $R^{\tau }_{K^{*}_L}$, the maximum deviations from the SM results arise for  S-V, S-VI and S-VIII. It can be noticed that all these $R^{\tau\mu }_i$'s i.e, $R^{\mu }_{K^{*}_L}$, $R^{\tau \mu}_{K^{*\tau}_T}$ and $R^{\tau\mu }_{K^{*}_{TT}}$ are less masked by the SM uncertainties, and also useful to disentangle the NP arises due to different beyond SM scenarios. 

To Summarize: the experimental data on the decays of $B$ mesons have revealed discrepancies from the predictions made by the SM, particularly in processes involving $\tau$ and $\mu$ leptons in the final state. To delve into these discrepancies, a variety of  different physical observables related to lepton universality $R^{\tau \mu}_{i}$  where the index $i$ represents different types of final state meson polarizations i.e, $i=K^*$, $K^*_L$, $K^*_T$, $K^*_{LT}$, $K^*_{TL}$, $K^*_{LL}$, $K^*_{TT}$ 
are studied.  Each of these observables provides unique insights into the decay processes and their potential deviations from SM predictions. To illustrate the impact of NP on the amplitude of these observables, we calculated their numerical values within the $q^2=14-s_{max}$ bin and showed them through the bar plots. Our findings reveal that these polarized observables are not only sensitive to the values of the NP WCs but are also useful tool for distinguishing among various NP scenarios.
Therefore, the aforementioned analysis demonstrates that the accurate measurements of the polarized and unpolarized ratios considered in this study will not only provide insights into potential NP but also be helpful to disentangle the tension among different beyond SM scenarios.

\begin{table*}[tbh]
\caption{\label{tab:table3}The numerical values of $\mathcal{R}^{\tau\mu(\tau)}_i$ in the $14\leq s\leq s_{max} \text{GeV}^2$ bin for D$>1$ NP scenarios.}
\renewcommand{\arraystretch}{1}
\centering
\scalebox{1}{
\begin{tabular}{c|cccccccc}
\hline\hline &&&&&&&&\\[-0.9em]
  $\mathcal{R}^{\tau\mu(\tau)}_{i}$&S-V&S-VI&S-VII&S-VIII&S-IX&S-X&S-XI&S-XIII  \\ 
 &&&&&&&&\\[-0.9em]\cline{1-9}\\[-0.9em]
 $\mathcal{R}_{K^*}^{\tau\mu}$ &$(0.31-0.71)$&$(0.44-0.52)$  & $(0.31-0.42)$ &$(0.31-0.40)$ &$(0.41-0.50)$ &$(0.45-0.53)$ &$(0.44-0.53)$&$(0.45-0.61)$   \\
&&&&&&&&\\[-0.9em]\cline{1-9}\\[-0.9em]
 $\mathcal{R}_{K_L^{*\tau}}^{\tau\mu }$ &$(0.14-0.31)$&$(0.17-0.25)$&$(0.14-0.22)$&$(0.14-0.22)$&$(0.16-0.23)$&$(0.17-0.25)$&$(0.17-0.25)$&$(0.18-0.29)$   \\
&&&&&&&&\\[-0.9em]\cline{1-9}\\[-0.9em]
$\mathcal{R}_{K_T^{*\tau}}^{\tau\mu }$ &$(0.22-0.40)$&$(0.270-0.266)$&$(0.18-0.20)$&$(0.19-0.18)$&$(0.25-0.26)$&$(0.279-0.282)$&$(0.27-0.28)$&$(0.27-0.32)$   \\
&&&&&&&&\\[-0.9em]\cline{1-9}\\[-0.9em]
$\mathcal{R}_{K_L^{*\mu}}^{\tau\mu }$  &$(1.00-1.75)$&$(1.289-1.294)$&$(0.94-1.06)$&$(0.90-1.02)$&$(1.00-1.11)$&$(1.25-1.38)$&$(1.31-1.28)$&$(1.28-1.54)$   \\
&&&&&&&&\\[-0.9em]\cline{1-9}\\[-0.9em]
$\mathcal{R}_{K_T^{*\mu}}^{\tau\mu }$  &$(0.55-1.18)$&$(0.67-0.86)$&$(0.49-0.70)$&$(0.48-0.67)$&$(0.63-0.83)$&$(0.70-0.89)$&$(0.68-0.89)$&$(0.69-1.00)$   \\
&&&&&&&&\\[-0.9em]\cline{1-9}\\[-0.9em]
   $\mathcal{R}_{K_{LL}^*}^{\tau\mu }$  &$(0.39-0.78)$&$(0.49-0.62)$&$(0.39-0.53)$&$(0.38-0.52)$&$(0.47-0.60)$&$(0.50-0.62)$&$(0.49-0.62)$&$(0.50,0.72)$  \\
&&&&&&&&\\[-0.9em]\cline{1-9}\\[-0.9em]
  $\mathcal{R}_{K_{TT}^*}^{\tau\mu }$ &$(0.17-0.36)$&$(0.29-0.32)$&$(0.22-0.31)$&$(0.24-0.29)$&$(0.31-0.35)$&$(0.26-0.30)$&$(0.25-0.29)$&$(0.20-0.27)$    \\
  &&&&&&&&\\[-0.9em]\cline{1-9}\\[-0.9em]
 $\mathcal{R}_{K_{LT}^*}^{\tau\mu }$ &$(0.61-0.99)$&$(0.67-0.79)$&$(0.50-0.51)$&$(0.47-0.50)$&$(0.66-0.73)$&$(0.69-0.81)$&$(0.69-0.79)$&$(0.78-0.81)$\\ 
  &&&&&&&&\\[-0.9em]\cline{1-9}\\[-0.9em]
  $\mathcal{R}_{K_{TL}^*}^{\tau\mu }$ &$(0.21-0.52)$&$(0.26-0.41)$&$(0.21-0.36)$&$(0.21-0.35)$&$(0.25-0.40)$&$(0.27-0.42)$&$(0.26-0.42)$&$(0.27-0.47)$  \\ 
 &&&&&&&&\\[-0.9em]\cline{1-9}\\[-0.9em]
 $\mathcal{R}_{K^*_L}^{\tau }$ &$(0.36-0.49)$&$(0.38-0.48)$&$(0.41-0.54)$&$(0.42-0.55)$&$(0.38-0.49)$&$(0.38-0.48)$&$(0.38-0.48)$&$(0.38-0.49)$ \\
&&&&&&&&\\[-0.9em]\cline{1-9}\\[-0.9em]
 $\mathcal{R}_{K^*_T}^{\tau }$ &$(0.56-0.61)$&$(0.52-0.62)$&$(0.49-0.57)$&$(0.48-0.57)$&$(0.52-0.61)$&$(0.53-0.62)$&$(0.53-0.61)$&$(0.53,0.61)$   \\
&&&&&&&&\\[-0.9em]\cline{1-9}\\[-0.9em]
$\mathcal{R}_{K^*_{TL}}^{\tau }$ &$(1.27-1.55)$&$(1.08-1.60)$&$(0.95-1.32)$&$(0.91-1.30)$&$(1.10-1.57)$&$(1.11-1.62)$&$(1.11-1.59)$&$(1.13-1.54)$   \\
 \hline \hline   \end{tabular}}
  \label{TableVa}
\end{table*}

\section*{Acknowledgments}

I.A. acknowledges the kind hospitality and financial support of CERN theory division during his stay at CERN in the summer of 2024.

\end{document}